\documentclass{article}

\PassOptionsToPackage{numbers,square}{natbib}
\usepackage[eandd, preprint]{neurips_2026}

\usepackage[utf8]{inputenc}
\usepackage[T1]{fontenc}
\usepackage{hyperref}
\usepackage{url}
\usepackage{booktabs}
\usepackage{amsfonts}
\usepackage{amsmath}
\usepackage{amssymb}
\usepackage{dsfont}
\usepackage[capitalize,nameinlink]{cleveref}
\crefname{appendix}{Appendix}{Appendices}
\Crefname{appendix}{Appendix}{Appendices}
\usepackage{nicefrac}
\usepackage{microtype}
\usepackage{xcolor}
\usepackage{graphicx}
\usepackage{multirow}
\usepackage{subcaption}
\usepackage{algorithm}
\usepackage{algpseudocode}
\usepackage{pifont}
\newcommand{\cmark}{\ding{51}}
\newcommand{\xmark}{\ding{55}}

\setcitestyle{numbers}

\definecolor{mydarkblue}{rgb}{0,0.08,0.45}

\definecolor{algcommentcolor}{rgb}{0.0, 0.35, 0.65}
\algrenewcommand{\algorithmicrequire}{\textbf{Input:}}
\algrenewcommand{\algorithmicensure}{\textbf{Output:}}
\hypersetup{%
    pdfborder=0 0 0,
    pdfpagemode=UseNone,
    colorlinks=true,
    linkcolor=mydarkblue,
    citecolor=mydarkblue,
    filecolor=mydarkblue,
    urlcolor=mydarkblue,
    pdfview=FitH
}

\makeatletter
  \newif\ifpreprint
  \if@preprint\preprinttrue\fi
\makeatother

\newcommand{\preprintonly}[1]{\ifpreprint#1\fi}

\newif\ifdraft
\draftfalse

\ifdraft
  \newcommand{\todo}[1]{\textcolor{red}{\textbf{[#1]}}}

\else
  \newcommand{\todo}[1]{}

\fi

\newcommand{\nToolCallQwenCoder}{5}
\newcommand{\nToolCallQwenCoderTotal}{20}

\newcommand{\Nretries}{2}         %
\newcommand{\Nharnessretries}{10} %

\newcommand{\bname}{StaminaBench}

\title{\bname{}: Stress-Testing Coding Agents over \\ 100 Interaction Turns}

\author{%
  Vlad Sobal\thanks{Correspondence to: \texttt{vsobal@amazon.com}} \quad
  Shuo Yang \quad
  Yuting Zhang \quad
  Wei Xia \quad
  Stefano Soatto \\[0.5em]
  AWS Agentic AI \\
}

\begin{document}

\maketitle

\begin{abstract}
    We introduce \textbf{\bname{}}, a benchmark that measures the \emph{stamina} of coding agents: how many consecutive interaction turns (change requests) they can handle before failing. Unlike the prevailing \emph{fraction-of-tasks-solved} metric, this matches real vibe-coding where sessions run dozens or hundreds of turns. In \bname{}, agents implement a REST API server and modify it across a tunable number of procedurally generated follow-up change requests---100 in our experiments, resulting in codebases of up to 6,000 lines. Tests are generated fully programmatically without LLM involvement, ensuring reproducibility and reliability; change sequences are drawn from either a hardcoded or LLM-driven sampler, both constrained to a structured action space to ensure changes are valid. The agent and the server run in an isolated environment and communicate with the benchmark through HTTP, making testing fully black-box and language-agnostic.
    We evaluate six agent harnesses paired with seven open-source LLMs across 20 scenarios of 100 turns each and find that: (1)~all the tested models fail within 5--6 turns, confirming that vibe-coding-style programming without thorough testing produces bugs; (2)~passing test feedback back to the agent and allowing it to retry improves passed turn count by up to \textbf{12$\times$}; and (3)~a good harness is required for strong performance: stronger models exhibit up to a \textbf{6$\times$} gap between their best and worst harness, while weaker models fail with any harness. We release the benchmark and the generated tasks to enable further research into multi-turn coding agent behavior. \preprintonly{\\ Benchmark code and data: \href{https://github.com/amazon-science/StaminaBench}{github.com/amazon-science/StaminaBench}.}
\end{abstract}

\section{Introduction}

Software development is highly iterative, with engineers building first a foundation prototype, then continuously refining and refactoring the existing code. This is also reflected in how popular coding agents \citep{claudecode, codex, geminicli, cursor} are used by engineers -- the interactions are rarely one-off, and instead span dozens or hundreds of interaction turns, where the developer may ask to build a certain feature, then modify it, then add related functionality, and so on \citep{wang2025aiagenticprogrammingsurvey}.
Nevertheless, most popular coding benchmarks like SWE-Bench~\citep{jimenez2024swebenchlanguagemodelsresolve}, SWE-BenchPro~\citep{deng2025swebenchproaiagents}, and TerminalBench~\citep{merrill2026terminalbenchbenchmarkingagentshard} all frame evaluation around a single, self-contained task -- the agent is given one problem to solve, it produces the answer, and is not asked to extend that work over subsequent requests.

Long-horizon, multi-turn\footnote{In this work, we use ``multi-turn'' to denote multiple interaction turns between the user and the agent, not the internal tool-use steps the agent takes within a single user request.} coding places qualitatively different demands on agents \citep{deng2026evoclawevaluatingaiagents, wu2025frontalkbenchmarkingfrontenddevelopment, wang2026codeflowbenchmultiturniterativebenchmark}. The agent must maintain a coherent model of an evolving codebase, apply changes that are consistent with prior instructions and modifications, and debug errors in the context of an accumulated history. As the conversation grows, the agent must manage increasing complexity within finite context windows, deciding what to focus on \citep{laban2025llmslostmultiturnconversation}, and what keep when compressing context \citep{packer2024memgptllmsoperatingsystems,jiang2023llmlinguacompressingpromptsaccelerated}. Multi-turn setting brings challenges that are orthogonal to single-turn abilities in coding \citep{kwa2026measuringaiabilitycomplete}.

\begin{figure}[t]
\centering
\includegraphics[width=0.99\linewidth]{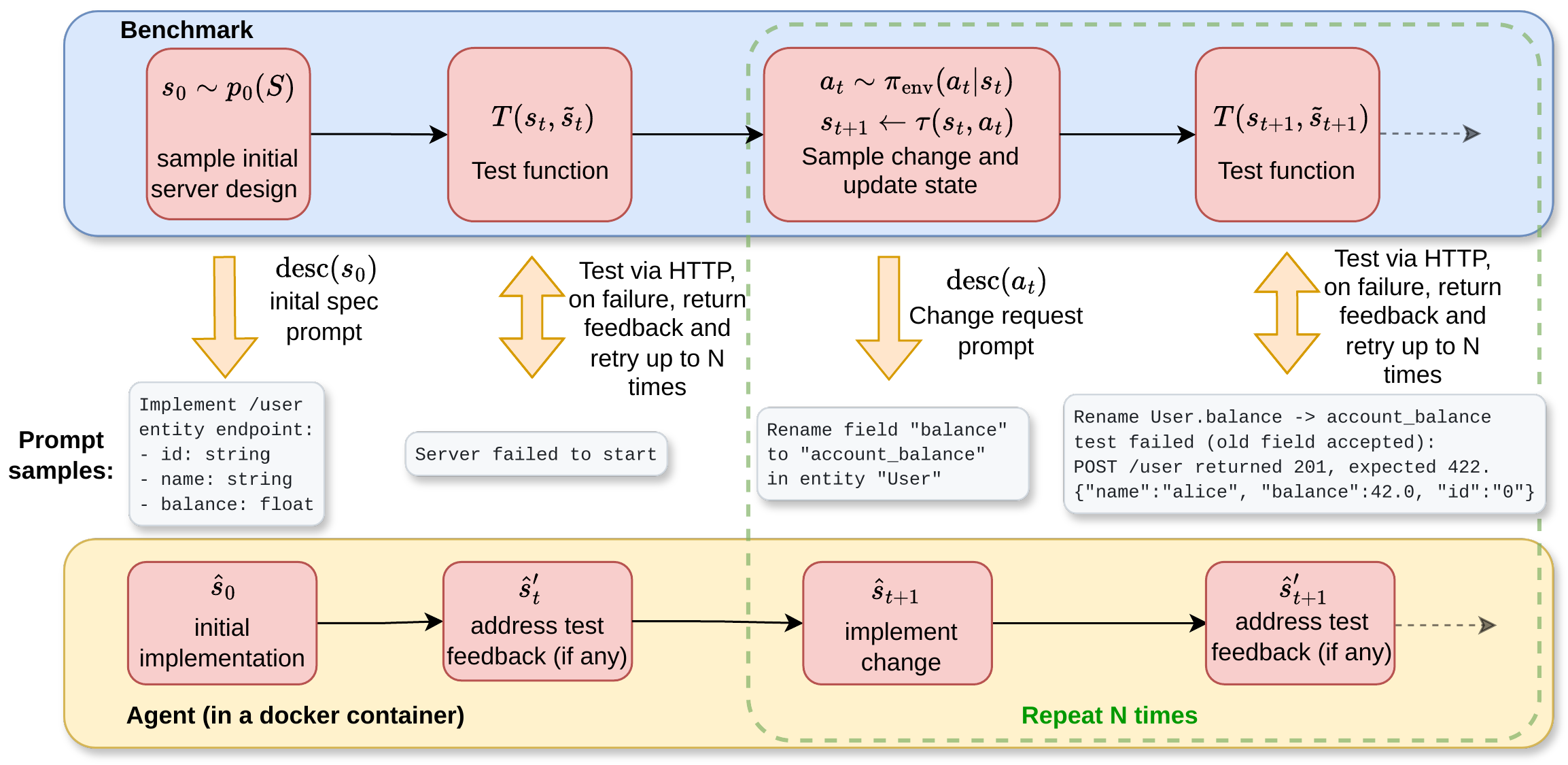}
\caption{Overview of \bname{}. The benchmark (blue) iteratively samples changes via $\pi_{\mathrm{env}}$ and advances the reference state. The agent (yellow) operates inside an isolated Docker container, receiving only NL descriptions and test feedback, and has to track the reference state. The full loop is formalized in \cref{alg:eval}.}
\label{fig:overview}
\end{figure}

However, creating multi-turn long-horizon benchmarks is highly complicated. While single-turn tasks can be naturally derived from pull requests of well-maintained repositories, deriving multiple turns this way is difficult due to the drift between the ground truth implementation and the agent's implementation. Both may be correct, but small differences accumulate, and make the ground truth tests inapplicable to the new code. However, testing is much simpler if there exists a clear interface that can be thoroughly tested while treating the implementation as a black box. Some existing works have already explored interface-based benchmarking~\citep{orlanski2026slopcodebenchbenchmarkingcodingagents}, but the number of user interaction turns still remains very low and nowhere near the number seen in real-world agent-assisted coding.

We introduce \textbf{\bname{}} to stress-test these capabilities. Each episode asks an agent to implement a REST (Representational State Transfer) API server (see \cref{app:rest-primer} for a brief description of the concept) from a generated specification and then iteratively modify it through a sequence of schema changes---entity additions, deletions, renames, field modifications, relationships, and analytics endpoints---while passing a benchmark-generated test suite at every turn. The agent runs inside an isolated environment (Docker container) without access to the tests, and observes only generated feature descriptions, and, optionally, test feedback text. REST API servers are a natural choice of domain: they are ubiquitous in real-world software, structurally complex enough to exercise inter-entity relationships, constraint validation, and aggregate queries, testable in a language-agnostic way via HTTP, and well-defined enough for fully procedural generation of specifications, changes, and tests without LLM involvement. Overall, our key contributions are:
\begin{enumerate}
    \item \textbf{\bname{}}, a procedural benchmark for long-horizon multi-turn coding evaluation, generating arbitrarily long sequences of evolving REST-API requirements with automatic language-agnostic verification and without LLM involvement in spec or test generation.
    \item \textbf{A general framework} underlying \bname{} for constructing long-horizon agent benchmarks with correctness guarantees (validity, infinite procedural generation, controlled difficulty), instantiable in any domain where initial design, change requests, and tests can be procedurally defined (\cref{sec:framework}).
    \item \textbf{A comprehensive empirical study}, with ablations, across 6 open source harnesses, and 7 open source models, spanning from 24B to 744B total parameters. We find that harness quality is a prerequisite for strong performance, even with a good model, and that all models fail very early --- within the first few turns. However, enabling test-feedback and retry loop substantially improves performance across the board.

\end{enumerate}

\section{Related Work}

\begin{table}[t]
\caption{Comparison of \bname{} with related benchmarks across key properties. \bname{} is unique in its ability to procedurally generate arbitrarily long follow-up change request sequences resulting in large codebases.}
\label{tab:positioning}
\centering
\small
\setlength{\tabcolsep}{4pt}
\begin{tabular}{llcrcr}
\toprule
Benchmark & Domain & Procedural & \shortstack{Interaction\\Turns} & \shortstack{Language\\Agnostic} & \shortstack{Lines of\\Code (Avg)} \\
\midrule
SWE-Bench~\citep{jimenez2024swebenchlanguagemodelsresolve}                 & Coding       & \xmark   & 1                  & \xmark      & $\sim$50 \\
SWE-EVO~\citep{thai2026sweevobenchmarkingcodingagents}                     & Coding       & \xmark   & 1                  & \xmark      & 610 \\
FronTalk~\citep{wu2025frontalkbenchmarkingfrontenddevelopment}             & Coding       & \xmark   & 10                 & \xmark      & ~1000 \\
Commit0~\citep{zhao2024commit0librarygenerationscratch}                    & Coding       & \xmark   & 1                  & \xmark      & 1000s \\
SlopCodeBench~\citep{orlanski2026slopcodebenchbenchmarkingcodingagents}    & Coding       & \xmark   & $\sim$5            & \cmark      & 1000+ \\
EvoClaw~\citep{deng2026evoclawevaluatingaiagents}                          & Coding       & \xmark   & up to 25           & \xmark      & 570 \\
$\tau^2$-Bench~\citep{barres2025tau2benchevaluatingconversationalagents}   & Customer svc.& \cmark   & $\sim$10                  & --     & -- \\
UserBench~\citep{qian2025userbenchinteractivegymenvironment} & User interaction & \cmark   & up to 20 & --          & -- \\
\midrule
    \textbf{\bname{} (ours)}                         & \textbf{Coding} & \cmark & \textbf{100-$\infty$} & \cmark & \textbf{1000s-$\infty$} \\
\bottomrule
\end{tabular}
\end{table}

\textbf{Single-interaction-turn coding benchmarks.} SWE-Bench~\citep{jimenez2024swebenchlanguagemodelsresolve} and its verified subset present real GitHub issues for agents to resolve and have become the de facto standard for evaluating coding agents. Most recent models achieve close to 80\% on it, prompting follow-ups \citep{deng2025swebenchproaiagents,yang2024swebenchmultimodalaisystems, yang2025swesmithscalingdatasoftware, zan2025multiswebenchmultilingualbenchmarkissue}. HumanEval~\citep{chen2021evaluatinglargelanguagemodels} and MBPP~\citep{austin2021programsynthesislargelanguage} evaluate function-level code generation. Recent work extends single-turn evaluation to larger scopes: LoCoBench~\citep{qiu2025locobenchbenchmarklongcontextlarge} evaluates long-context code understanding across several software-engineering task categories (architectural understanding, cross-file refactoring, bug investigation, and others) at contexts up to 1M tokens.  R2E-Gym and SWE-Gym \citep{jain2025r2egymproceduralenvironmentshybrid,pan2025trainingsoftwareengineeringagents} build executable environments for RL training.
Many works target longer coding horizons: SWE-EVO~\citep{thai2026sweevobenchmarkingcodingagents} tests agents on real release-note changes; SWE-CI~\citep{chen2026swecievaluatingagentcapabilities} builds evaluation around continuous-integration; NL2Repo-Bench~\citep{ding2026nl2repobenchlonghorizonrepositorygeneration} and Commit0~\citep{zhao2024commit0librarygenerationscratch} ask agents to build entire repos; LongCLI-Bench~\citep{feng2026longclibenchpreliminarybenchmarkstudy} curates 20 long-horizon CLI tasks with step-level scoring; \citet{merrill2026terminalbenchbenchmarkingagentshard} propose Terminal-Bench, a set of long-horizon tasks in a terminal. 

\textbf{Long-horizon and long-context LLM reasoning evaluation.}
A growing body of work shows that LLM performance degrades over extended interactions. The Illusion of Diminishing Returns~\citep{sinha2026illusiondiminishingreturnsmeasuring} isolates long-horizon \emph{execution} by giving the model the plan and knowledge upfront, and shows per-step accuracy degrades with horizon length due to a self-conditioning effect; GSM-Infinite~\citep{zhou2025gsminfinitellmsbehaveinfinitely} builds on the GSM-8K formulation to generate problems of increasing reasoning complexity and context length, and show how LLMs degrade as reasoning depth increases. \citet{hsieh2024rulerwhatsrealcontext} show that extending needle in a haystack \citep{kamradt2023needle} with tracing and aggregation operations makes models struggle.

\textbf{Multi-turn Agent benchmarks beyond coding.}
WebArena~\citep{zhou2024webarenarealisticwebenvironment}, AgentBench~\citep{liu2025agentbenchevaluatingllmsagents}, TravelPlanner~\citep{xie2024travelplannerbenchmarkrealworldplanning}, OSWorld~\citep{xie2024osworldbenchmarkingmultimodalagents}, and AlfWorld~\citep{shridhar2021alfworldaligningtextembodied} test agents in areas beyond coding (web navigation, OS/DB/games, planning, embodied household tasks). $\tau$-Bench \citep{barres2025tau2benchevaluatingconversationalagents,yao2024taubenchbenchmarktoolagentuserinteraction} and UserBench \citep{qian2025userbenchinteractivegymenvironment} evaluate agents in multi-turn interactive setting with a simulated user. MINT \citep{wang2024mintevaluatingllmsmultiturn} simulate a user with an LLM for benchmarking tool-calling.
Vending-Bench~\citep{backlund2025vendingbenchbenchmarklongtermcoherence} evaluates long-term coherence via a vending-machine management task spanning very long (multi-million-token) runs. These establish multi-turn performance as a distinct capability, but evaluate LLMs/agents on non-coding tasks.

\textbf{Multi-interaction-turn coding benchmarks.}
Many works have already explored multi-turn setting in coding: \citet{wu2025frontalkbenchmarkingfrontenddevelopment} test agents on front-end tasks with up to 10 follow-up instruction turns; \citet{deng2026evoclawevaluatingaiagents} evaluate agents' ability to solve a series of subtasks with complex dependency patterns, with up to 23 subtasks. \citet{han2025convcodeworldbenchmarkingconversationalcode} explores simulating a user to provide feedback on function-level code generation task. Closely related to our work, SlopCodeBench~\citep{orlanski2026slopcodebenchbenchmarkingcodingagents} tracks code-quality degradation as agents extend their own solutions over multiple interaction turns across 20 problems, and use language-agnostic testing through command-line interface or web-server APIs.

\cref{tab:positioning} compares \bname{} to existing works across key dimensions. Overall, we note that existing coding benchmarks only evaluate a very small number of turns (or sequential subtasks), and do not simulate extended interactions with the user~\citep{deng2026evoclawevaluatingaiagents,orlanski2026slopcodebenchbenchmarkingcodingagents}. Other benchmarks with multi-turn user interactions exist, but those usually address non-coding domains~\citep{barres2025tau2benchevaluatingconversationalagents,yao2024taubenchbenchmarktoolagentuserinteraction,qian2025userbenchinteractivegymenvironment}.

\section{Benchmark Design}
\label{sec:design}

\begin{algorithm}[t]
\caption{Long-Horizon Agent Evaluation. {\color{algcommentcolor}Blue comments show how each step is realized in \bname{} (REST API).}}\label{alg:eval}
\begin{algorithmic}[1]
\Require initial distribution $p_0$;\ \ environment policy $\pi_{\mathrm{env}}$;\ \ transition function $\tau$;\ \ agent policy $\pi$;\ \ test function $T$;\ \ horizon $N$;\ \ retry budget $R$
\Ensure number of turns passed (in $\{0, 1, \ldots, N+1\}$)
\State $s_0 \sim p_0$ \Comment{sample an OpenAPI-like schema (2--3 entities)}
\State $\hat{s}_0 \leftarrow \pi(\hat{s}_\varnothing, \mathrm{desc}_{\mathrm{state}}(s_0))$ \Comment{agent reads spec + README, writes \texttt{server.py}}
\State $r_0 \leftarrow T(s_0, \hat{s}_0)$ \Comment{launch server on Docker port; run HTTP test suite}
\For{$i \leftarrow 1$ \textbf{to} $R$ \textbf{while} $\neg \mathrm{pass}(r_0)$} \Comment{if failed, retry initial implementation up to $R$ times}
    \State $\hat{s}_0 \leftarrow \pi(\hat{s}_0, \mathrm{desc}_{\mathrm{fb}}(r_0))$ \Comment{agent reads test failures, edits files}
    \State $r_0 \leftarrow T(s_0, \hat{s}_0)$ \Comment{re-run tests}
\EndFor
\State \textbf{if} $\neg \mathrm{pass}(r_0)$ \textbf{then return} $0$ \Comment{fail if initial implementation never passed}
\For{$t \leftarrow 1$ \textbf{to} $N$} \Comment{for each of $N=100$ change turns}
    \State $a_t \sim \pi_{\mathrm{env}}(\cdot \mid s_{t-1})$ \Comment{sample a change (rename, add field, \ldots)}
    \State $s_t \leftarrow \tau(s_{t-1}, a_t)$ \Comment{apply change to reference spec}
    \State $\hat{s}_t \leftarrow \pi(\hat{s}_{t-1}, \mathrm{desc}_{\mathrm{act}}(a_t))$ \Comment{agent edits server to match new spec}
    \State $r_t \leftarrow T(s_t, \hat{s}_t)$ \Comment{run refreshed test suite}
    \For{$i \leftarrow 1$ \textbf{to} $R$ \textbf{while} $\neg \mathrm{pass}(r_t)$} \Comment{retry this turn with feedback}
        \State $\hat{s}_t \leftarrow \pi(\hat{s}_t, \mathrm{desc}_{\mathrm{fb}}(r_t))$ \Comment{agent edits files using feedback}
        \State $r_t \leftarrow T(s_t, \hat{s}_t)$ \Comment{re-run tests}
    \EndFor
    \State \textbf{if} $\neg \mathrm{pass}(r_t)$ \textbf{then break} \Comment{terminate scenario on unrecovered failure}
\EndFor
\State \Return $\sum_{t=0}^{N} \mathrm{pass}(r_t)$ \Comment{number of turns passed}
\end{algorithmic}
\end{algorithm}

\subsection{A Framework for Multi-Turn Agent Benchmarks}
\label{sec:framework}

We first propose a framework for designing multi-turn agent benchmarks. A multi-turn agent benchmark consists of two coupled systems: a \textbf{reference system} whose state evolves through a known trajectory, and an \textbf{agent system} that must track it. At each step, a \textbf{test function} checks whether the agent's output matches the reference state. The benchmark then measures how well the agent tracks the reference over a long horizon. In the context of coding, one might think of the reference system as the engineer's \emph{mental model} of the project. During development, this mental model evolves as the engineer comes up with new features. The coding agent's job is to keep the codebase matching this mental model, working from the engineer's prompts. The engineer then verifies the match through a test suite, or through manual interaction with the program. More formally, a \textbf{reference system} is defined by:
\begin{itemize}
    \item \textbf{State space} $\mathcal{S}$: the set of all valid states.
    \item \textbf{Initial state distribution} $p_0(s)$: how the starting state is sampled. Analogous to the initial state distribution in a Markov Decision Process (MDP).
    \item \textbf{Action space} $\mathcal{A}(s)$: the set of valid actions at state $s$. Actions can be state-dependent --- not all actions are valid in all states.
    \item \textbf{Transition function} $\tau: \mathcal{S} \times \mathcal{A} \to \mathcal{S}$: a deterministic function $s_{t+1} = \tau(s_t, a_t)$ producing the next reference state. Unlike an MDP, the ``policy'' selecting actions is part of the environment, not the agent.
    \item \textbf{Action selection} $\pi_{\mathrm{env}}(a \mid s)$: how the environment selects the next action. This can be uniform over $\mathcal{A}(s)$, difficulty-weighted, or \emph{LLM-driven} (\cref{sec:sampling}).
\end{itemize}

This produces a reference trajectory $s_0, a_0, s_1, a_1, \ldots, s_N$. The reference system is a deterministic process once the seed and action-selection policy are fixed.

\textbf{Agent system.}~ The agent is a policy $\pi: \hat{\mathcal{S}} \times \mathcal{D} \to \hat{\mathcal{S}}$ that maintains its own state $\hat{s}_t \in \hat{\mathcal{S}}$ (e.g., a codebase, a configuration, a set of files) and updates it in response to text observations $d \in \mathcal{D}$. The agent starts from an empty state $\hat{s}_\varnothing$ and does \emph{not} observe the reference state $s_t$ directly --- only through description functions:
\begin{itemize}
    \item $\mathrm{desc}_{\mathrm{state}}(s_0)$: initial state in natural language.
    \item $\mathrm{desc}_{\mathrm{action}}(a_t)$: description of the action applied at turn $t$.
    \item $\mathrm{desc}_{\mathrm{feedback}}(r_t)$ (optional): natural-language rendering of the test result (discussed below, configurable from detailed error messages to a binary pass/fail signal).
\end{itemize}
The details provided in descriptions define the difficulty. We can fully specify the state explicitly, and let the agent copy it directly, or we can specify only the minimal amount of detail necessary to reconstruct the state.

\textbf{Test function.}~ A test function $T: \mathcal{S} \times \hat{\mathcal{S}} \to \mathcal{R}$ evaluates the agent's state against the reference state, producing a result $r_t \in \mathcal{R}$. A predicate $\mathrm{pass}: \mathcal{R} \to \{0, 1\}$ extracts the binary outcome, and $\mathrm{desc}_{\mathrm{feedback}}: \mathcal{R} \to \mathcal{D}$ extracts NL description.

Overall, the agent computes $\hat{s}_0 = \pi(\hat{s}_\varnothing, \mathrm{desc}_{\mathrm{state}}(s_0))$ at turn 0, and $\hat{s}_t = \pi(\hat{s}_{t-1}, \mathrm{desc}_{\mathrm{action}}(a_t))$ at subsequent turns. If the retry loop is enabled, then on failed tests, $\hat{s}_t = \pi(\hat{s}_t, \mathrm{desc}_{\mathrm{feedback}}(r_t))$ updates the state using feedback. The full algorithm is presented in \cref{alg:eval}.

\subsection{\bname{}: A REST-API Instantiation}
\label{sec:restbench}

\bname{} instantiates the framework (\cref{sec:framework}) for REST API servers:
\begin{itemize}
    \item \textbf{State space $\mathcal{S}$}: OpenAPI-like~\citep{openapi} schemas with typed entities (e.g.\ a \texttt{User} with \texttt{id:int}, \texttt{name:str}, \texttt{balance:float}), inter-entity relationships (e.g.\ groups containing references to users), analytics endpoints (e.g.\ user counts), and \emph{business logic}---stateful workflows where actions transition enum fields between states, guarded by conditions and triggering side effects (e.g.\ \texttt{Order.submit} moves the order to \texttt{submitted} and clears the cart, only if the cart was non-empty). $S$ is represented by a strongly typed data structure.
    \item \textbf{Initial distribution $p_0$}: a seeded procedural generator or an LLM that outputs $s_0$. Since $S$ is highly structured, this amounts to filling out fields of a data structure with random field names, types, operations, etc.
    \item \textbf{Action space $\mathcal{A}(s)$}: a fixed set of change types (add entity/field, rename, delete, \ldots; full list in \cref{tab:changes}), each defined by a strict schema for easier generation (sampling becomes selecting a change type and populating its fields) and validation (i.e. making sure field values are valid and checking for conflicts with other changes). We sample 5 individual changes and concatenate them into one bigger change at each interaction turn;
    \item \textbf{Transition $\tau$}: applies a change to produce the next schema. Implemented purely in Python, leveraging the highly structured state and action spaces.
    \item \textbf{Test function $T$}: runs HTTP-based tests against the agent's server. Again, due to highly structured state space $S$, tests are generated programmatically. Tests include reading/writing for each endpoint and field, validation, actions, etc. Feedback is configurable to binary, test-names-only, or detailed assertion info (see \cref{sec:feedback}).
\end{itemize}
Server complexity grows monotonically: early turns involve 2--3 entities, later turns have dozens of entities with relationships, cascade deletions, and aggregate analytics, and total thousands lines of code, see \cref{fig:loc_growth}. Instructions are passed to the agent via a \texttt{README.md} that contains all information needed to pass all tests without retries. Full component details are in \cref{app:benchmark-details}; concrete prompts in \crefrange{app:readme}{app:feedback-example}.

A \bname{} evaluation consists of $K$ independent \emph{scenarios} (each with a different seed), executed via \cref{alg:eval}. Since scenarios are generated, $K$ and $N$ are bounded only by compute. The agent and its code run inside a Docker container; the benchmark communicates only through one exposed port. Throughout the paper, we focus on two metrics: average turns passed and pass rate. Let $r_{k,t} = T(s_t, \hat{s}_t)$ denote the test result for scenario $k$ at turn $t$ (after up to $R$ retries). The \textbf{average turns passed @R} $\frac{1}{K} \sum_{k=1}^{K} \sum_{t=0}^{N} \mathrm{pass}(r_{k,t})$ measures how long agents maintain correctness before failing, and the \textbf{pass rate @R} $\frac{1}{K} \sum_{k=1}^{K} \mathds{1}\!\left[\sum_{t=0}^{N} \mathrm{pass}(r_{k,t}) = N+1\right]$ is the fraction of scenarios completed end-to-end. @R denotes the retry attempt budget, and, unless stated otherwise, $R=\Nretries$.

\begin{table}[t]
\centering
\small
\caption{Average turns passed with no feedback loop (@R=0) $\pm$ SE per model--harness combination.}
\label{tab:first-failure}
\begin{tabular}{lcccc}
\toprule
Model & OpenCode & Mini-SWE & OpenHands & \shortstack{Model Provider\\Agent} \\
\midrule
Devstral 2 & 0.7\,${\scriptstyle\pm\,0.3}$ & 1.4\,${\scriptstyle\pm\,0.6}$ & 0.9\,${\scriptstyle\pm\,0.4}$ & 0.8\,${\scriptstyle\pm\,0.3}$ \\
Devstral Small 2 & 0.8\,${\scriptstyle\pm\,0.3}$ & 0.7\,${\scriptstyle\pm\,0.2}$ & 0.5\,${\scriptstyle\pm\,0.2}$ & 0.9\,${\scriptstyle\pm\,0.4}$ \\
GLM-5 & 4.5\,${\scriptstyle\pm\,1.0}$ & 6.2\,${\scriptstyle\pm\,0.7}$ & 1.8\,${\scriptstyle\pm\,0.3}$ & --- \\
Kimi K2.5 & 2.3\,${\scriptstyle\pm\,0.7}$ & 3.8\,${\scriptstyle\pm\,1.0}$ & 0.9\,${\scriptstyle\pm\,0.3}$ & 3.2\,${\scriptstyle\pm\,0.8}$ \\
Nemotron Super & 0.3\,${\scriptstyle\pm\,0.2}$ & 0.1\,${\scriptstyle\pm\,0.0}$ & 0.3\,${\scriptstyle\pm\,0.1}$ & --- \\
Qwen3-Coder-Next & 0.9\,${\scriptstyle\pm\,0.5}$ & 0.9\,${\scriptstyle\pm\,0.3}$ & 0.6\,${\scriptstyle\pm\,0.2}$ & 0.8\,${\scriptstyle\pm\,0.3}$ \\
Qwen3.5-122B & 1.1\,${\scriptstyle\pm\,0.4}$ & 3.2\,${\scriptstyle\pm\,0.8}$ & 1.9\,${\scriptstyle\pm\,0.4}$ & 0.9\,${\scriptstyle\pm\,0.4}$
 \\
\bottomrule
\end{tabular}
\end{table}

\section{Experiments}
\label{sec:setup}

\textbf{Models.}~ We evaluate a range of open-source and open-weight LLMs spanning different scales and architectures: \textbf{Devstral 2}~\citep{mistral2025devstral2} (Mistral, 123B dense), \textbf{Devstral Small 2}~\citep{mistral2025devstral2} (Mistral, 24B dense), \textbf{GLM-5}~\citep{glm5team2026glm5vibecodingagentic} (Zhipu, 744B/40B MoE), \textbf{Kimi K2.5}~\citep{kimiteam2026kimik25visualagentic} (Moonshot, 1T/32B MoE), \textbf{Nemotron Super}~\citep{nvidia2026nemotron3superopen} (NVIDIA, 120B/12B hybrid Mamba-Transformer), \textbf{Qwen3-Coder-Next}~\citep{cao2026qwen3codernexttechnicalreport} (Alibaba, 80B/3B MoE), and \textbf{Qwen3.5-122B}~\citep{qwen2026qwen35} (Alibaba, 122B/10B MoE).

\textbf{Agents.}~ For each model, we evaluate three open-source coding agent harnesses --- \textbf{OpenCode}~\citep{opencode}, \textbf{Mini-SWE}~\citep{miniswe} (modified to enable context compaction), and \textbf{OpenHands}~\citep{wang2025openhandsopenplatformai,openhands} --- as well as the harness developed by the model provider when available: \textbf{Mistral Vibe}~\citep{mistralvibe,mistral2025devstral2} for Mistral (Devstral) models, \textbf{Kimi CLI}~\citep{kimicli} for Kimi K2.5, and \textbf{QwenCode}~\citep{qwencode} for Qwen models. All agents run inside isolated Docker containers with full shell access. Further details are in \cref{app:agents}.

\textbf{Experiment settings.}~ All experiments use the following configuration unless otherwise noted: $K=20$ scenarios, $N=100$ turns per scenario, and, if the feedback loop is enabled, $R=\Nretries{}$ retries per turn. Each configuration thus produces up to $20 \times 101 = 2{,}020$ user-agent interaction turns. By default, the agents are instructed to use Python, and are free to install and use new packages. For fair comparison, the $K=20$ scenarios are the same across all the experiments. Across five independent runs on the same scenarios, per-configuration averages are stable to within ${\sim}$6--7 turns (see \cref{app:variance} for variance analysis). Because the same scenarios are used across configurations, we test claims with paired Wilcoxon signed-rank tests \citep{wilcoxon} scenario-by-scenario (turn counts are bounded integer outcomes). We apply Holm correction \citep{holm} when a single hypothesis is tested across many cells (e.g.\ retry across all 26 model$\times$harness pairs); for ablations that vary one factor across models, we report uncorrected per-model $p$-values.

\label{sec:results}

\subsection{Results}

\begin{figure}[t]
\centering
\begin{subfigure}[t]{0.32\linewidth}
\centering
\includegraphics[width=\linewidth]{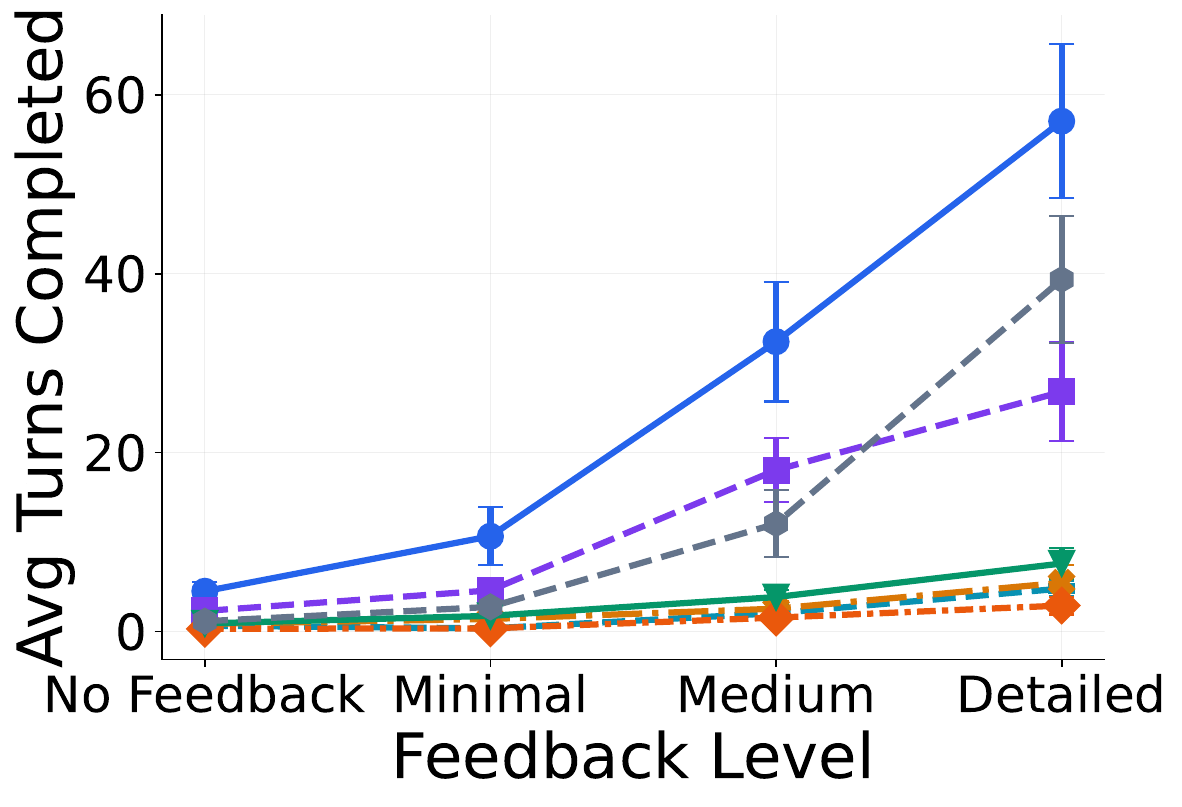}
\end{subfigure}\hfill
\begin{subfigure}[t]{0.32\linewidth}
\centering
\includegraphics[width=\linewidth]{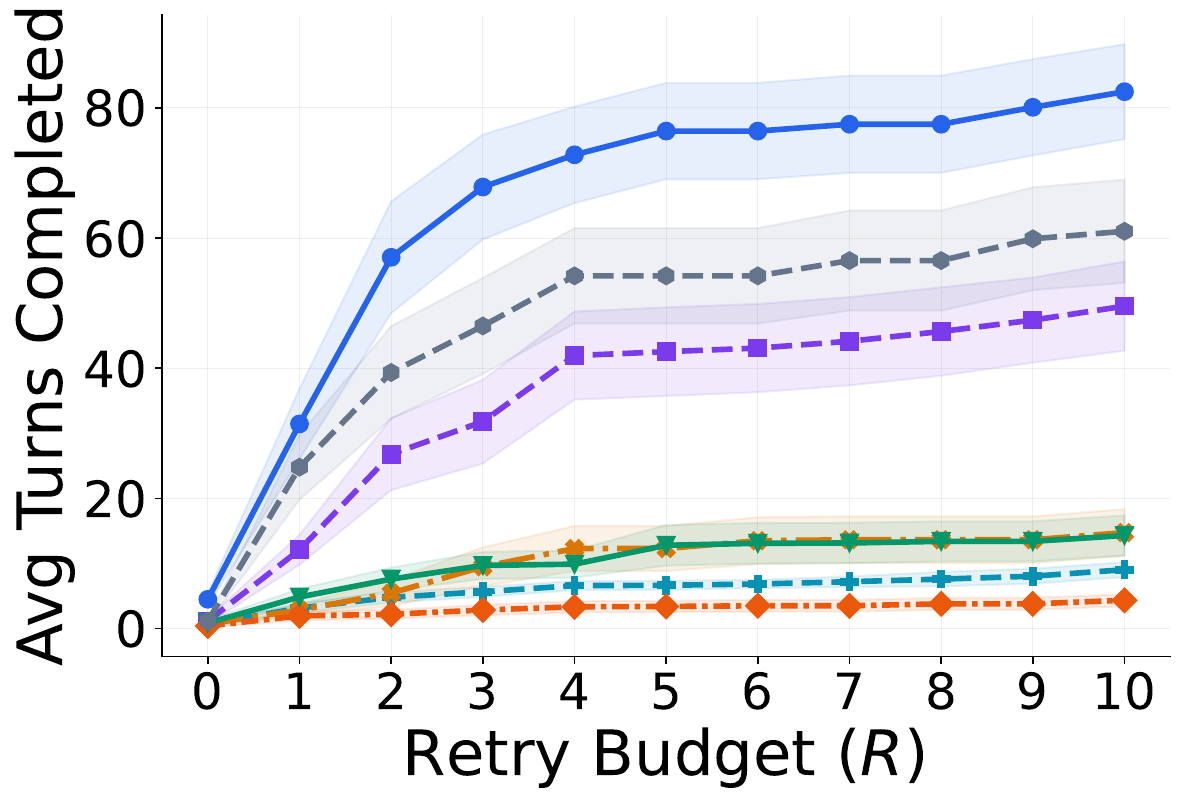}
\end{subfigure}\hfill
\begin{subfigure}[t]{0.32\linewidth}
\centering
\includegraphics[width=\linewidth]{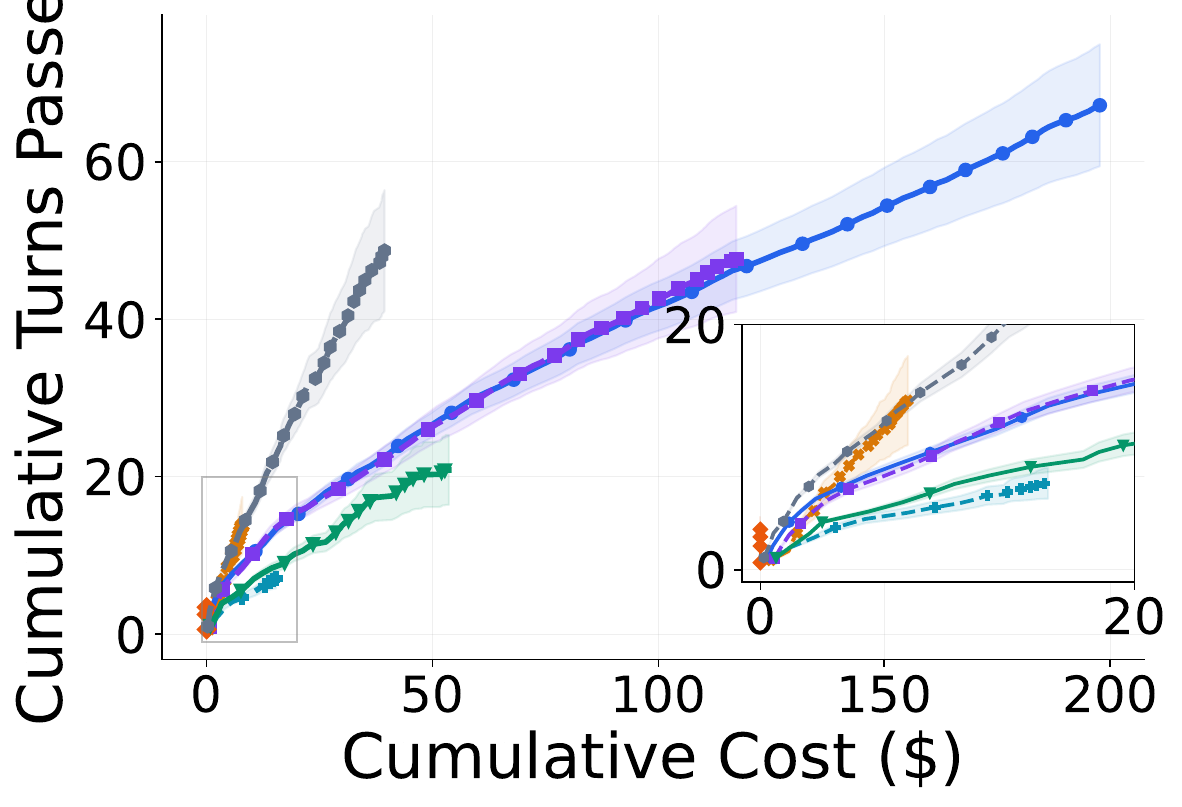}
\end{subfigure}\\[2pt]
\includegraphics[width=0.98\linewidth]{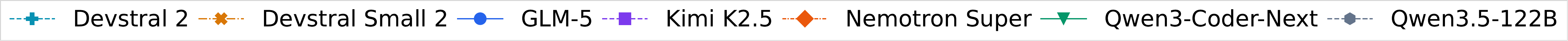}\\[-6pt]
\begin{subfigure}[t]{0.32\linewidth}
\centering
\caption{Turns vs.\ feedback level.}
\label{fig:feedback_ablation}
\end{subfigure}\hfill
\begin{subfigure}[t]{0.32\linewidth}
\centering
\caption{Turns vs.\ attempt budget.}
\label{fig:perf_vs_attempts}
\end{subfigure}\hfill
\begin{subfigure}[t]{0.32\linewidth}
\centering
\caption{Turns passed vs.\ cost (\$).}
\label{fig:cost}
\end{subfigure}
\caption{Scaling and ablation dynamics (OpenCode), averaged over scenarios with $\pm$1 SE. \textbf{(a)}~Feedback ablation: all models improve sharply from no feedback to detailed; stronger models gain the most in absolute terms. \textbf{(b)}~Retry budget: performance vs.\ maximum allowed attempts per turn; most gains come from the first 3--5 attempts. \textbf{(c)}~Cost efficiency. We vary max cost per scenario, and measure how many avg turns are reached within that budget.}
\label{fig:scaling}
\end{figure}

\textbf{No feedback loop.} First, we test all agents without the feedback loop -- if the tests fail, the scenario is terminated (i.e. $R=0$ in \cref{alg:eval}). We show results in \cref{tab:first-failure}. We see that even the best model we tested, GLM-5, failed just after 6.2 turns on average with the best harness, highlighting that multi-turn reliability is still a challenge even for the strongest models. Further, this shows that vibe-coding is prone to generate bugs even when instructions are crystal clear and precise.

\textbf{Enabling retry loop.} In all following experiments, we enable the retry loop with $R=\Nretries{}$ retries on test failure (\cref{tab:oss-coverage}). All 26 model$\times$harness cells improve significantly over $R=0$ (Wilcoxon signed-rank, Holm-corrected, $p<0.05$). Harness choice matters substantially, echoing \citep{kapoor2025holisticagentleaderboardmissing}: OpenCode is the best or statistically indistinguishable from the best for all 7 models; OpenHands is the worst for 6/7; Mini-SWE is competitive despite exposing only a single bash tool. Provider-built harnesses do not reliably help: QwenCode is significantly \emph{worse} than OpenCode for both Qwen3-Coder-Next ($p=0.014$) and Qwen3.5-122B ($p=0.011$), while other provider pairings show good averages but no significant difference from OpenCode.

Model rankings also shift across harnesses: GLM-5 leads on OpenCode with 57.0 turns but drops to 15.1 on Mini-SWE, where Qwen3.5-122B leads numerically (33.1). On OpenCode, GLM-5 is significantly above every other model (Wilcoxon $p<0.05$) except Qwen3.5-122B, which is not reliably distinguishable ($p=0.07$). On Mini-SWE, Qwen3.5-122B's lead over GLM-5 and Kimi K2.5 does not reach significance ($p=0.34$ and $p=0.30$ respectively), due to the high variance of scores (SE$=9.0$ for Qwen3.5-122B). Overall, these complex interaction effects show that \bname{} tests the model and harness interplay rather than model capability alone.

We also study the cost efficiency of the models in \cref{fig:cost}, and see that Kimi K25 and GLM5 exhibit similar cost performance, while Qwen3.5-122B is vastly more efficient for the price.

\begin{table}[t]
\centering
\small
\caption{Avg turns passed $\pm$ SE (pass rate \%) per model--harness combination, @$R=2$ retries. \textbf{Bold} are the best per row, \underline{underlined} are the results that are not significantly worse than the best under an uncorrected Wilcoxon signed-rank test ($p > 0.05$). We annotate only this table because it is the main cross-comparison; ablation tables are interpreted via inline $p$-values in the text.}
\label{tab:oss-coverage}
\begin{tabular}{lcccc}
\toprule
Model & OpenCode & Mini-SWE & OpenHands & \shortstack{Model Provider\\Agent} \\
\midrule
Devstral 2 & \underline{\phantom{0}4.8\,${\scriptstyle\pm\,0.8}$ (\phantom{0}0\%)} & \underline{\phantom{0}5.0\,${\scriptstyle\pm\,0.7}$ (\phantom{0}0\%)} & 2.8\,${\scriptstyle\pm\,0.6}$ (0\%) & \textbf{\phantom{0}8.9\,${\scriptstyle\pm\,2.3}$ (0\%)} \\
Devstral Small 2 & \underline{\phantom{0}5.5\,${\scriptstyle\pm\,2.0}$ (\phantom{0}0\%)} & \underline{\phantom{0}2.3\,${\scriptstyle\pm\,0.6}$ (\phantom{0}0\%)} & 0.9\,${\scriptstyle\pm\,0.3}$ (0\%) & \textbf{\phantom{0}9.2\,${\scriptstyle\pm\,3.1}$ (0\%)} \\
GLM-5 & \textbf{57.0\,${\scriptstyle\pm\,8.6}$ (25\%)} & 15.1\,${\scriptstyle\pm\,1.8}$ (\phantom{0}0\%) & 8.7\,${\scriptstyle\pm\,1.3}$ (0\%) & --- \\
Kimi K2.5 & \underline{26.8\,${\scriptstyle\pm\,5.5}$ (\phantom{0}0\%)} & 14.9\,${\scriptstyle\pm\,2.7}$ (\phantom{0}0\%) & 4.9\,${\scriptstyle\pm\,1.0}$ (0\%) & \textbf{33.7\,${\scriptstyle\pm\,6.4}$ (0\%)} \\
Nemotron Super & \textbf{\phantom{0}2.9\,${\scriptstyle\pm\,1.0}$ (\phantom{0}0\%)} & \underline{\phantom{0}1.6\,${\scriptstyle\pm\,0.6}$ (\phantom{0}0\%)} & 0.8\,${\scriptstyle\pm\,0.2}$ (0\%) & --- \\
Qwen3-Coder-Next & \textbf{\phantom{0}7.6\,${\scriptstyle\pm\,1.8}$ (\phantom{0}0\%)} & \underline{\phantom{0}6.2\,${\scriptstyle\pm\,2.3}$ (\phantom{0}0\%)} & 3.2\,${\scriptstyle\pm\,0.5}$ (0\%) & \phantom{0}3.0\,${\scriptstyle\pm\,0.5}$ (0\%) \\
Qwen3.5-122B & \textbf{39.4\,${\scriptstyle\pm\,7.1}$ (10\%)} & \underline{33.1\,${\scriptstyle\pm\,9.0}$ (15\%)} & 7.3\,${\scriptstyle\pm\,1.1}$ (0\%) & 19.0\,${\scriptstyle\pm\,4.8}$ (0\%)
 \\
\bottomrule
\end{tabular}
\end{table}

\textbf{Scaling retry turns.} As we observed, adding retry loop drastically improves performance. In \cref{fig:perf_vs_attempts}, we scale attempt budget for all models with OpenCode harness to up to 10 retries. We see that all stronger models like GLM-5, Kimi K2.5 and Qwen3.5-122B improve drastically over the first 5 attempts, and plateau after. Weaker models improve too, but much less.

\subsection{Ablations}

\textbf{Feedback.}~
\label{sec:feedback}
To isolate the role of error feedback, we repeat the evaluation under three feedback levels: \textbf{detailed} (specific assertion failure messages), \textbf{medium} (pass/fail per test, no error details), and \textbf{minimal} (only told that tests failed, no further information). \cref{fig:feedback_ablation} (per-model numbers in \cref{tab:feedback-ablation}, \cref{app:additional-results}) shows that detailed feedback yields 6--12$\times$ more turns completed than minimal feedback across all models, with the drop steepest for the strongest models (GLM-5 falls from 57 to 10.7 turns; Qwen3.5 from 39.4 to 2.8). Under minimal feedback all models converge to fewer than 11 turns. The detailed-vs-minimal gap is significant for 6 of 7 models (Wilcoxon $p<0.05$; Devstral Small 2 marginal at $p=0.058$). \emph{This shows coding agents need precise feedback to correct their mistakes.}

\textbf{Implementation language.}~
\label{sec:language}
The benchmark's language-agnostic evaluation (testing via HTTP requests) allows us to measure how implementation language affects long-horizon performance. \cref{tab:lang-sampling-ablation} compares Python, JavaScript, and Rust using OpenCode. We observe positive trends for JavaScript on GLM-5 and Kimi K2.5, though these differences are not statistically significant at $n=20$. All models trend lower with Rust, which is not a common choice for REST API servers; the drop is significant for Kimi K2.5, Nemotron Super, Qwen3-Coder-Next, and Qwen3.5-122B (Wilcoxon $p<0.05$). We hypothesize that the differences in performance are due to the training distribution, with less data for REST APIs available in Rust compared to Python and JavaScript.

\textbf{Sampling strategy.}~
\label{sec:sampling}
The environment policy $\pi_{\mathrm{env}}$ can select actions either via an LLM (which produces coherent development narratives) or via a programmatic sampler (which draws uniformly from valid changes; see \cref{app:programmatic-spec} for an example). \cref{tab:lang-sampling-ablation} (rightmost and leftmost columns) compares these two strategies. Overall we see roughly similar performance: deterministic sampling does not significantly differ from LLM-driven sampling for any model except Nemotron Super, whose absolute scores are very small (Wilcoxon $p>0.05$ for the other 6). GLM-5 trends $+17$ turns higher under deterministic sampling but does not reach significance at $n=20$ ($p=0.21$). 

\begin{table}[t]
\centering
\small
\caption{Avg turns passed $\pm$ SE (pass rate \%) for language and sampling strategy ablations (OpenCode, @$R=2$ retries). Python column uses LLM sampling, and also serves as the baseline for Programmatic sampling.}
\label{tab:lang-sampling-ablation}
\begin{tabular}{lcccc}
\toprule
 & \multicolumn{3}{c}{Language} & \multicolumn{1}{c}{Scenario Gen.} \\
\cmidrule(lr){2-4} \cmidrule(lr){5-5}
Model & Python & JavaScript & Rust & Programmatic \\
\midrule
Devstral 2 & \phantom{0}4.8\,${\scriptstyle\pm\,0.8}$ (\phantom{0}0\%) & \phantom{0}3.8\,${\scriptstyle\pm\,1.2}$ (\phantom{0}0\%) & \phantom{0}4.0\,${\scriptstyle\pm\,1.2}$ (\phantom{0}0\%) & \phantom{0}4.8\,${\scriptstyle\pm\,1.0}$ (\phantom{0}0\%) \\
Devstral Small 2 & \phantom{0}5.5\,${\scriptstyle\pm\,2.0}$ (\phantom{0}0\%) & \phantom{0}5.8\,${\scriptstyle\pm\,1.6}$ (\phantom{0}0\%) & \phantom{0}3.1\,${\scriptstyle\pm\,1.3}$ (\phantom{0}0\%) & \phantom{0}7.3\,${\scriptstyle\pm\,1.9}$ (\phantom{0}0\%) \\
GLM-5 & 57.0\,${\scriptstyle\pm\,8.6}$ (25\%) & 70.7\,${\scriptstyle\pm\,7.1}$ (30\%) & 50.1\,${\scriptstyle\pm\,8.3}$ (20\%) & 74.1\,${\scriptstyle\pm\,6.9}$ (45\%) \\
Kimi K2.5 & 26.8\,${\scriptstyle\pm\,5.5}$ (\phantom{0}0\%) & 43.6\,${\scriptstyle\pm\,8.1}$ (15\%) & 13.7\,${\scriptstyle\pm\,3.3}$ (\phantom{0}0\%) & 26.9\,${\scriptstyle\pm\,6.4}$ (\phantom{0}5\%) \\
Nemotron Super & \phantom{0}2.9\,${\scriptstyle\pm\,1.0}$ (\phantom{0}0\%) & \phantom{0}1.5\,${\scriptstyle\pm\,0.4}$ (\phantom{0}0\%) & \phantom{0}0.3\,${\scriptstyle\pm\,0.1}$ (\phantom{0}0\%) & \phantom{0}6.0\,${\scriptstyle\pm\,1.0}$ (\phantom{0}0\%) \\
Qwen3-Coder-Next & \phantom{0}7.6\,${\scriptstyle\pm\,1.8}$ (\phantom{0}0\%) & \phantom{0}9.4\,${\scriptstyle\pm\,2.0}$ (\phantom{0}0\%) & \phantom{0}2.5\,${\scriptstyle\pm\,1.0}$ (\phantom{0}0\%) & \phantom{0}7.1\,${\scriptstyle\pm\,1.3}$ (\phantom{0}0\%) \\
Qwen3.5-122B & 39.4\,${\scriptstyle\pm\,7.1}$ (10\%) & 26.8\,${\scriptstyle\pm\,5.4}$ (\phantom{0}5\%) & \phantom{0}8.6\,${\scriptstyle\pm\,2.4}$ (\phantom{0}0\%) & 25.4\,${\scriptstyle\pm\,3.8}$ (\phantom{0}0\%)
 \\
\bottomrule
\end{tabular}
\end{table}

\section{Failure Analysis}
\label{sec:analysis}

\begin{figure}[t]
\centering
\begin{subfigure}[t]{0.62\linewidth}
\centering
\includegraphics[width=\linewidth]{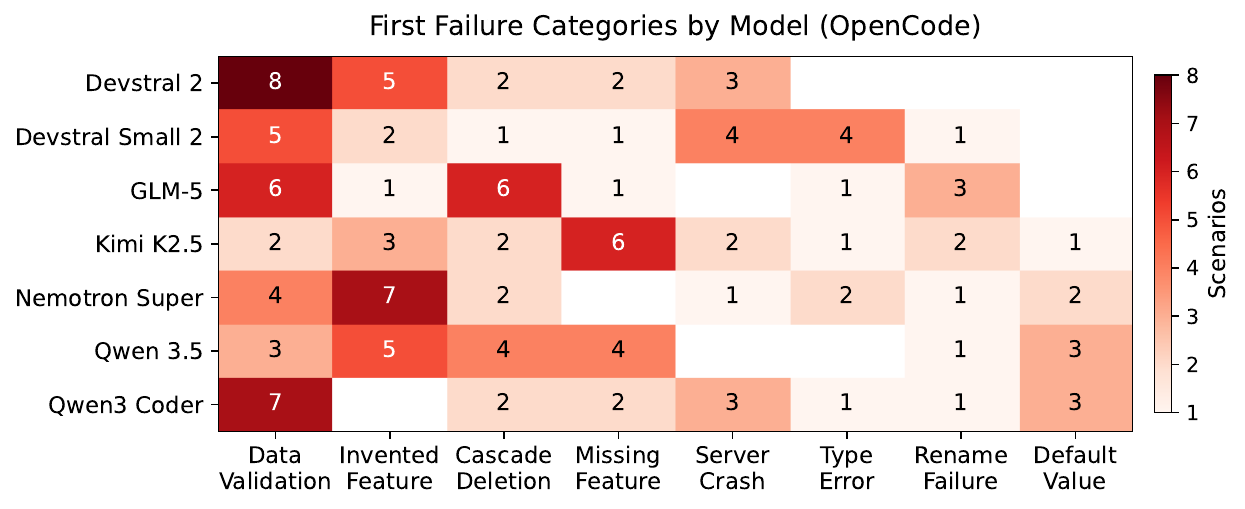}
\caption{Failure types with $R=0$.}
\label{fig:heatmap-teaser}
\end{subfigure}\hfill
\begin{subfigure}[t]{0.35\linewidth}
\centering
\includegraphics[width=\linewidth]{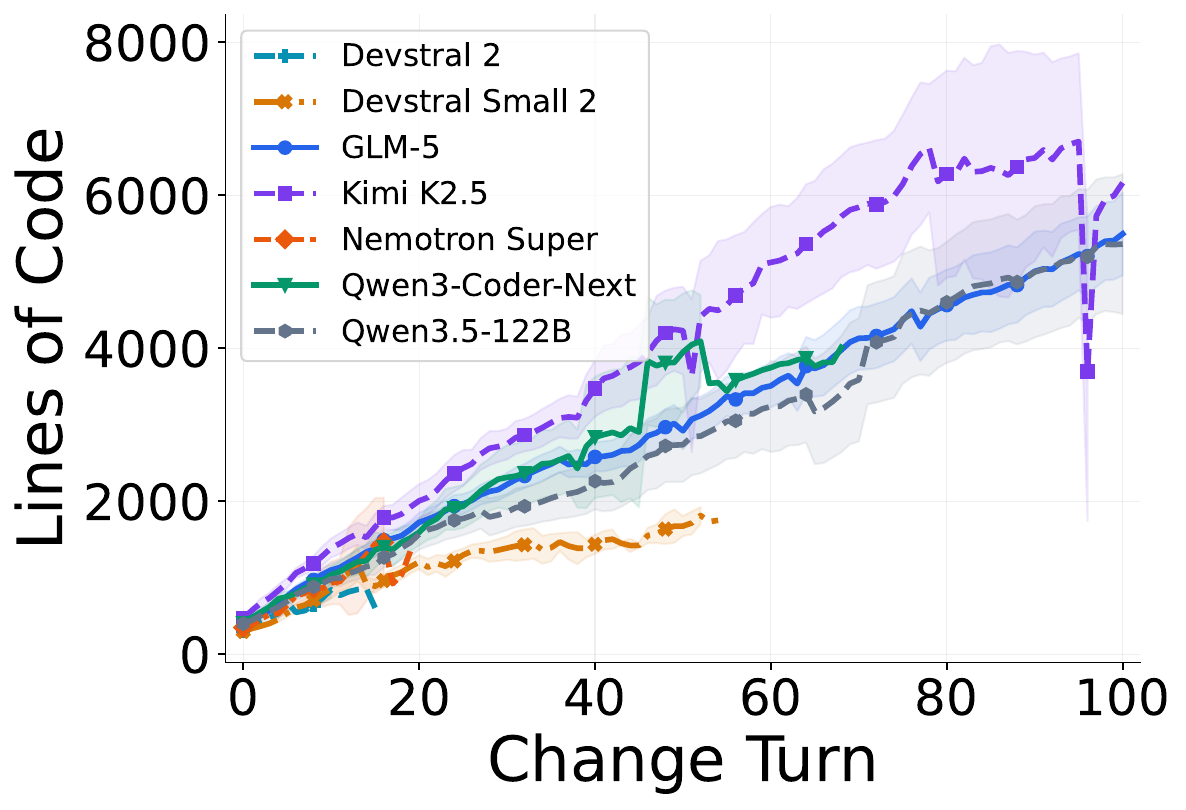}
\caption{Codebase size vs.\ turn.}
\label{fig:loc_growth}
\end{subfigure}
    \caption{Failure characterization on OpenCode, averaged over scenarios with $\pm$1 SE. \textbf{(a)}~Distribution of failure types at the turn each scenario first fails, restricted to the most common categories. Rows are models, columns are failure categories; each cell shows the number of scenarios (out of 20) in which that category caused the first failure. Full breakdowns across all harnesses and retry budgets are in \cref{app:heatmaps}. \textbf{(b)}~Codebase size (lines of code) as a function of change turn; stronger models reach ~6{,}000 LoC, while curves of weaker models end early as no scenarios reach higher turn counts.}
\label{fig:analysis-overview}
\end{figure}

Beyond aggregate pass rates, we inspect the \emph{types} of failures agents encounter. First, we split failures into types, such as data validation issues, cascade deletion bugs, renames issues, etc. For the full list with brief descriptions, see \cref{app:heatmaps}. Then, we run an agent and instruct it to analyze the logs of each experiment, and classify each run into one of these buckets. We build a histogram of the resulting failures distribution with no retries ($R=0$) in \cref{fig:heatmap-teaser} (and in \cref{fig:heatmap-first-failure} with more detail), for ($R=2$) in \cref{fig:heatmap-3-attempts}, and for $R=10$ in \cref{fig:heatmap-final}.

\textbf{Implementation bugs.}~
With retries budget $R=0$ and $R=2$, we observe that most agents make mistakes by not following the provided instructions exactly, and make data validation too strict or too loose. For example, a common mistake is accepting \texttt{null} when a field is explicitly non-nullable, and instructions say ``Fields can be null if they are defined as nullable, otherwise they cannot be.''. Although the instructions were designed to give agent the required information to solve the task perfectly without having to retry, agents likely pay less and less attention to them as the context grows across turns. This confirms a phenomenon observed in prior work~\citep{liu2023lostmiddlelanguagemodels,hong2025context}. This highlights the importance of testing agents in multi-turn settings where context needs to be compressed multiple times over the session, and instructions may be forgotten or disregarded. We note that the instructions file is always present in the agents' working directory, so even when context is large, the agent may read the file again to refresh memory, but this does not fully prevent the failures.
Agents also tend to hallucinate new unrequested changes. For example, MiniSWE with Devstral 2 Small was instructed to add a string field \texttt{activity\_type}, but instead hallucinated a new \texttt{enum} instead of the string, which is not unreasonable, but is not what the prompt requested. For more example failures with code excerpts, see \cref{app:failure-examples}.

\textbf{Failures Unrelated to Coding.}~
In the heatmap for $R=10$ in \cref{fig:heatmap-final}, we find that infrastructure issues come up more, with agents using wrong tool format, calling overly general \texttt{pkill}, calling tools during compaction, or getting stuck in a loop. With feedback and enough retries, agents are able to eventually figure out a lot of the implementation issues, and are bottlenecked by infrastructure more. Notably, Qwen3-Coder-Next and Nemotron Super struggle whenever OpenCode triggers context compaction: they tend to call tools \emph{during} compaction, and OpenCode immediately raises an error when this happens (Nemotron Super: 8/20 scenarios killed; Qwen3-Coder-Next: \nToolCallQwenCoder/\nToolCallQwenCoderTotal). OpenHands exhibits a pathology with its strict loop-detection logic, which raises an error whenever the agent sends the same message four times in a row; once triggered, the error pollutes the session state, so all follow-up requests fail and the scenario is effectively killed (worst: GLM-5 with 13/20, Devstral 2 with 10/20). A third common issue is agent self-kill via \texttt{pkill}: although \texttt{README.md} explicitly instructs the agent to \texttt{pkill} by PID only, agents occasionally \texttt{pkill} by regex with too loose a pattern and end up killing their own harness process (worst: Nemotron Super + OpenHands - 11/20). In all these cases, to let the agent recover, we retry the agent call up to 10 times if the return code is not 0, but the agents still fail. These failures demonstrate the ability of \bname{} to stress-test not just the model, but also its interplay with the harness, making it a useful tool for researchers.

\section{Discussion}
\label{sec:discussion}

We introduced \bname{}, a procedurally generated benchmark for evaluating performance of coding agents across a large number of interaction turns, grounded in a general framework for constructing such benchmarks (\cref{sec:framework}). Through evaluation of six agent harnesses and seven open-source LLMs across up to 2,000 agent-turns per configuration, we find that even the strongest agents fail very early and are unable to complete even 10 interaction turns without making mistakes.
Agents perform much better when detailed feedback is enabled, but such feedback is not available in most contexts. Our experiments highlight the importance of improving reliability of coding agents beyond single one-off tasks commonly tested in other benchmarks. Multi-turn performance puts more pressure on reliability of not just the model, but the harness, and tests its ability to compress context without losing important instructions, and to recover from infrastructure issues like malformed tool calls. We release the implementation of \bname{} to enable the community to generate new scenarios, possibly spanning far more than 100 turns, and release the data we used in our experiments to ensure reproducibility and transparency.

\textbf{Limitations.}
Multi-turn evaluation is inherently expensive, and this benchmark is no exception: a single full-grid configuration consumes billions of input tokens (e.g., GLM-5 + OpenCode used 4.5B input / 7.5M output tokens for one 20-scenario sweep). At frontier closed-source pricing, replicating this on Claude Sonnet 4.6 would cost roughly \$13.6K and on GPT-5.5 roughly \$22.7K \emph{per configuration}, putting a full closed-source comparison out of reach for us; see \cref{app:cost} for the per-model breakdown. Additionally, our experiments use $K=20$ scenarios. More scenarios may be needed for more statistical power. Another limitation is that \bname{} tests one specific task (REST API server implementation). While we believe the iterative modification paradigm generalizes, specific findings may not hold in other domains. We would like to note however that REST here is just a widely used interface convention that allows us to test code behavior in a black-box manner, and we argue that REST is broad enough to encompass a large portion of software engineering problems.

\textbf{Broader Impact.}
\bname{} is an evaluation benchmark with no direct deployment risk. Our results may influence how agent developers allocate research effort; we believe this redirection is net positive, as it encourages investment in multi-turn reliability and robust agent harnesses. We release all code and configurations to support transparency and reproducibility.

\bibliographystyle{plainnat}
\bibliography{references,references_manual}

\begin{thebibliography}{64}
\providecommand{\natexlab}[1]{#1}
\providecommand{\url}[1]{\texttt{#1}}
\expandafter\ifx\csname urlstyle\endcsname\relax
  \providecommand{\doi}[1]{doi: #1}\else
  \providecommand{\doi}{doi: \begingroup \urlstyle{rm}\Url}\fi

\bibitem[{Anomaly}()]{opencode}
{Anomaly}.
\newblock Opencode.
\newblock \url{https://github.com/anomalyco/opencode}.
\newblock Accessed: 2026-05-05.

\bibitem[{Anthropic}(2025)]{claudecode}
{Anthropic}.
\newblock Claude code.
\newblock \url{https://www.anthropic.com/claude-code}, 2025.
\newblock Accessed: 2026-05-06.

\bibitem[{Anysphere}(2025)]{cursor}
{Anysphere}.
\newblock Cursor: The ai code editor.
\newblock \url{https://cursor.com/}, 2025.
\newblock Accessed: 2026-05-06.

\bibitem[Austin et~al.(2021)Austin, Odena, Nye, Bosma, Michalewski, Dohan,
  Jiang, Cai, Terry, Le, and Sutton]{austin2021programsynthesislargelanguage}
Jacob Austin, Augustus Odena, Maxwell Nye, Maarten Bosma, Henryk Michalewski,
  David Dohan, Ellen Jiang, Carrie Cai, Michael Terry, Quoc Le, and Charles
  Sutton.
\newblock Program synthesis with large language models, 2021.
\newblock URL \url{https://arxiv.org/abs/2108.07732}.

\bibitem[Backlund and
  Petersson(2025)]{backlund2025vendingbenchbenchmarklongtermcoherence}
Axel Backlund and Lukas Petersson.
\newblock Vending-bench: A benchmark for long-term coherence of autonomous
  agents, 2025.
\newblock URL \url{https://arxiv.org/abs/2502.15840}.

\bibitem[Barres et~al.(2025)Barres, Dong, Ray, Si, and
  Narasimhan]{barres2025tau2benchevaluatingconversationalagents}
Victor Barres, Honghua Dong, Soham Ray, Xujie Si, and Karthik Narasimhan.
\newblock $\tau^2$-bench: Evaluating conversational agents in a dual-control
  environment, 2025.
\newblock URL \url{https://arxiv.org/abs/2506.07982}.

\bibitem[Cao et~al.(2026)Cao, Chen, Chen, Cui, Feng, Hui, Jing, Li, Li, Lin,
  Ma, Shum, Wang, Wei, Yang, Zhang, Zhang, Zhang, Zhao, and
  Zhou]{cao2026qwen3codernexttechnicalreport}
Ruisheng Cao, Mouxiang Chen, Jiawei Chen, Zeyu Cui, Yunlong Feng, Binyuan Hui,
  Yuheng Jing, Kaixin Li, Mingze Li, Junyang Lin, Zeyao Ma, Kashun Shum, Xuwu
  Wang, Jinxi Wei, Jiaxi Yang, Jiajun Zhang, Lei Zhang, Zongmeng Zhang, Wenting
  Zhao, and Fan Zhou.
\newblock Qwen3-coder-next technical report, 2026.
\newblock URL \url{https://arxiv.org/abs/2603.00729}.

\bibitem[Chen et~al.(2026)Chen, Xu, Wei, Chen, and
  Zhao]{chen2026swecievaluatingagentcapabilities}
Jialong Chen, Xander Xu, Hu~Wei, Chuan Chen, and Bing Zhao.
\newblock Swe-ci: Evaluating agent capabilities in maintaining codebases via
  continuous integration, 2026.
\newblock URL \url{https://arxiv.org/abs/2603.03823}.

\bibitem[Chen et~al.(2021)Chen, Tworek, Jun, Yuan, de~Oliveira~Pinto, Kaplan,
  Edwards, Burda, Joseph, Brockman, Ray, Puri, Krueger, Petrov, Khlaaf, Sastry,
  Mishkin, Chan, Gray, Ryder, Pavlov, Power, Kaiser, Bavarian, Winter, Tillet,
  Such, Cummings, Plappert, Chantzis, Barnes, Herbert-Voss, Guss, Nichol,
  Paino, Tezak, Tang, Babuschkin, Balaji, Jain, Saunders, Hesse, Carr, Leike,
  Achiam, Misra, Morikawa, Radford, Knight, Brundage, Murati, Mayer, Welinder,
  McGrew, Amodei, McCandlish, Sutskever, and
  Zaremba]{chen2021evaluatinglargelanguagemodels}
Mark Chen, Jerry Tworek, Heewoo Jun, Qiming Yuan, Henrique~Ponde
  de~Oliveira~Pinto, Jared Kaplan, Harri Edwards, Yuri Burda, Nicholas Joseph,
  Greg Brockman, Alex Ray, Raul Puri, Gretchen Krueger, Michael Petrov, Heidy
  Khlaaf, Girish Sastry, Pamela Mishkin, Brooke Chan, Scott Gray, Nick Ryder,
  Mikhail Pavlov, Alethea Power, Lukasz Kaiser, Mohammad Bavarian, Clemens
  Winter, Philippe Tillet, Felipe~Petroski Such, Dave Cummings, Matthias
  Plappert, Fotios Chantzis, Elizabeth Barnes, Ariel Herbert-Voss,
  William~Hebgen Guss, Alex Nichol, Alex Paino, Nikolas Tezak, Jie Tang, Igor
  Babuschkin, Suchir Balaji, Shantanu Jain, William Saunders, Christopher
  Hesse, Andrew~N. Carr, Jan Leike, Josh Achiam, Vedant Misra, Evan Morikawa,
  Alec Radford, Matthew Knight, Miles Brundage, Mira Murati, Katie Mayer, Peter
  Welinder, Bob McGrew, Dario Amodei, Sam McCandlish, Ilya Sutskever, and
  Wojciech Zaremba.
\newblock Evaluating large language models trained on code, 2021.
\newblock URL \url{https://arxiv.org/abs/2107.03374}.

\bibitem[Deng et~al.(2026)Deng, Chen, Yu, Fan, Liu, Yang, Parikh, Kannan, Cong,
  Wang, Zhang, Prasanna, Tang, and Wang]{deng2026evoclawevaluatingaiagents}
Gangda Deng, Zhaoling Chen, Zhongming Yu, Haoyang Fan, Yuhong Liu, Yuxin Yang,
  Dhruv Parikh, Rajgopal Kannan, Le~Cong, Mengdi Wang, Qian Zhang, Viktor
  Prasanna, Xiangru Tang, and Xingyao Wang.
\newblock Evoclaw: Evaluating ai agents on continuous software evolution, 2026.
\newblock URL \url{https://arxiv.org/abs/2603.13428}.

\bibitem[Deng et~al.(2025)Deng, Da, Pan, He, Ide, Garg, Lauffer, Park, Pasari,
  Rane, Sampath, Krishnan, Kundurthy, Hendryx, Wang, Bharadwaj, Holm, Aluri,
  Zhang, Jacobson, Liu, and Kenstler]{deng2025swebenchproaiagents}
Xiang Deng, Jeff Da, Edwin Pan, Yannis~Yiming He, Charles Ide, Kanak Garg,
  Niklas Lauffer, Andrew Park, Nitin Pasari, Chetan Rane, Karmini Sampath, Maya
  Krishnan, Srivatsa Kundurthy, Sean Hendryx, Zifan Wang, Vijay Bharadwaj, Jeff
  Holm, Raja Aluri, Chen Bo~Calvin Zhang, Noah Jacobson, Bing Liu, and Brad
  Kenstler.
\newblock Swe-bench pro: Can ai agents solve long-horizon software engineering
  tasks?, 2025.
\newblock URL \url{https://arxiv.org/abs/2509.16941}.

\bibitem[Ding et~al.(2026)Ding, Long, Pu, Zhou, Gao, Gao, He, Hou, Hu, Li, Shi,
  Wang, Zan, Zhang, Zhang, Chen, Cheng, Deng, Gu, Hua, Lin, Liu, Li, Pan, Peng,
  Qin, Shan, Tan, Xie, Wang, Yuan, Zhang, Zhao, Zhao, Zhu, Zhu, Zou, Ding,
  Jiao, Liu, Liu, Liu, Tao, Yang, Yang, Zhang, Chen, Huang, and
  Zhang]{ding2026nl2repobenchlonghorizonrepositorygeneration}
Jingzhe Ding, Shengda Long, Changxin Pu, Huan Zhou, Hongwan Gao, Xiang Gao,
  Chao He, Yue Hou, Fei Hu, Zhaojian Li, Weiran Shi, Zaiyuan Wang, Daoguang
  Zan, Chenchen Zhang, Xiaoxu Zhang, Qizhi Chen, Xianfu Cheng, Bo~Deng,
  Qingshui Gu, Kai Hua, Juntao Lin, Pai Liu, Mingchen Li, Xuanguang Pan, Zifan
  Peng, Yujia Qin, Yong Shan, Zhewen Tan, Weihao Xie, Zihan Wang, Yishuo Yuan,
  Jiayu Zhang, Enduo Zhao, Yunfei Zhao, He~Zhu, Liya Zhu, Chenyang Zou, Ming
  Ding, Jianpeng Jiao, Jiaheng Liu, Minghao Liu, Qian Liu, Chongyang Tao, Jian
  Yang, Tong Yang, Zhaoxiang Zhang, Xinjie Chen, Wenhao Huang, and Ge~Zhang.
\newblock Nl2repo-bench: Towards long-horizon repository generation evaluation
  of coding agents, 2026.
\newblock URL \url{https://arxiv.org/abs/2512.12730}.

\bibitem[Feng et~al.(2026)Feng, Sun, Yang, Ai, Li, Li, Zhang, He, Ma, Lin, Sun,
  Xiao, Zhou, Wu, Liu, Liu, Qiao, Zhang, and
  Zhang]{feng2026longclibenchpreliminarybenchmarkstudy}
Yukang Feng, Jianwen Sun, Zelai Yang, Jiaxin Ai, Chuanhao Li, Zizhen Li, Fanrui
  Zhang, Kang He, Rui Ma, Jifan Lin, Jie Sun, Yang Xiao, Sizhuo Zhou, Wenxiao
  Wu, Yiming Liu, Pengfei Liu, Yu~Qiao, Shenglin Zhang, and Kaipeng Zhang.
\newblock Longcli-bench: A preliminary benchmark and study for long-horizon
  agentic programming in command-line interfaces, 2026.
\newblock URL \url{https://arxiv.org/abs/2602.14337}.

\bibitem[Fielding(2000)]{fielding2000architectural}
Roy~Thomas Fielding.
\newblock \emph{Architectural styles and the design of network-based software
  architectures}.
\newblock Publication, University of California, Irvine, 2000.
\newblock URL
  \url{https://www.ics.uci.edu/~fielding/pubs/dissertation/top.htm}.

\bibitem[GLM-5-Team et~al.(2026)GLM-5-Team, :, Zeng, Lv, Hou, Du, Zheng, Chen,
  Yin, Ge, Huang, Xie, Zhu, Yin, Wang, Pan, Zeng, Zhang, Wang, Chen, Zhang,
  Jiao, Guo, Wang, Du, Wu, Wang, Li, Fan, Zhong, Liu, Zhao, Du, Dong, Lu,
  Shuang-Li, Cao, Liu, Jiang, Chen, Zhang, Huang, Dong, Xu, Wei, An, Niu, Zhu,
  Wen, Cen, Bai, Qiao, Wang, Wang, Zhu, Liu, Li, Wang, Wen, Huang, Cai, Yu, Li,
  Hu, Zhang, Zhang, Lin, Yang, Wang, Ai, Zhu, Yi, Chen, Wen, Sun, Zhao, Hu,
  Zhang, Liu, Zhang, Peng, Tai, Zhang, Liu, Wang, Yan, Ge, Liu, Chu, Zhao,
  Wang, Zhao, Ren, Wang, Zhang, Gui, Zhao, Li, An, Li, Yuan, Du, Liu, Zhi,
  Duan, Zhou, Wei, Wang, Luo, Zhang, Sha, Xu, Wu, Ding, Chen, Li, Lin, Ta, Zou,
  Song, Yang, Tu, Yang, Wu, Zhang, Li, Li, Fan, Qin, Tian, Zhang, Yu, Liang,
  Kuang, Cheng, Li, Yan, Hu, Ling, Fan, Xia, Zhang, Zhang, Pan, Zou, Zhang,
  Liu, Wu, Li, Wang, Zhu, Tan, Zhou, Pan, Zhang, Su, Geng, Yan, Tan, Bi, Shen,
  Yang, Li, Liu, Wang, Li, Wu, Zhang, Duan, Zhang, Liu, Jiang, Yan, Zhang, Wei,
  Chen, Feng, Yao, Chai, Wang, Zhang, Xu, Huang, Wang, Li, Dong, and
  Tang]{glm5team2026glm5vibecodingagentic}
GLM-5-Team, :, Aohan Zeng, Xin Lv, Zhenyu Hou, Zhengxiao Du, Qinkai Zheng, Bin
  Chen, Da~Yin, Chendi Ge, Chenghua Huang, Chengxing Xie, Chenzheng Zhu,
  Congfeng Yin, Cunxiang Wang, Gengzheng Pan, Hao Zeng, Haoke Zhang, Haoran
  Wang, Huilong Chen, Jiajie Zhang, Jian Jiao, Jiaqi Guo, Jingsen Wang,
  Jingzhao Du, Jinzhu Wu, Kedong Wang, Lei Li, Lin Fan, Lucen Zhong, Mingdao
  Liu, Mingming Zhao, Pengfan Du, Qian Dong, Rui Lu, Shuang-Li, Shulin Cao,
  Song Liu, Ting Jiang, Xiaodong Chen, Xiaohan Zhang, Xuancheng Huang, Xuezhen
  Dong, Yabo Xu, Yao Wei, Yifan An, Yilin Niu, Yitong Zhu, Yuanhao Wen, Yukuo
  Cen, Yushi Bai, Zhongpei Qiao, Zihan Wang, Zikang Wang, Zilin Zhu, Ziqiang
  Liu, Zixuan Li, Bojie Wang, Bosi Wen, Can Huang, Changpeng Cai, Chao Yu, Chen
  Li, Chengwei Hu, Chenhui Zhang, Dan Zhang, Daoyan Lin, Dayong Yang, Di~Wang,
  Ding Ai, Erle Zhu, Fangzhou Yi, Feiyu Chen, Guohong Wen, Hailong Sun, Haisha
  Zhao, Haiyi Hu, Hanchen Zhang, Hanrui Liu, Hanyu Zhang, Hao Peng, Hao Tai,
  Haobo Zhang, He~Liu, Hongwei Wang, Hongxi Yan, Hongyu Ge, Huan Liu, Huanpeng
  Chu, Jia'ni Zhao, Jiachen Wang, Jiajing Zhao, Jiamin Ren, Jiapeng Wang,
  Jiaxin Zhang, Jiayi Gui, Jiayue Zhao, Jijie Li, Jing An, Jing Li, Jingwei
  Yuan, Jinhua Du, Jinxin Liu, Junkai Zhi, Junwen Duan, Kaiyue Zhou, Kangjian
  Wei, Ke~Wang, Keyun Luo, Laiqiang Zhang, Leigang Sha, Liang Xu, Lindong Wu,
  Lintao Ding, Lu~Chen, Minghao Li, Nianyi Lin, Pan Ta, Qiang Zou, Rongjun
  Song, Ruiqi Yang, Shangqing Tu, Shangtong Yang, Shaoxiang Wu, Shengyan Zhang,
  Shijie Li, Shuang Li, Shuyi Fan, Wei Qin, Wei Tian, Weining Zhang, Wenbo Yu,
  Wenjie Liang, Xiang Kuang, Xiangmeng Cheng, Xiangyang Li, Xiaoquan Yan,
  Xiaowei Hu, Xiaoying Ling, Xing Fan, Xingye Xia, Xinyuan Zhang, Xinze Zhang,
  Xirui Pan, Xu~Zou, Xunkai Zhang, Yadi Liu, Yandong Wu, Yanfu Li, Yidong Wang,
  Yifan Zhu, Yijun Tan, Yilin Zhou, Yiming Pan, Ying Zhang, Yinpei Su, Yipeng
  Geng, Yong Yan, Yonglin Tan, Yuean Bi, Yuhan Shen, Yuhao Yang, Yujiang Li,
  Yunan Liu, Yunqing Wang, Yuntao Li, Yurong Wu, Yutao Zhang, Yuxi Duan, Yuxuan
  Zhang, Zezhen Liu, Zhengtao Jiang, Zhenhe Yan, Zheyu Zhang, Zhixiang Wei,
  Zhuo Chen, Zhuoer Feng, Zijun Yao, Ziwei Chai, Ziyuan Wang, Zuzhou Zhang, Bin
  Xu, Minlie Huang, Hongning Wang, Juanzi Li, Yuxiao Dong, and Jie Tang.
\newblock Glm-5: from vibe coding to agentic engineering, 2026.
\newblock URL \url{https://arxiv.org/abs/2602.15763}.

\bibitem[{Google}(2025)]{geminicli}
{Google}.
\newblock Gemini cli.
\newblock \url{https://github.com/google-gemini/gemini-cli}, 2025.
\newblock Accessed: 2026-05-06.

\bibitem[Han et~al.(2025)Han, won Hwang, Samdani, and
  He]{han2025convcodeworldbenchmarkingconversationalcode}
Hojae Han, Seung won Hwang, Rajhans Samdani, and Yuxiong He.
\newblock Convcodeworld: Benchmarking conversational code generation in
  reproducible feedback environments, 2025.
\newblock URL \url{https://arxiv.org/abs/2502.19852}.

\bibitem[Holm(1979)]{holm}
Sture Holm.
\newblock A simple sequentially rejective multiple test procedure.
\newblock \emph{Scandinavian Journal of Statistics}, 6\penalty0 (2):\penalty0
  65--70, 1979.
\newblock ISSN 03036898, 14679469.
\newblock URL \url{http://www.jstor.org/stable/4615733}.

\bibitem[Hong et~al.(2025)Hong, Troynikov, and Huber]{hong2025context}
Kelly Hong, Anton Troynikov, and Jeff Huber.
\newblock Context rot: How increasing input tokens impacts llm performance.
\newblock Technical report, Chroma, July 2025.
\newblock URL \url{https://trychroma.com/research/context-rot}.

\bibitem[Hsieh et~al.(2024)Hsieh, Sun, Kriman, Acharya, Rekesh, Jia, Zhang, and
  Ginsburg]{hsieh2024rulerwhatsrealcontext}
Cheng-Ping Hsieh, Simeng Sun, Samuel Kriman, Shantanu Acharya, Dima Rekesh, Fei
  Jia, Yang Zhang, and Boris Ginsburg.
\newblock Ruler: What's the real context size of your long-context language
  models?, 2024.
\newblock URL \url{https://arxiv.org/abs/2404.06654}.

\bibitem[Jain et~al.(2025)Jain, Singh, Shetty, Zheng, Sen, and
  Stoica]{jain2025r2egymproceduralenvironmentshybrid}
Naman Jain, Jaskirat Singh, Manish Shetty, Liang Zheng, Koushik Sen, and Ion
  Stoica.
\newblock R2e-gym: Procedural environments and hybrid verifiers for scaling
  open-weights swe agents, 2025.
\newblock URL \url{https://arxiv.org/abs/2504.07164}.

\bibitem[Jiang et~al.(2023)Jiang, Wu, Lin, Yang, and
  Qiu]{jiang2023llmlinguacompressingpromptsaccelerated}
Huiqiang Jiang, Qianhui Wu, Chin-Yew Lin, Yuqing Yang, and Lili Qiu.
\newblock Llmlingua: Compressing prompts for accelerated inference of large
  language models, 2023.
\newblock URL \url{https://arxiv.org/abs/2310.05736}.

\bibitem[Jimenez et~al.(2024)Jimenez, Yang, Wettig, Yao, Pei, Press, and
  Narasimhan]{jimenez2024swebenchlanguagemodelsresolve}
Carlos~E. Jimenez, John Yang, Alexander Wettig, Shunyu Yao, Kexin Pei, Ofir
  Press, and Karthik Narasimhan.
\newblock Swe-bench: Can language models resolve real-world github issues?
\newblock In \emph{International Conference on Learning Representations
  (ICLR)}, 2024.
\newblock URL \url{https://arxiv.org/abs/2310.06770}.

\bibitem[Kamradt(2023)]{kamradt2023needle}
Gregory Kamradt.
\newblock Needle in a haystack - pressure testing llms.
\newblock \url{https://github.com/gkamradt/LLMTest_NeedleInAHaystack}, 2023.
\newblock Accessed: 2026-05-06.

\bibitem[Kapoor et~al.(2025)Kapoor, Stroebl, Kirgis, Nadgir, Siegel, Wei, Xue,
  Chen, Chen, Utpala, Ndzomga, Oruganty, Luskin, Liu, Yu, Arora, Hahm, Trivedi,
  Sun, Lee, Jin, Mai, Zhou, Zhu, Bommasani, Kang, Song, Henderson, Su, Liang,
  and Narayanan]{kapoor2025holisticagentleaderboardmissing}
Sayash Kapoor, Benedikt Stroebl, Peter Kirgis, Nitya Nadgir, Zachary~S Siegel,
  Boyi Wei, Tianci Xue, Ziru Chen, Felix Chen, Saiteja Utpala, Franck Ndzomga,
  Dheeraj Oruganty, Sophie Luskin, Kangheng Liu, Botao Yu, Amit Arora, Dongyoon
  Hahm, Harsh Trivedi, Huan Sun, Juyong Lee, Tengjun Jin, Yifan Mai, Yifei
  Zhou, Yuxuan Zhu, Rishi Bommasani, Daniel Kang, Dawn Song, Peter Henderson,
  Yu~Su, Percy Liang, and Arvind Narayanan.
\newblock Holistic agent leaderboard: The missing infrastructure for ai agent
  evaluation, 2025.
\newblock URL \url{https://arxiv.org/abs/2510.11977}.

\bibitem[Kwa et~al.(2026)Kwa, West, Becker, Deng, Garcia, Hasin, Jawhar,
  Kinniment, Rush, Arx, Bloom, Broadley, Du, Goodrich, Jurkovic, Miles, Nix,
  Lin, Parikh, Rein, Sato, Wijk, Ziegler, Barnes, and
  Chan]{kwa2026measuringaiabilitycomplete}
Thomas Kwa, Ben West, Joel Becker, Amy Deng, Katharyn Garcia, Max Hasin, Sami
  Jawhar, Megan Kinniment, Nate Rush, Sydney~Von Arx, Ryan Bloom, Thomas
  Broadley, Haoxing Du, Brian Goodrich, Nikola Jurkovic, Luke~Harold Miles,
  Seraphina Nix, Tao Lin, Neev Parikh, David Rein, Lucas Jun~Koba Sato, Hjalmar
  Wijk, Daniel~M. Ziegler, Elizabeth Barnes, and Lawrence Chan.
\newblock Measuring ai ability to complete long software tasks, 2026.
\newblock URL \url{https://arxiv.org/abs/2503.14499}.

\bibitem[Laban et~al.(2025)Laban, Hayashi, Zhou, and
  Neville]{laban2025llmslostmultiturnconversation}
Philippe Laban, Hiroaki Hayashi, Yingbo Zhou, and Jennifer Neville.
\newblock Llms get lost in multi-turn conversation, 2025.
\newblock URL \url{https://arxiv.org/abs/2505.06120}.

\bibitem[Liu et~al.(2023)Liu, Lin, Hewitt, Paranjape, Bevilacqua, Petroni, and
  Liang]{liu2023lostmiddlelanguagemodels}
Nelson~F. Liu, Kevin Lin, John Hewitt, Ashwin Paranjape, Michele Bevilacqua,
  Fabio Petroni, and Percy Liang.
\newblock Lost in the middle: How language models use long contexts, 2023.
\newblock URL \url{https://arxiv.org/abs/2307.03172}.

\bibitem[Liu et~al.(2024)Liu, Yu, Zhang, Xu, Lei, Lai, Gu, Ding, Men, Yang,
  Zhang, Deng, Zeng, Du, Zhang, Shen, Zhang, Su, Sun, Huang, Dong, and
  Tang]{liu2025agentbenchevaluatingllmsagents}
Xiao Liu, Hao Yu, Hanchen Zhang, Yifan Xu, Xuanyu Lei, Hanyu Lai, Yu~Gu,
  Hangliang Ding, Kaiwen Men, Kejuan Yang, Shudan Zhang, Xiang Deng, Aohan
  Zeng, Zhengxiao Du, Chenhui Zhang, Sheng Shen, Tianjun Zhang, Yu~Su, Huan
  Sun, Minlie Huang, Yuxiao Dong, and Jie Tang.
\newblock Agentbench: Evaluating llms as agents.
\newblock In \emph{International Conference on Learning Representations
  (ICLR)}, 2024.
\newblock URL \url{https://arxiv.org/abs/2308.03688}.

\bibitem[Merrill et~al.(2026)Merrill, Shaw, Carlini, Li, Raj, Bercovich, Shi,
  Shin, Walshe, Buchanan, Shen, Ye, Lin, Poulos, Wang, Nezhurina, Jitsev, Lu,
  Mastromichalakis, Xu, Chen, Liu, Zhang, Chen, Kashyap, Uslu, Li, Wu, Yan,
  Bian, Sharma, Sun, Dillmann, Anand, Lanpouthakoun, Koopah, Hu, Guha, Dreiman,
  Zhu, Krauth, Zhong, Muennighoff, Amanfu, Tan, Pimpalgaonkar, Aggarwal, Lin,
  Lan, Zhao, Liang, Wang, Wang, Zhou, Heineman, Liu, Trivedi, Yang, Lin,
  Shetty, Yang, Omi, Raoof, Li, Zhuo, Lin, Dai, Wang, Chai, Zhou, Wahdany, She,
  Hu, Dong, Zhu, Cui, Saiyed, Kolbeinsson, Hu, Rytting, Marten, Wang, Dimakis,
  Konwinski, and Schmidt]{merrill2026terminalbenchbenchmarkingagentshard}
Mike~A. Merrill, Alexander~G. Shaw, Nicholas Carlini, Boxuan Li, Harsh Raj,
  Ivan Bercovich, Lin Shi, Jeong~Yeon Shin, Thomas Walshe, E.~Kelly Buchanan,
  Junhong Shen, Guanghao Ye, Haowei Lin, Jason Poulos, Maoyu Wang, Marianna
  Nezhurina, Jenia Jitsev, Di~Lu, Orfeas~Menis Mastromichalakis, Zhiwei Xu,
  Zizhao Chen, Yue Liu, Robert Zhang, Leon~Liangyu Chen, Anurag Kashyap,
  Jan-Lucas Uslu, Jeffrey Li, Jianbo Wu, Minghao Yan, Song Bian, Vedang Sharma,
  Ke~Sun, Steven Dillmann, Akshay Anand, Andrew Lanpouthakoun, Bardia Koopah,
  Changran Hu, Etash Guha, Gabriel H.~S. Dreiman, Jiacheng Zhu, Karl Krauth,
  Li~Zhong, Niklas Muennighoff, Robert Amanfu, Shangyin Tan, Shreyas
  Pimpalgaonkar, Tushar Aggarwal, Xiangning Lin, Xin Lan, Xuandong Zhao, Yiqing
  Liang, Yuanli Wang, Zilong Wang, Changzhi Zhou, David Heineman, Hange Liu,
  Harsh Trivedi, John Yang, Junhong Lin, Manish Shetty, Michael Yang, Nabil
  Omi, Negin Raoof, Shanda Li, Terry~Yue Zhuo, Wuwei Lin, Yiwei Dai, Yuxin
  Wang, Wenhao Chai, Shang Zhou, Dariush Wahdany, Ziyu She, Jiaming Hu, Zhikang
  Dong, Yuxuan Zhu, Sasha Cui, Ahson Saiyed, Arinbjörn Kolbeinsson, Jesse Hu,
  Christopher~Michael Rytting, Ryan Marten, Yixin Wang, Alex Dimakis, Andy
  Konwinski, and Ludwig Schmidt.
\newblock Terminal-bench: Benchmarking agents on hard, realistic tasks in
  command line interfaces, 2026.
\newblock URL \url{https://arxiv.org/abs/2601.11868}.

\bibitem[{Mistral AI}()]{mistralvibe}
{Mistral AI}.
\newblock Mistral vibe.
\newblock \url{https://github.com/mistralai/mistral-vibe}.
\newblock Accessed: 2026-05-05.

\bibitem[{Mistral AI}(2025)]{mistral2025devstral2}
{Mistral AI}.
\newblock Devstral 2 and vibe cli.
\newblock \url{https://mistral.ai/news/devstral-2-vibe-cli}, 2025.
\newblock Accessed: 2026-05-05.

\bibitem[{Moonshot AI}()]{kimicli}
{Moonshot AI}.
\newblock Kimi cli.
\newblock \url{https://github.com/MoonshotAI/kimi-cli}.
\newblock Accessed: 2026-05-05.

\bibitem[NVIDIA et~al.(2026)NVIDIA, :, Chandiramani, Blakeman, Olaoye, Gupta,
  Somasamudramath, Khattar, Adesoba, Renduchintala, Asif, Agrawal, Vavre,
  Kiswani, Padmakumar, Hotchandani, Shukla, Bercovich, Ficek, Shaposhnikov,
  Gronskiy, Kondratenko, Neefus, Steiner, Yang, Bukharin, Young, Hatamizadeh,
  Taghibakhshi, Galiautdinova, Liu, Kumar, Mahabaleshwarkar, Klein, Zuker,
  Geifman, Bhiwandiwalla, Subramaniam, Tao, Shrivastava, Agrusa, Srivastava,
  Verma, Guan, Shors, Chockalingam, Mandarwal, Ramani, Mehta, Jain, Venkatesan,
  Anoosheh, Aithal, Poojary, Ahamed, Mishra, Demiroz, Thekkumpate,
  Sohrabizadeh, Kaur, Dattagupta, Anandan, Sadeghi, Simkin, Lanir, Schifferer,
  Chislett, Nushi, Kartal, Thiede, Rouhani, Chen, Ginsburg, Norick, Kisacanin,
  Yu, Catanzaro, Mani, del Mundo, Lee, Kim, Hwang, Ni, Wang, Truong, Hsieh, Yu,
  Luo, Wang, Mungekar, Patel, Alexiuk, Holguin, Wing, Munley, Parisien, Desai,
  Sheng, Neale, Meurillon, Kumar, Gil, Su, Corneil, Afrimi, Triana, Egert,
  Fatade, Lo, Rohrer, Serebrenik, Sorokin, Gitman, Levy, Stosic, Edelsohn,
  Messina, Mosallanezhad, Tamok, Donia, Narayanan, O'Kelly, Peri, Nathawani,
  Wu, Rekesh, Yared, Kakwani, Tuttle, Ahn, Jiang, Poorkay, O'Flaherty, Riach,
  Stosic, Stee, Minasyan, Lin, Long, Segal, Lantz, Lewis, Evans, Ning, Chung,
  Harper, Pham-Hung, Tramel, Galinkin, Pounds, Etrog, Briones, Wu, Bakhturina,
  Tsykunov, Dobrowolska, Movahed, Memarian, Wang, Jia, Soares, Frujeri, Chen,
  Lin, Galko, Zhang, Siino, Hou, Bhatt, Prasad, Venkataramani, Gupta,
  Armstrong, Shen, Borghesi, Neskovic, Batmaz, Lam, Wu, Pauloski, Davis,
  Nalbandyan, Zhang, Farber, Huang, Qian, Kumar, Kim, Sharma, Iso, Ross, Hum,
  Sahota, Wang, Soni, Upadhyay, Nguyen, Cunningham, Galil, Shahaf, Padovani,
  Gitman, Shovkun, Dhillon, Loshchilov, Kelly, Schen, Levy, Moshkov, Golan,
  Putterman, Tu, Baczek, Kautz, Scowcroft, Rosenberg, Casper, Pflum, Grant,
  Sewall, Mitra, Glick, Chen, Oliver, Xu, Zhu, Song, Zhang, Zeng, Lou, Milton,
  Chow, Zhang, Choi, Huang, Huang, Caruso, Conway, Guman, Jatko, Kamalu, Greco,
  Cohen, Raiman, Jennings, Daw, Yu, Tapia, Yi, Parmar, Achar, Briski, Mattoo,
  Cheung, Luna, Wyss, Shih, Kong, Nguyen, Bhardwaj, Buryak, Sivamani,
  Krommydas, Murphy, Puvvada, Pawelec, Anik, Tewari, Sleiman, Du, Derczynski,
  Ding, Ilan, Wu, Wei, Vega, Su, Segbroeck, de~Melo, Zhang, Fathi, Sreedhar,
  Sreedhar, Chandran, Gomez, Ashkenazi, Cuevas, Romeijn, Zhang, Cai, Gabel,
  Kliegl, Patelka, Moosaei, Varacalli, Novikov, Ferrato, Samadi, Corpuz, Xin,
  Wang, Wang, Price, Schaffer, Andersch, Boone, Evans, Wang, Martinez, Khona,
  Chrzanowski, Hollinger, Ma, Lee, Dabbah, Shoeybi, Patwary, Mulepati, Khalil,
  Nabwani, Agarwal, Balasubramaniam, Hennouni, Kodukula, Hereth, Pinckney,
  Assaf, Habibi, Qin, Zmora, Haber, Reamaroon, Quak, Bhatia, Jukar, Pope,
  Ludwig, Tajbakhsh, Ailon, Juluru, De, Pitt, Rybakov, Hrinchuk, Kuchaiev,
  Delalleau, Olabiyi, Argov, Almog, Puny, Tropp, Padovani, Xie, Chadha, Shamis,
  Gibbons, Molchanov, Belcak, Jin, Xu, Januszewski, Jannaty, Shevate, Thalasta,
  Thombre, Varshney, Gambhir, Gundecha, Tredak, Miao, Wan, Minh, Mahabadi,
  Oberman, Garg, Kandu, Zhong, El-Yaniv, Zilberstein, Shafipour, Yao, Pi,
  Mazzarese, Wang, Izzo, Singla, Shahbazyan, Garg, Borkar, Gala, Islam, Clark,
  Hesse, Waleffe, Kalidindi, Watve, Koren, Fan, Kharwar, Cai, Zhang, Hewett,
  Prenger, Timbrook, Egashira, Mahdavi, Joshi, Modi, Kriman, Pombra, Kariyappa,
  Satheesh, Pombo, Kaji, Pasumarthi, Mishra, Muralidharan, Hara, Narenthiran,
  Rogawski, Na, Bak, Sameni, Poulos, Mor, Acharya, Lord, Sreenivas, Kotek,
  Gharghabi, Thomas, Lin, Likhite, Fan, Chen, Gopal, Prabhumoye, Pachori,
  Toshniwal, Zhang, Ding, Renjith, Prayaga, Jain, Sun, Rella, Das, Ithape, S,
  Majumdar, Singhal, Singudasu, Niverty, Sergienko, Gloginic, Alborghetti, Ge,
  McCullough, Devare, Velury, Rao, Barua, Gai, Panguluri, Koundinyan, Patnam,
  Priyadarshi, Bhendigeri, Akter, Arunagiri, Yuan, Abramovich, Bui, Yu, Kong,
  Do, Gburek, Marques, Moore, Blankevoort, Moon, Ma, Mitra, Grzegorzek, Asida,
  Natan, Keren, Ronen, Rebedea, Starkey, Konuk, Vashishth, Condensa, Karpas,
  De, Noorozi, Noroozi, Shah, Vaidyanathan, Srinivasan, Elango, Cui,
  Korthikanti, Mehta, Adams, Wu, Kurin, Lavrukhin, Anisimov, Seo, Jiang, Ahmad,
  Du, Ping, Chen, Quan, Dai, Gao, Jennings, Zhang, Ren, Xin, Li, Yu, Chen,
  Galron, Karnati, Choi, Meyer, Wu, Zhang, Lin, Geifman, Fu, Suhara, Kwon,
  Zhang, Huang, Moshe, Wang, Cheng, Zhu, Yang, Liu, Chen, Yan, and
  Ahmed]{nvidia2026nemotron3superopen}
NVIDIA, :, Aakshita Chandiramani, Aaron Blakeman, Abdullahi Olaoye, Abhibha
  Gupta, Abhilash Somasamudramath, Abhinav Khattar, Adeola Adesoba, Adi
  Renduchintala, Adil Asif, Aditya Agrawal, Aditya Vavre, Ahmad Kiswani,
  Aishwarya Padmakumar, Ajay Hotchandani, Akanksha Shukla, Akhiad Bercovich,
  Aleksander Ficek, Aleksandr Shaposhnikov, Alex Gronskiy, Alex Kondratenko,
  Alex Neefus, Alex Steiner, Alex Yang, Alexander Bukharin, Alexander Young,
  Ali Hatamizadeh, Ali Taghibakhshi, Alina Galiautdinova, Alisa Liu, Alok
  Kumar, Ameya~Sunil Mahabaleshwarkar, Amir Klein, Amit Zuker, Amnon Geifman,
  Anahita Bhiwandiwalla, Ananth Subramaniam, Andrew Tao, Anjaney Shrivastava,
  Anjulie Agrusa, Ankur Srivastava, Ankur Verma, Ann Guan, Anna Shors,
  Annamalai Chockalingam, Anubhav Mandarwal, Aparnaa Ramani, Arham Mehta, Arti
  Jain, Arun Venkatesan, Asha Anoosheh, Ashwath Aithal, Ashwin Poojary, Asif
  Ahamed, Asit Mishra, Asli~Sabanci Demiroz, Asma~Kuriparambil Thekkumpate,
  Atefeh Sohrabizadeh, Avinash Kaur, Ayush Dattagupta, Barath~Subramaniam
  Anandan, Bardiya Sadeghi, Barnaby Simkin, Ben Lanir, Benedikt Schifferer,
  Benjamin Chislett, Besmira Nushi, Bilal Kartal, Bill Thiede, Bita~Darvish
  Rouhani, Bobby Chen, Boris Ginsburg, Brandon Norick, Branislav Kisacanin,
  Brian Yu, Bryan Catanzaro, Buvaneswari Mani, Carlo del Mundo, Chankyu Lee,
  Chanran Kim, Chantal Hwang, Chao Ni, Charles Wang, Charlie Truong, Cheng-Ping
  Hsieh, Chenhan Yu, Chenjie Luo, Cherie Wang, Chetan Mungekar, Chintan Patel,
  Chris Alexiuk, Chris Holguin, Chris Wing, Christian Munley, Christopher
  Parisien, Chuck Desai, Chunyang Sheng, Collin Neale, Cyril Meurillon, Dakshi
  Kumar, Dan Gil, Dan Su, Dane Corneil, Daniel Afrimi, Daniel Burkhardt~Eliuth
  Triana, Daniel Egert, Daniel Fatade, Daniel Lo, Daniel Rohrer, Daniel
  Serebrenik, Daniil Sorokin, Daria Gitman, Daria Levy, Darko Stosic, David
  Edelsohn, David Messina, David Mosallanezhad, David Tamok, Deena Donia,
  Deepak Narayanan, Devin O'Kelly, Dheeraj Peri, Dhruv Nathawani, Di~Wu, Dima
  Rekesh, Dina Yared, Divyanshu Kakwani, Dmitry Konyagin~Brandon Tuttle, Dong
  Ahn, Dongfu Jiang, Dorrin Poorkay, Douglas O'Flaherty, Duncan Riach, Dusan
  Stosic, Dustin~Van Stee, Edgar Minasyan, Edward Lin, Eileen~Peters Long, Elad
  Segal, Elena Lantz, Elena Lewis, Ellie Evans, Elliott Ning, Eric Chung, Eric
  Harper, Eric Pham-Hung, Eric~W. Tramel, Erick Galinkin, Erik Pounds, Esti
  Etrog, Evan Briones, Evan Wu, Evelina Bakhturina, Evgeny Tsykunov, Ewa
  Dobrowolska, Farshad~Saberi Movahed, Farzan Memarian, Fay Wang, Fei Jia,
  Felipe Soares, Felipe~Vieira Frujeri, Feng Chen, Fengguang Lin, Ferenc Galko,
  Fortuna Zhang, Frankie Siino, Frida Hou, Gantavya Bhatt, Gargi Prasad,
  Geethapriya Venkataramani, Geetika Gupta, George Armstrong, Gerald Shen,
  Giulio Borghesi, Gordana Neskovic, Gorkem Batmaz, Grace Lam, Grace Wu, Greg
  Pauloski, Greyson Davis, Grigor Nalbandyan, Guoming Zhang, Guy Farber, Guyue
  Huang, Haifeng Qian, Haran Kumar~Shiv Kumar, Harry Kim, Harsh Sharma, Hayate
  Iso, Hayley Ross, Herbert Hum, Herman Sahota, Hexin Wang, Himanshu Soni,
  Hiren Upadhyay, Huy Nguyen, Iain Cunningham, Ido Galil, Ido Shahaf, Igino
  Padovani, Igor Gitman, Igor Shovkun, Ikroop Dhillon, Ilya Loshchilov, Ingrid
  Kelly, Itamar Schen, Itay Levy, Ivan Moshkov, Izik Golan, Izzy Putterman,
  Jain Tu, Jan Baczek, Jan Kautz, Jane~Polak Scowcroft, Janica Rosenberg, Jared
  Casper, Jarrod Pflum, Jason Grant, Jason Sewall, Jatin Mitra, Jeffrey Glick,
  Jenny Chen, Jesse Oliver, Jiacheng Xu, Jiafan Zhu, Jialin Song, Jian Zhang,
  Jiaqi Zeng, Jie Lou, Jill Milton, Jim Chow, Jimmy Zhang, Jinhang Choi, Jining
  Huang, Jocelyn Huang, Joel Caruso, Joey Conway, Joey Guman, Johan Jatko, John
  Kamalu, Johnny Greco, Jonathan Cohen, Jonathan Raiman, Joseph Jennings,
  Joyjit Daw, Juan Yu, Julio Tapia, Junkeun Yi, Jupinder Parmar, Jyothi Achar,
  Kari Briski, Kartik Mattoo, Katherine Cheung, Katherine Luna, Keith Wyss,
  Kevin Shih, Kezhi Kong, Khanh Nguyen, Khushi Bhardwaj, Kirill Buryak,
  Kirthi~Shankar Sivamani, Konstantinos Krommydas, Kris Murphy, Krishna~C.
  Puvvada, Krzysztof Pawelec, Kumar Anik, Laikh Tewari, Laya Sleiman, Leo Du,
  Leon Derczynski, Li~Ding, Lilach Ilan, Lingjie Wu, Lizzie Wei, Luis Vega, Lun
  Su, Maarten~Van Segbroeck, Maer~Rodrigues de~Melo, Magaret Zhang, Mahan
  Fathi, Makesh~Narsimhan Sreedhar, Makesh Sreedhar, Makesh~Tarun Chandran,
  Manuel~Reyes Gomez, Maor Ashkenazi, Marc Cuevas, Marc Romeijn, Margaret
  Zhang, Mark Cai, Mark Gabel, Markus Kliegl, Martyna Patelka, Maryam Moosaei,
  Matthew Varacalli, Matvei Novikov, Mauricio Ferrato, Mehrzad Samadi, Melissa
  Corpuz, Meng Xin, Mengdi Wang, Mengru Wang, Meredith Price, Micah Schaffer,
  Michael Andersch, Michael Boone, Michael Evans, Michael~Z Wang, Miguel
  Martinez, Mikail Khona, Mike Chrzanowski, Mike Hollinger, Mingyuan Ma,
  Minseok Lee, Mohammad Dabbah, Mohammad Shoeybi, Mostofa Patwary, Nabin
  Mulepati, Nader Khalil, Najeeb Nabwani, Nancy Agarwal, Nanthini
  Balasubramaniam, Narimane Hennouni, Narsi Kodukula, Natalie Hereth, Nathaniel
  Pinckney, Nave Assaf, Negar Habibi, Nestor Qin, Neta Zmora, Netanel Haber,
  Nick Reamaroon, Nickson Quak, Nidhi Bhatia, Nikhil Jukar, Nikki Pope, Nikolai
  Ludwig, Nima Tajbakhsh, Nir Ailon, Nirmal Juluru, Nirmalya De, Nowel Pitt,
  Oleg Rybakov, Oleksii Hrinchuk, Oleksii Kuchaiev, Olivier Delalleau,
  Oluwatobi Olabiyi, Omer~Ullman Argov, Omri Almog, Omri Puny, Oren Tropp,
  Otavio Padovani, Ouye Xie, Parth Chadha, Pasha Shamis, Paul Gibbons, Pavlo
  Molchanov, Peter Belcak, Peter Jin, Pinky Xu, Piotr Januszewski, Pooya
  Jannaty, Prachi Shevate, Pradeep Thalasta, Pranav~Prashant Thombre, Prasoon
  Varshney, Prerana Gambhir, Pritam Gundecha, Przemek Tredak, Qing Miao, Qiyu
  Wan, Quan~Tran Minh, Rabeeh~Karimi Mahabadi, Rachel Oberman, Rachit Garg,
  Rahul Kandu, Raina Zhong, Ran El-Yaniv, Ran Zilberstein, Rasoul Shafipour,
  Renee Yao, Renjie Pi, Richard Mazzarese, Richard Wang, Rick Izzo, Ridhima
  Singla, Rima Shahbazyan, Rishabh Garg, Ritika Borkar, Ritu Gala, Riyad Islam,
  Robert Clark, Robert Hesse, Roger Waleffe, Rohit~Varma Kalidindi, Rohit
  Watve, Roi Koren, Ron Fan, Ruchika Kharwar, Ruisi Cai, Ruoxi Zhang,
  Russell~J. Hewett, Ryan Prenger, Ryan Timbrook, Ryota Egashira, Sadegh
  Mahdavi, Sagar Singh~Ashutosh Joshi, Sahil Modi, Samuel Kriman, Sandeep
  Pombra, Sanjay Kariyappa, Sanjeev Satheesh, Santiago Pombo, Saori Kaji,
  Satish Pasumarthi, Saurav Mishra, Saurav Muralidharan, Scott Hara, Sean
  Narenthiran, Sebastian Rogawski, Seonjin Na, Seonmyeong Bak, Sepehr Sameni,
  Seth Poulos, Shahar Mor, Shantanu Acharya, Shaona Ghosh~Adam Lord,
  Sharath~Turuvekere Sreenivas, Shaun Kotek, Shaya Gharghabi, Shelby Thomas,
  Sheng-Chieh Lin, Shibani Likhite, Shiqing Fan, Shiyang Chen, Shreya Gopal,
  Shrimai Prabhumoye, Shubham Pachori, Shubham Toshniwal, Shuo Zhang, Shuoyang
  Ding, Shyam Renjith, Shyamala Prayaga, Siddhartha Jain, Simeng Sun, Sirisha
  Rella, Sirshak Das, Smita Ithape, Sneha~Harishchandra S, Somshubra Majumdar,
  Soumye Singhal, Sri~Harsha Singudasu, Sriharsha Niverty, Stas Sergienko,
  Stefana Gloginic, Stefania Alborghetti, Stephen Ge, Stephen McCullough,
  Sugam~Dipak Devare, Suguna~Varshini Velury, Sukrit Rao, Sumeet~Kumar Barua,
  Sunny Gai, Suseella Panguluri, Sushil Koundinyan, Swathi Patnam, Sweta
  Priyadarshi, Swetha Bhendigeri, Syeda~Nahida Akter, Sylendran Arunagiri,
  Tailling Yuan, Talor Abramovich, Tan Bui, Tan Yu, Terry Kong, Thanh Do,
  Thomas Gburek, Thorgane Marques, Tiffany Moore, Tijmen Blankevoort, Tim Moon,
  Timothy Ma, Tiyasa Mitra, Tomasz Grzegorzek, Tomer Asida, Tomer~Bar Natan,
  Tomer Keren, Tomer Ronen, Traian Rebedea, Trenton Starkey, Tugrul Konuk,
  Twinkle Vashishth, Tyler Condensa, Udi Karpas, Ushnish De, Vahid Noorozi,
  Vahid Noroozi, Vanshil~Atul Shah, Veena Vaidyanathan, Venkat Srinivasan,
  Venmugil Elango, Victor Cui, Vijay Korthikanti, Vikas Mehta, Virginia Adams,
  Virginia Wu, Vitaly Kurin, Vitaly Lavrukhin, Vladimir Anisimov, Wan Seo,
  Wanli Jiang, Wasi~Uddin Ahmad, Wei Du, Wei Ping, Wei-Ming Chen, Wendy Quan,
  Wenliang Dai, Wenwen Gao, Will Jennings, William Zhang, Xiaowei Ren, Xiaowen
  Xin, Xin Li, Yang Yu, Yangyi Chen, Yaniv Galron, Yashaswi Karnati, Yejin
  Choi, Yev Meyer, Yi-Fu Wu, Yian Zhang, Ying Lin, Yonatan Geifman, Yonggan Fu,
  Yoshi Suhara, Youngeun Kwon, Yuan Zhang, Yuki Huang, Zach Moshe, Zhilin Wang,
  Zhiyu Cheng, Zhongbo Zhu, Zhuolin Yang, Zihan Liu, Zijia Chen, Zijie Yan, and
  Zuhair Ahmed.
\newblock Nemotron 3 super: Open, efficient mixture-of-experts hybrid
  mamba-transformer model for agentic reasoning, 2026.
\newblock URL \url{https://arxiv.org/abs/2604.12374}.

\bibitem[{OpenAI}(2025)]{codex}
{OpenAI}.
\newblock Openai codex.
\newblock \url{https://openai.com/codex/}, 2025.
\newblock Accessed: 2026-05-06.

\bibitem[{OpenAPI Initiative}()]{openapi}
{OpenAPI Initiative}.
\newblock Openapi initiative.
\newblock \url{https://www.openapis.org/}.
\newblock Accessed: 2026-05-06.

\bibitem[{OpenHands}()]{openhands}
{OpenHands}.
\newblock Openhands.
\newblock \url{https://github.com/OpenHands/OpenHands}.
\newblock Accessed: 2026-05-05.

\bibitem[Orlanski et~al.(2026)Orlanski, Roy, Yun, Shin, Gu, Ge, Adila, Sala,
  and Albarghouthi]{orlanski2026slopcodebenchbenchmarkingcodingagents}
Gabriel Orlanski, Devjeet Roy, Alexander Yun, Changho Shin, Alex Gu, Albert Ge,
  Dyah Adila, Frederic Sala, and Aws Albarghouthi.
\newblock Slopcodebench: Benchmarking how coding agents degrade over
  long-horizon iterative tasks, 2026.
\newblock URL \url{https://arxiv.org/abs/2603.24755}.

\bibitem[Packer et~al.(2024)Packer, Wooders, Lin, Fang, Patil, Stoica, and
  Gonzalez]{packer2024memgptllmsoperatingsystems}
Charles Packer, Sarah Wooders, Kevin Lin, Vivian Fang, Shishir~G. Patil, Ion
  Stoica, and Joseph~E. Gonzalez.
\newblock Memgpt: Towards llms as operating systems, 2024.
\newblock URL \url{https://arxiv.org/abs/2310.08560}.

\bibitem[Pan et~al.(2025)Pan, Wang, Neubig, Jaitly, Ji, Suhr, and
  Zhang]{pan2025trainingsoftwareengineeringagents}
Jiayi Pan, Xingyao Wang, Graham Neubig, Navdeep Jaitly, Heng Ji, Alane Suhr,
  and Yizhe Zhang.
\newblock Training software engineering agents and verifiers with swe-gym,
  2025.
\newblock URL \url{https://arxiv.org/abs/2412.21139}.

\bibitem[Qian et~al.(2025)Qian, Liu, Prabhakar, Liu, Zhang, Chen, Ji, Yao,
  Heinecke, Savarese, Xiong, and
  Wang]{qian2025userbenchinteractivegymenvironment}
Cheng Qian, Zuxin Liu, Akshara Prabhakar, Zhiwei Liu, Jianguo Zhang, Haolin
  Chen, Heng Ji, Weiran Yao, Shelby Heinecke, Silvio Savarese, Caiming Xiong,
  and Huan Wang.
\newblock Userbench: An interactive gym environment for user-centric agents,
  2025.
\newblock URL \url{https://arxiv.org/abs/2507.22034}.

\bibitem[Qiu et~al.(2025)Qiu, Liu, Liu, Murthy, Zhang, Chen, Wang, Zhu, Yang,
  Tan, Cen, Qian, Heinecke, Yao, Savarese, Xiong, and
  Wang]{qiu2025locobenchbenchmarklongcontextlarge}
Jielin Qiu, Zuxin Liu, Zhiwei Liu, Rithesh Murthy, Jianguo Zhang, Haolin Chen,
  Shiyu Wang, Ming Zhu, Liangwei Yang, Juntao Tan, Zhepeng Cen, Cheng Qian,
  Shelby Heinecke, Weiran Yao, Silvio Savarese, Caiming Xiong, and Huan Wang.
\newblock Locobench: A benchmark for long-context large language models in
  complex software engineering, 2025.
\newblock URL \url{https://arxiv.org/abs/2509.09614}.

\bibitem[{Qwen Team}()]{qwencode}
{Qwen Team}.
\newblock Qwen code.
\newblock \url{https://github.com/QwenLM/qwen-code}.
\newblock Accessed: 2026-05-05.

\bibitem[{Qwen Team}(2026)]{qwen2026qwen35}
{Qwen Team}.
\newblock Qwen3.5.
\newblock \url{https://qwen.ai/blog?id=qwen3.5}, 2026.
\newblock Accessed: 2026-05-05.

\bibitem[Shridhar et~al.(2021)Shridhar, Yuan, Côté, Bisk, Trischler, and
  Hausknecht]{shridhar2021alfworldaligningtextembodied}
Mohit Shridhar, Xingdi Yuan, Marc-Alexandre Côté, Yonatan Bisk, Adam
  Trischler, and Matthew Hausknecht.
\newblock Alfworld: Aligning text and embodied environments for interactive
  learning.
\newblock In \emph{International Conference on Learning Representations
  (ICLR)}, 2021.
\newblock URL \url{https://arxiv.org/abs/2010.03768}.

\bibitem[Sinha et~al.(2026)Sinha, Arun, Goel, Staab, and
  Geiping]{sinha2026illusiondiminishingreturnsmeasuring}
Akshit Sinha, Arvindh Arun, Shashwat Goel, Steffen Staab, and Jonas Geiping.
\newblock The illusion of diminishing returns: Measuring long horizon execution
  in llms.
\newblock In \emph{International Conference on Learning Representations
  (ICLR)}, 2026.
\newblock URL \url{https://arxiv.org/abs/2509.09677}.

\bibitem[{SWE-agent Team}()]{miniswe}
{SWE-agent Team}.
\newblock Mini-swe-agent.
\newblock \url{https://github.com/SWE-agent/mini-swe-agent}.
\newblock Accessed: 2026-05-05.

\bibitem[Team et~al.(2026)Team, Bai, Bai, Bao, Cai, Cao, Charles, Che, Chen,
  Chen, Chen, Chen, Chen, Chen, Chen, Chen, Chen, Chen, Chen, Chen, Chen, Chen,
  Chen, Chen, Chen, Chen, Chen, Chen, Chen, Cheng, Chu, Cui, Deng, Diao, Ding,
  Dong, Dong, Dong, Dong, Du, Du, Du, Du, Du, Fan, Fang, Feng, Feng, Fu, Fu,
  Gao, Gao, Ge, Geng, Gong, Gong, Gongque, Gu, Gu, Gu, Guan, Guo, Hao, He, He,
  He, Hong, Hu, Hu, Hu, Hu, Huang, Huang, Huang, Huang, Jiang, Jiang, Jin,
  Jing, Lai, Li, Li, Li, Li, Li, Li, Li, Li, Li, Li, Li, Li, Li, Li, Li, Li,
  Li, Li, Li, Li, Li, Li, Li, Liao, Lin, Lin, Lin, Lin, Liu, Liu, Liu, Liu,
  Liu, Liu, Liu, Liu, Liu, Liu, Liu, Liu, Liu, Liu, Liu, Liu, Liu, Liu, Lu, Lu,
  Lu, Luo, Luo, Luo, Ma, Ma, Mao, Mei, Men, Meng, Meng, Miao, Ni, Ouyang, Pan,
  Pang, Qian, Qin, Qin, Qiu, Qu, Shang, Shao, Shen, Shen, Shi, Shi, Shi, Song,
  Song, Song, Song, Su, Su, Su, Sui, Sun, Sun, Sun, Sung, Tai, Tang, Tang,
  Tang, Tang, Tao, Teng, Tian, Tian, Wang, Wang, Wang, Wang, Wang, Wang, Wang,
  Wang, Wang, Wang, Wang, Wang, Wang, Wang, Wang, Wang, Wang, Wang, Wang, Wang,
  Wang, Wang, Wang, Wang, Wang, Wang, Wang, Wang, Wang, Wang, Wang, Wang, Wang,
  Wang, Wang, Wang, Wang, Wang, Wei, Wei, Wen, Wen, Wu, Wu, Wu, Wu, Wu, Wu, Wu,
  Wu, Wu, Xiao, Xie, Xie, Xie, Xin, Xing, Xu, Xu, Xu, Xu, Xu, Xu, Xu, Xu, Xu,
  Xu, Xu, Xu, Xu, Xu, Xu, Yan, Yan, Yang, Yang, Yang, Yang, Yang, Yang, Yang,
  Yang, Yang, Yang, Yang, Yang, Yang, Yang, Yao, Ye, Ye, Ye, Yin, Yu, Yu, Yu,
  Yu, Yuan, Yuan, Yuan, Yue, Zeng, Zha, Zhan, Zhang, Zhang, Zhang, Zhang,
  Zhang, Zhang, Zhang, Zhang, Zhang, Zhang, Zhang, Zhang, Zhang, Zhang, Zhang,
  Zhang, Zhang, Zhang, Zhao, Zhao, Zhao, Zhao, Zhao, Zhao, Zhao, Zheng, Zheng,
  Zheng, Zheng, Zhong, Zhong, Zhong, Zhou, Zhou, Zhou, Zhou, Zhu, Zhu, Zhu,
  Zhu, Zhu, Zhuang, Zhuang, Zou, and Zu]{kimiteam2026kimik25visualagentic}
Kimi Team, Tongtong Bai, Yifan Bai, Yiping Bao, S.~H. Cai, Yuan Cao,
  Y.~Charles, H.~S. Che, Cheng Chen, Guanduo Chen, Huarong Chen, Jia Chen,
  Jiahao Chen, Jianlong Chen, Jun Chen, Kefan Chen, Liang Chen, Ruijue Chen,
  Xinhao Chen, Yanru Chen, Yanxu Chen, Yicun Chen, Yimin Chen, Yingjiang Chen,
  Yuankun Chen, Yujie Chen, Yutian Chen, Zhirong Chen, Ziwei Chen, Dazhi Cheng,
  Minghan Chu, Jialei Cui, Jiaqi Deng, Muxi Diao, Hao Ding, Mengfan Dong,
  Mengnan Dong, Yuxin Dong, Yuhao Dong, Angang Du, Chenzhuang Du, Dikang Du,
  Lingxiao Du, Yulun Du, Yu~Fan, Shengjun Fang, Qiulin Feng, Yichen Feng,
  Garimugai Fu, Kelin Fu, Hongcheng Gao, Tong Gao, Yuyao Ge, Shangyi Geng,
  Chengyang Gong, Xiaochen Gong, Zhuoma Gongque, Qizheng Gu, Xinran Gu, Yicheng
  Gu, Longyu Guan, Yuanying Guo, Xiaoru Hao, Weiran He, Wenyang He, Yunjia He,
  Chao Hong, Hao Hu, Jiaxi Hu, Yangyang Hu, Zhenxing Hu, Ke~Huang, Ruiyuan
  Huang, Weixiao Huang, Zhiqi Huang, Tao Jiang, Zhejun Jiang, Xinyi Jin,
  Yu~Jing, Guokun Lai, Aidi Li, C.~Li, Cheng Li, Fang Li, Guanghe Li, Guanyu
  Li, Haitao Li, Haoyang Li, Jia Li, Jingwei Li, Junxiong Li, Lincan Li, Mo~Li,
  Weihong Li, Wentao Li, Xinhang Li, Xinhao Li, Yang Li, Yanhao Li, Yiwei Li,
  Yuxiao Li, Zhaowei Li, Zheming Li, Weilong Liao, Jiawei Lin, Xiaohan Lin,
  Zhishan Lin, Zichao Lin, Cheng Liu, Chenyu Liu, Hongzhang Liu, Liang Liu,
  Shaowei Liu, Shudong Liu, Shuran Liu, Tianwei Liu, Tianyu Liu, Weizhou Liu,
  Xiangyan Liu, Yangyang Liu, Yanming Liu, Yibo Liu, Yuanxin Liu, Yue Liu,
  Zhengying Liu, Zhongnuo Liu, Enzhe Lu, Haoyu Lu, Zhiyuan Lu, Junyu Luo,
  Tongxu Luo, Yashuo Luo, Long Ma, Yingwei Ma, Shaoguang Mao, Yuan Mei, Xin
  Men, Fanqing Meng, Zhiyong Meng, Yibo Miao, Minqing Ni, Kun Ouyang, Siyuan
  Pan, Bo~Pang, Yuchao Qian, Ruoyu Qin, Zeyu Qin, Jiezhong Qiu, Bowen Qu, Zeyu
  Shang, Youbo Shao, Tianxiao Shen, Zhennan Shen, Juanfeng Shi, Lidong Shi,
  Shengyuan Shi, Feifan Song, Pengwei Song, Tianhui Song, Xiaoxi Song, Hongjin
  Su, Jianlin Su, Zhaochen Su, Lin Sui, Jinsong Sun, Junyao Sun, Tongyu Sun,
  Flood Sung, Yunpeng Tai, Chuning Tang, Heyi Tang, Xiaojuan Tang, Zhengyang
  Tang, Jiawen Tao, Shiyuan Teng, Chaoran Tian, Pengfei Tian, Ao~Wang, Bowen
  Wang, Chensi Wang, Chuang Wang, Congcong Wang, Dingkun Wang, Dinglu Wang,
  Dongliang Wang, Feng Wang, Hailong Wang, Haiming Wang, Hengzhi Wang, Huaqing
  Wang, Hui Wang, Jiahao Wang, Jinhong Wang, Jiuzheng Wang, Kaixin Wang, Linian
  Wang, Qibin Wang, Shengjie Wang, Shuyi Wang, Si~Wang, Wei Wang, Xiaochen
  Wang, Xinyuan Wang, Yao Wang, Yejie Wang, Yipu Wang, Yiqin Wang, Yucheng
  Wang, Yuzhi Wang, Zhaoji Wang, Zhaowei Wang, Zhengtao Wang, Zhexu Wang, Zihan
  Wang, Zizhe Wang, Chu Wei, Ming Wei, Chuan Wen, Zichen Wen, Chengjie Wu,
  Haoning Wu, Junyan Wu, Rucong Wu, Wenhao Wu, Yuefeng Wu, Yuhao Wu, Yuxin Wu,
  Zijian Wu, Chenjun Xiao, Jin Xie, Xiaotong Xie, Yuchong Xie, Yifei Xin, Bowei
  Xing, Boyu Xu, Jianfan Xu, Jing Xu, Jinjing Xu, L.~H. Xu, Lin Xu, Suting Xu,
  Weixin Xu, Xinbo Xu, Xinran Xu, Yangchuan Xu, Yichang Xu, Yuemeng Xu, Zelai
  Xu, Ziyao Xu, Junjie Yan, Yuzi Yan, Guangyao Yang, Hao Yang, Junwei Yang, Kai
  Yang, Ningyuan Yang, Ruihan Yang, Xiaofei Yang, Xinlong Yang, Ying Yang,
  Yi~Yang, Yi~Yang, Zhen Yang, Zhilin Yang, Zonghan Yang, Haotian Yao, Dan Ye,
  Wenjie Ye, Zhuorui Ye, Bohong Yin, Chengzhen Yu, Longhui Yu, Tao Yu,
  Tianxiang Yu, Enming Yuan, Mengjie Yuan, Xiaokun Yuan, Yang Yue, Weihao Zeng,
  Dunyuan Zha, Haobing Zhan, Dehao Zhang, Hao Zhang, Jin Zhang, Puqi Zhang,
  Qiao Zhang, Rui Zhang, Xiaobin Zhang, Y.~Zhang, Yadong Zhang, Yangkun Zhang,
  Yichi Zhang, Yizhi Zhang, Yongting Zhang, Yu~Zhang, Yushun Zhang, Yutao
  Zhang, Yutong Zhang, Zheng Zhang, Chenguang Zhao, Feifan Zhao, Jinxiang Zhao,
  Shuai Zhao, Xiangyu Zhao, Yikai Zhao, Zijia Zhao, Huabin Zheng, Ruihan Zheng,
  Shaojie Zheng, Tengyang Zheng, Junfeng Zhong, Longguang Zhong, Weiming Zhong,
  M.~Zhou, Runjie Zhou, Xinyu Zhou, Zaida Zhou, Jinguo Zhu, Liya Zhu, Xinhao
  Zhu, Yuxuan Zhu, Zhen Zhu, Jingze Zhuang, Weiyu Zhuang, Ying Zou, and Xinxing
  Zu.
\newblock Kimi k2.5: Visual agentic intelligence, 2026.
\newblock URL \url{https://arxiv.org/abs/2602.02276}.

\bibitem[Thai et~al.(2026)Thai, Le, Manh, Nhat, and
  Bui]{thai2026sweevobenchmarkingcodingagents}
Minh V.~T. Thai, Tue Le, Dung~Nguyen Manh, Huy~Phan Nhat, and Nghi D.~Q. Bui.
\newblock Swe-evo: Benchmarking coding agents in long-horizon software
  evolution scenarios, 2026.
\newblock URL \url{https://arxiv.org/abs/2512.18470}.

\bibitem[Wang et~al.(2025{\natexlab{a}})Wang, Gong, Zhang, Xu, and
  Wang]{wang2025aiagenticprogrammingsurvey}
Huanting Wang, Jingzhi Gong, Huawei Zhang, Jie Xu, and Zheng Wang.
\newblock Ai agentic programming: A survey of techniques, challenges, and
  opportunities, 2025{\natexlab{a}}.
\newblock URL \url{https://arxiv.org/abs/2508.11126}.

\bibitem[Wang et~al.(2026)Wang, Wang, Ma, Yu, Ling, Li, Xiong, and
  Zhang]{wang2026codeflowbenchmultiturniterativebenchmark}
Sizhe Wang, Zhengren Wang, Dongsheng Ma, Yongan Yu, Rui Ling, Zhiyu Li, Feiyu
  Xiong, and Wentao Zhang.
\newblock Codeflowbench: A multi-turn, iterative benchmark for complex code
  generation, 2026.
\newblock URL \url{https://arxiv.org/abs/2504.21751}.

\bibitem[Wang et~al.(2024)Wang, Wang, Liu, Chen, Yuan, Peng, and
  Ji]{wang2024mintevaluatingllmsmultiturn}
Xingyao Wang, Zihan Wang, Jiateng Liu, Yangyi Chen, Lifan Yuan, Hao Peng, and
  Heng Ji.
\newblock Mint: Evaluating llms in multi-turn interaction with tools and
  language feedback, 2024.
\newblock URL \url{https://arxiv.org/abs/2309.10691}.

\bibitem[Wang et~al.(2025{\natexlab{b}})Wang, Li, Song, Xu, Tang, Zhuge, Pan,
  Song, Li, Singh, Tran, Li, Ma, Zheng, Qian, Shao, Muennighoff, Zhang, Hui,
  Lin, Brennan, Peng, Ji, and Neubig]{wang2025openhandsopenplatformai}
Xingyao Wang, Boxuan Li, Yufan Song, Frank~F. Xu, Xiangru Tang, Mingchen Zhuge,
  Jiayi Pan, Yueqi Song, Bowen Li, Jaskirat Singh, Hoang~H. Tran, Fuqiang Li,
  Ren Ma, Mingzhang Zheng, Bill Qian, Yanjun Shao, Niklas Muennighoff, Yizhe
  Zhang, Binyuan Hui, Junyang Lin, Robert Brennan, Hao Peng, Heng Ji, and
  Graham Neubig.
\newblock Openhands: An open platform for ai software developers as generalist
  agents, 2025{\natexlab{b}}.
\newblock URL \url{https://arxiv.org/abs/2407.16741}.

\bibitem[Wilcoxon(1945)]{wilcoxon}
Frank Wilcoxon.
\newblock Individual comparisons by ranking methods.
\newblock \emph{Biometrics Bulletin}, 1\penalty0 (6):\penalty0 80--83, 1945.
\newblock ISSN 00994987.
\newblock URL \url{http://www.jstor.org/stable/3001968}.

\bibitem[Wu et~al.(2025)Wu, Xue, Yin, Zhou, Chang, Peng, and
  Wen]{wu2025frontalkbenchmarkingfrontenddevelopment}
Xueqing Wu, Zihan Xue, Da~Yin, Shuyan Zhou, Kai-Wei Chang, Nanyun Peng, and
  Yeming Wen.
\newblock Frontalk: Benchmarking front-end development as conversational code
  generation with multi-modal feedback, 2025.
\newblock URL \url{https://arxiv.org/abs/2601.04203}.

\bibitem[Xie et~al.(2024{\natexlab{a}})Xie, Zhang, Chen, Zhu, Lou, Tian, Xiao,
  and Su]{xie2024travelplannerbenchmarkrealworldplanning}
Jian Xie, Kai Zhang, Jiangjie Chen, Tinghui Zhu, Renze Lou, Yuandong Tian,
  Yanghua Xiao, and Yu~Su.
\newblock Travelplanner: A benchmark for real-world planning with language
  agents.
\newblock In \emph{International Conference on Machine Learning (ICML)},
  2024{\natexlab{a}}.
\newblock URL \url{https://arxiv.org/abs/2402.01622}.

\bibitem[Xie et~al.(2024{\natexlab{b}})Xie, Zhang, Chen, Li, Zhao, Cao, Hua,
  Cheng, Shin, Lei, Liu, Xu, Zhou, Savarese, Xiong, Zhong, and
  Yu]{xie2024osworldbenchmarkingmultimodalagents}
Tianbao Xie, Danyang Zhang, Jixuan Chen, Xiaochuan Li, Siheng Zhao, Ruisheng
  Cao, Toh~Jing Hua, Zhoujun Cheng, Dongchan Shin, Fangyu Lei, Yitao Liu,
  Yiheng Xu, Shuyan Zhou, Silvio Savarese, Caiming Xiong, Victor Zhong, and Tao
  Yu.
\newblock Osworld: Benchmarking multimodal agents for open-ended tasks in real
  computer environments, 2024{\natexlab{b}}.
\newblock URL \url{https://arxiv.org/abs/2404.07972}.

\bibitem[Yang et~al.(2024)Yang, Jimenez, Zhang, Lieret, Yang, Wu, Press,
  Muennighoff, Synnaeve, Narasimhan, Yang, Wang, and
  Press]{yang2024swebenchmultimodalaisystems}
John Yang, Carlos~E. Jimenez, Alex~L. Zhang, Kilian Lieret, Joyce Yang, Xindi
  Wu, Ori Press, Niklas Muennighoff, Gabriel Synnaeve, Karthik~R. Narasimhan,
  Diyi Yang, Sida~I. Wang, and Ofir Press.
\newblock Swe-bench multimodal: Do ai systems generalize to visual software
  domains?, 2024.
\newblock URL \url{https://arxiv.org/abs/2410.03859}.

\bibitem[Yang et~al.(2025)Yang, Lieret, Jimenez, Wettig, Khandpur, Zhang, Hui,
  Press, Schmidt, and Yang]{yang2025swesmithscalingdatasoftware}
John Yang, Kilian Lieret, Carlos~E. Jimenez, Alexander Wettig, Kabir Khandpur,
  Yanzhe Zhang, Binyuan Hui, Ofir Press, Ludwig Schmidt, and Diyi Yang.
\newblock Swe-smith: Scaling data for software engineering agents, 2025.
\newblock URL \url{https://arxiv.org/abs/2504.21798}.

\bibitem[Yao et~al.(2024)Yao, Shinn, Razavi, and
  Narasimhan]{yao2024taubenchbenchmarktoolagentuserinteraction}
Shunyu Yao, Noah Shinn, Pedram Razavi, and Karthik Narasimhan.
\newblock $\tau$-bench: A benchmark for tool-agent-user interaction in
  real-world domains, 2024.
\newblock URL \url{https://arxiv.org/abs/2406.12045}.

\bibitem[Zan et~al.(2025)Zan, Huang, Liu, Chen, Zhang, Xin, Chen, Liu, Zhong,
  Li, Liu, Xiao, Chen, Zhang, Su, Liu, Long, Shen, and
  Xiang]{zan2025multiswebenchmultilingualbenchmarkissue}
Daoguang Zan, Zhirong Huang, Wei Liu, Hanwu Chen, Linhao Zhang, Shulin Xin,
  Lu~Chen, Qi~Liu, Xiaojian Zhong, Aoyan Li, Siyao Liu, Yongsheng Xiao,
  Liangqiang Chen, Yuyu Zhang, Jing Su, Tianyu Liu, Rui Long, Kai Shen, and
  Liang Xiang.
\newblock Multi-swe-bench: A multilingual benchmark for issue resolving, 2025.
\newblock URL \url{https://arxiv.org/abs/2504.02605}.

\bibitem[Zhao et~al.(2024)Zhao, Jiang, Lee, Chiu, Cardie, Gallé, and
  Rush]{zhao2024commit0librarygenerationscratch}
Wenting Zhao, Nan Jiang, Celine Lee, Justin~T Chiu, Claire Cardie, Matthias
  Gallé, and Alexander~M Rush.
\newblock Commit0: Library generation from scratch, 2024.
\newblock URL \url{https://arxiv.org/abs/2412.01769}.

\bibitem[Zhou et~al.(2024)Zhou, Xu, Zhu, Zhou, Lo, Sridhar, Cheng, Ou, Bisk,
  Fried, Alon, and Neubig]{zhou2024webarenarealisticwebenvironment}
Shuyan Zhou, Frank~F. Xu, Hao Zhu, Xuhui Zhou, Robert Lo, Abishek Sridhar,
  Xianyi Cheng, Tianyue Ou, Yonatan Bisk, Daniel Fried, Uri Alon, and Graham
  Neubig.
\newblock Webarena: A realistic web environment for building autonomous agents.
\newblock In \emph{International Conference on Learning Representations
  (ICLR)}, 2024.
\newblock URL \url{https://arxiv.org/abs/2307.13854}.

\bibitem[Zhou et~al.(2025)Zhou, Liu, Chen, Tian, and
  Chen]{zhou2025gsminfinitellmsbehaveinfinitely}
Yang Zhou, Hongyi Liu, Zhuoming Chen, Yuandong Tian, and Beidi Chen.
\newblock Gsm-infinite: How do your llms behave over infinitely increasing
  context length and reasoning complexity?, 2025.
\newblock URL \url{https://arxiv.org/abs/2502.05252}.

\end{thebibliography}

\appendix
\crefalias{section}{appendix}

\section{Experiment Cost}
\label{app:cost}

Per-token pricing used to compute the costs reported in \cref{tab:cost-grid} is shown in \cref{tab:model-pricing}.

\begin{table}[h]
\centering
\small
\caption{Per-token pricing used in cost estimates.}
\label{tab:model-pricing}
\begin{tabular}{lcc}
\toprule
Model & Input (\$/MTok) & Output (\$/MTok) \\
\midrule
Devstral 2          & 0.40 & 2.00 \\
Devstral Small 2    & 0.10 & 0.30 \\
GLM-5               & 0.60 & 2.08 \\
Kimi K2.5           & 0.44 & 2.00 \\
Nemotron Super      & 0.09 & 0.45 \\
Qwen3-Coder-Next    & 0.12 & 0.80 \\
Qwen3.5-122B        & 0.26 & 2.08 \\
\bottomrule
\end{tabular}
\end{table}

\begin{table}[h]
\centering
\small
\caption{Estimated API cost per experiment (20 scenarios, 101 turns each). We were unable to gather cost data with KimiCLI and QwenCode harnesses.}
\label{tab:cost-grid}
\begin{tabular}{lcccc}
\toprule
Model & OpenCode & Mini-SWE & OpenHands & \shortstack{Model Provider\\Agent} \\
\midrule
Devstral 2 & \$342 & \$123 & \$106 & \$2 \\
Devstral Small 2 & \$130 & \$26 & \$42 & \$2 \\
GLM-5 & \$2,713 & \$308 & \$176 & --- \\
Kimi K2.5 & \$1,739 & \$274 & \$158 & N/A$^\dagger$ \\
Nemotron Super & \$92 & \$43 & \$28 & --- \\
Qwen3-Coder-Next & \$122 & \$83 & \$46 & N/A$^\dagger$ \\
Qwen3.5-122B & \$862 & \$310 & \$81 & N/A$^\dagger$
 \\
\bottomrule
\end{tabular}
\end{table}

Assuming GLM-5-level token consumption (4.5B input tokens, 7.5M output tokens) via OpenCode, running the full benchmark (20 scenarios, 101 turns) with frontier closed-source models would cost approximately \$13,600 for Claude Sonnet 4.6 (\$3/\$15 per MTok input/output) and \$22,700 for GPT-5.5 (\$5/\$30 per MTok input/output), making these experiments too expensive to run for us.

In terms of wall-clock time, a single scenario takes roughly 4--8 hours to complete, depending on how many of the 101 turns the agent reaches before failing (scenarios that terminate early finish faster). The benchmark supports running multiple scenarios in parallel against the same agent--harness configuration, so the 20-scenario sweeps in this paper complete in under a day on commodity infrastructure when fully parallelized. The bottleneck is usually the LLM API calls.

\section{Framework Instantiations}
\label{app:instantiations}

The framework is domain-agnostic: any system where $\mathcal{S}$, $\mathcal{A}(s)$, $\tau$, and $T$ can be procedurally defined yields a valid benchmark. \cref{tab:instantiations} illustrates this with four instantiations of increasing complexity, from a trivial running-sum calculator to \bname{}. All share \cref{alg:eval}'s evaluation loop and compounding dynamics; they differ only in what the components contain. We implement the last column; the others are hypothetical examples showing the framework's generality.

\begin{table}[h]
\caption{Example instantiations of the general framework. Each column maps the abstract components to a concrete domain. The last column (\textbf{bold}) is the instantiation evaluated in this paper.}
\label{tab:instantiations}
\centering
\small
\setlength{\tabcolsep}{3pt}
\begin{tabular}{@{}l p{2.4cm} p{2.8cm} p{2.8cm} p{3.2cm}@{}}
\toprule
& Addition Calculator & Office Assistant & CLI Tool & \textbf{REST API (Ours)} \\
\midrule
$\mathcal{S}$ & Running sum & Emails, calendar events, contacts & Command grammar (subcommands, flags, args) & \textbf{API schema (entities, fields, relationships, analytics)} \\[4pt]
$\mathcal{A}(s)$ & New value to add  & New email, meeting request, reschedule & Add subcommand, add flag, change output format & \textbf{Add/delete/rename entity, add field, change constraints (\cref{tab:changes})} \\[4pt]
$\tau$ & Add the value to the sum & Update mailbox / calendar state & Extend grammar definition & \textbf{Apply schema change} \\[4pt]
$\hat{s}_t$ & Predicted sum & Tool calls (send email, create event) & CLI binary / script & \textbf{Running HTTP server} \\[4pt]
$T$ & Compare to the true sum & Check mailbox \& calendar state & Run commands, check output \& exit codes & \textbf{HTTP test suite against server} \\[4pt]
$\mathrm{desc}_{\mathrm{spec}}$ & ``Start with value 0'' & ``Here is your inbox and calendar'' & ``Implement this CLI spec'' & \textbf{NL specification document} \\[4pt]
$\mathrm{desc}_{\mathrm{action}}$ & ``Add 7'' & ``Schedule a meeting with Alice at 3pm'' & ``Add a \texttt{--verbose} flag to \texttt{list}'' & \textbf{NL change description} \\[4pt]
$\mathrm{desc}_{\mathrm{feedback}}$ & ``Wrong: expected 7, got 6'' & ``Meeting conflicts with existing event'' & ``\texttt{list --verbose} returned exit code 1'' & \textbf{Test failure messages (configurable detail)} \\
\bottomrule
\end{tabular}
\end{table}

\section{Background: REST API Servers}
\label{app:rest-primer}

A REST (Representational State Transfer)~\citep{fielding2000architectural} API server exposes a set of typed \emph{resources} (entities) over HTTP, with operations on each resource specified by the HTTP method and URL path. The standard CRUD (Create, Read, Update, Delete) mapping is: \texttt{POST /entity} creates a new entity, \texttt{GET /entity} lists existing entities, \texttt{GET /entity/\{id\}} reads a specific entity, \texttt{PATCH /entity/\{id\}} updates one, and \texttt{DELETE /entity/\{id\}} removes one. Requests and responses carry JSON payloads, and the server signals outcomes through HTTP status codes (e.g.\ 200 for success, 400 for malformed input, 404 for missing resources, 422 for validation failures).

Because the contract is fully expressed through URL paths, JSON bodies, and status codes, a REST server can be evaluated as a black box: any HTTP client can issue requests and check responses without inspecting source code. This makes REST APIs a natural choice for language-agnostic evaluation---the agent can implement the server in any language or framework, and the same test suite still applies.

REST servers are also structurally rich enough to exercise non-trivial coding behaviors. Beyond simple CRUD, they typically include input validation (rejecting malformed or out-of-range values), inter-entity \emph{references} (one entity pointing at another by ID, with cascade behavior on deletion), \emph{analytics endpoints} that compute aggregates (sum, average, count) over stored data, and \emph{actions}---state-transition endpoints that mutate an entity's state subject to guard conditions and trigger side effects. \bname{} exercises all of these in its evolving specifications.

\section{Benchmark Component Details}
\label{app:benchmark-details}

\subsection{Schema Generation}

Each scenario begins by sampling an initial state $s_0 \sim p_0$ using the scenario seed $\sigma_k$. A state $s \in \mathcal{S}$ is an OpenAPI-like \citep{openapi} schema (design) consisting of: \textbf{entities} with typed fields (string, integer, float, boolean, enum) and constraints (min/max values, max length, required/optional, nullable); \textbf{inter-entity relationships} (single references, reference lists); \textbf{analytics endpoints} computing aggregates (avg, min, max, sum, count) over entity fields; and \textbf{business logic} --- actions that transition enum fields between states, guarded by conditions (comparisons, null checks, arithmetic expressions, aggregate predicates like \texttt{count(this.items) > 0}, and quantifiers like \texttt{any}/\texttt{all} over list references) and triggering effects (set field, increment field, arithmetic computation, cross-entity modification, for-each over list references). The initial state typically contains 2--3 entities with 4--5 fields each and 1--2 analytics endpoints. As the transition function $\tau$ applies successive actions, states grow monotonically in complexity, reaching 10+ entities with complex inter-entity relationships and multi-step workflows by turn 20--30.

\subsection{Change Generation}

At each turn $t$, the action-selection policy $\pi_{\mathrm{env}}$ samples an action $a_t \in \mathcal{A}(s_{t-1})$ consisting of one or more schema modifications drawn from the types in \cref{tab:changes}. The action is rendered as $\mathrm{desc}_{\mathrm{action}}(a_t)$ and delivered to the agent. An action includes 6 changes (configurable), ranging from simple field additions to complex operations like entity renames that require updating all downstream references, or adding guarded actions with conditions over related entities and side effects that propagate through reference chains.

\begin{table}[h]
\caption{Schema change types supported by \bname{}, organized by category.}
\label{tab:changes}
\centering
\small
\begin{tabular}{lll}
\toprule
Category & Change Type & Description \\
\midrule
\multirow{3}{*}{Entity} & Add entity & New entity with fields and CRUD endpoints \\
& Delete entity & Remove entity and all dependent operations \\
& Rename entity & Change name, update all endpoint paths \\
\midrule
\multirow{3}{*}{Field} & Add field & New required or optional field on existing entity \\
& Rename field & Change field name, update dependent analytics \\
& Change constraints & Modify min/max bounds or max length \\
\midrule
\multirow{3}{*}{Analytics} & Add analytics & New aggregate query (avg, min, max, sum, count) \\
& Delete analytics & Remove an existing aggregate endpoint \\
\midrule
\multirow{5}{*}{\shortstack[l]{Business\\Logic}} & Add action & New state-transition endpoint with guards/effects \\
& Remove action & Delete an action endpoint \\
& Add/remove condition & Modify guard conditions on an action \\
& Add/remove effect & Modify side effects triggered by an action \\
& Add enum value & Extend the state space of an enum field \\
\bottomrule
\end{tabular}
\end{table}

\subsection{Test Generation}

For each state $s_t$, the test function $T$ generates a test suite that validates the current specification by sending HTTP requests to the running server. Tests cover: CRUD operations (create, read, update, delete for every entity); bulk operations; field validation (type checking, constraint enforcement, required field presence); analytics correctness (aggregates over empty data, single values, and multiple values); relationship handling (reference validation, cascade deletion); deletion verification (confirming deleted entities return 404); and business logic (action happy paths verifying state transitions and effects, wrong-state calls expecting 409, guard failures expecting 400, and effect verification via subsequent GET requests). A typical initial test suite contains 20--30 tests; by turn 20--30, suites grow to 100+ tests.

\subsection{Agent Interaction Protocol}

The agent $\pi$ operates inside an isolated Docker container with full shell access. The protocol maps directly onto \cref{alg:eval}: (1)~the agent receives $\mathrm{desc}_{\mathrm{state}}(s_0)$ as a specification document and README; (2)~$\pi$ produces $\hat{s}_t$ and signals completion by writing to a marker file; (3)~the harness starts the server via \texttt{bash run\_server.sh <port>} and computes $r_t = T(s_t, \hat{s}_t)$; (4)~if $\mathrm{pass}(r_t) = 0$ and retries remain, $\pi$ receives $\mathrm{desc}_{\mathrm{feedback}}(r_t)$ and updates $\hat{s}_t$ (up to $R$ retries); (5)~on success or retry exhaustion, $\mathrm{desc}_{\mathrm{action}}(a_{t+1})$ is delivered.

The feedback function $\mathrm{desc}_{\mathrm{feedback}}$ is configurable: \textbf{detailed} feedback includes specific failure messages (e.g., ``POST /entity returned 404, expected 201'' or ``POST /order/1/submit returned 200, expected 400 because condition \texttt{total > 0} is not satisfied''); \textbf{minimal} feedback states only that failures occurred (``One or more issues were detected, no further information is available'').

\section{Concrete Examples}

\subsection{Agent Instructions (README)}
\label{app:readme}

The following is the README provided to the agent at the start of each scenario. It defines the interface contract, endpoint conventions, error handling rules, and constraints. The agent receives this alongside the initial specification.

\begin{small}
\begin{verbatim}
Requirements:
1. Implement a REST API server that matches the specification and apply user requested changes;
2. All entities must support CRUD operations (Create, Read, Update, Delete);
3. All operations/analytics endpoints must work correctly;
4. You MUST create a script called `run_server.sh` that starts the server;
5. The script must accept a port number as the first argument;
6. You must use python programming language, but you can use ANY web framework as
long as it is python;
7. You should test your implementation during development to ensure correctness.

Required Script:
Create a file called `run_server.sh` that:
- Takes a port number as the first argument (e.g., `bash run_server.sh 8001`)
- Starts your REST API server on that port
- The server should listen on 0.0.0.0 (all interfaces)

Example run_server.sh:
  #!/bin/bash
  PORT=$1
  # Your command to start the server, e.g.:
  # python <your script>.py --port $PORT

Expected CRUD Endpoints for each entity ({entity} denotes entity name).
Note: {entity} is the entity name converted to lowercase with no separators.
For example, entity "UserProfile" maps to /userprofile, /userprofile/{id}, etc.
- POST /{entity} - Create a new entity
- GET /{entity} - List all entities
- GET /{entity}/{id} - Get a specific entity by ID
- PATCH /{entity}/{id} - Update a specific entity (partial update)
- DELETE /{entity}/{id} - Delete a specific entity

Required Testing Endpoint:
- POST /reset - Reset the database to empty state
  * Must clear all entities from all collections
  * Return: {"status": "reset"} with status code 200 or 204
  * This endpoint is REQUIRED for the automated test suite

Testing:
- You can run the server and make test requests during development;
- The final evaluation will run automated tests against your server. The
  system is fully automated, and will give you info regarding failures if
  tests don't pass. You are only given a limited number of attempts.
  You are not talking to a human, so you cannot ask for clarifications.
- You will be asked to modify the code later on if the tests pass. You must
  remember these instructions, and adhere to them when modifying code.

Constraints:
- Do not modify `initial_specification.md`;
- For storing the data, the server must use any database solution you'd like,
  including in-memory;
- Must use python programming language;
- Must handle all CRUD operations correctly;
- Must implement all analytics operations from the spec;
- Must implement all action endpoints from the spec, if any. Actions are
  POST endpoints at /{entity}/{id}/{action_name} that transition an enum
  field between allowed values. Status codes must be:
  * 404 -- entity not found
  * 400 -- the entity's current enum value is not in the allowed "from"
    values for this action (i.e., wrong state for this transition)
  * 409 -- the enum value is valid for the action, but a guard condition
    is not satisfied (e.g., a numeric field doesn't meet a threshold,
    count of items in a list is too low, sum of a field across referenced
    entities doesn't meet a requirement, or any/all items in a list
    don't satisfy a sub-condition). This is distinct from 400: the
    state transition is allowed, but additional preconditions are not met.
  * 200 -- success. Apply all effects listed in the spec (field updates,
    increments, cross-entity field updates on referenced entities, and
    for-each effects on list-referenced entities), update the enum field
    to the target value, and return the updated entity.
  These status codes must be checked in order: 404, then 400, then 409;
Notation used in conditions and effects:
- `this.field_name` refers to a field on the current entity being acted upon
  (e.g., `this.total > 0` means the entity's `total` field must be > 0)
- `'value'` (quoted) is a literal string value; unquoted numbers are literal
- `count(list_field)` counts the number of referenced entities in a list field
- `sum(list_field.field)` sums a numeric field across all entities referenced
  by a list field
- Cross-entity effects like "on the referenced X (via ref_field), ..." mean:
  look up the entity referenced by `ref_field` on the current entity, then
  apply the effect to that referenced entity
- For-each effects like "for each X in list_field, ..." mean: iterate over all
  entities referenced in the list field and apply the effect to each one
- `any`/`all` conditions like "all item in list_field satisfies: ..." mean:
  check the sub-condition against every entity referenced in the list field

- Analytics operations should return a JSON object {"result": <value>}.
  When result is impossible to compute (like a min of an empty list), return
  {"result": null}. Sum of an empty list is expected to be {"result": 0};
- When an entity or field is renamed, any operations endpoints should be
  updated. E.g., /analytics/user/age/max endpoint should become
  /analytics/userprofile/age_cache/max if User was renamed to UserProfile
  and `age` was renamed to `age_cache`;
- When an entity is renamed, do not rename other entities' fields that refer
  to the renamed entity. For example, if Dog has a field "cats" referencing
  Cat, and Cat is renamed to CatSuper, the Dog field stays named "cats" --
  only the endpoint changes from /cat to /catsuper;
- When an entity is renamed, all enum types, their values, and field
  constraints (min, max, defaults) must be preserved exactly. Do not
  regenerate or guess enum values -- copy them from the existing
  implementation;
- Handle errors appropriately (404 for not found, etc.);
- datetime fields store values as strings in YYYY-MM-DD hh:mm:ss format
  (e.g., "2024-03-15 14:30:00"). Validate format and calendar correctness
  on input. Range constraints (before/after) use string comparison, which
  works because the format is fixed-width and lexicographically ordered;
- Required values must be specified when passed to POST, unless the field
  has a default value. If a required field has a default and is omitted
  from the POST body, the server should use the default value. Fields can
  be null if they are nullable, otherwise they cannot be. If a field is
  non-required and non-nullable, it can be omitted, but when not omitted
  it cannot be null. If these requirements do not hold in the passed data,
  return appropriate failure status codes;
- When adding or modifying data, handle the missing required values, values
  of incorrect type, values out of range, references to non-existent
  objects, etc correctly by returning proper error codes;
- Deleting an object from the database should also update any objects which
  have a reference or reference list referring to the deleted object, with
  references set to null, and reference lists updated by removing the
  non-existent reference;
- Deleting an entity should delete fields of other entities referring to
  such entity;
- Return proper HTTP status codes, including 404 for deleted entities;
- When killing processes with pkill, use exact PID instead of regexes
  whenever possible to avoid killing your own coding agent process.
- Remember to test your code and make sure it adheres to any requirements
  mentioned in this file;
- PATCH requests may include read-only fields like `id`. If the value
  matches the current value, accept the request. Do not reject PATCH
  requests just because they contain fields that cannot be changed.
- Reference fields store the ID of the referenced entity, not the full
  embedded object. When reading an entity, reference fields should return
  the referenced entity's ID as a string.
- The `id` field in each entity is provided by the client in POST requests.
  Use it as the primary key. Do not generate server-side UUIDs.
- The `id` field is always the primary key and endpoint identifier for each
  entity (e.g., GET /{entity}/{id}). Other fields ending in `_id` (like
  `user_id`, `order_id`) are different identifiers, and should not be used
  for endpoints;

When you are done, ensure the server starts successfully with:
`bash run_server.sh <port>`
\end{verbatim}
\end{small}

\subsection{Initial Specification Example}
\label{app:spec-example}

At turn 0, the agent receives the following prompt:

\begin{small}
\begin{verbatim}
Implement a REST API server based on the specification in
`initial_specification.md`, following instructions in `README.md`.
\end{verbatim}
\end{small}

\noindent The \texttt{initial\_specification.md} file describes the API the agent must implement. Here is a real example from a scenario with 2 entities, 4 enum types, and 2 analytics endpoints:

\begin{small}
\begin{verbatim}
# Specifications

The system has 2 entities, 0 connections (relationships) between entities,
and 2 analytics endpoints.

# Enum Types

Enum values are case-sensitive and must match exactly as listed.

- CoffeeBatchRoastLevel: allowed values are light, medium, dark, espresso
- CoffeeBatchStatus: allowed values are processing, roasted,
  quality_checked, ready_for_sale
- SubscriptionRoastPreference: allowed values are light, medium,
  dark, espresso
- SubscriptionStatus: allowed values are active, paused, cancelled

# Entities

## CoffeeBatch
(/coffeebatch endpoint)
CoffeeBatch entity has the following fields:
- batch_number of type required, non-nullable string
  (max length 20 inclusive, min length 5 inclusive)
- cupping_score of type optional, non-nullable float
- id of type required, non-nullable string, unique
- origin_country of type required, non-nullable string
  (max length 50 inclusive, min length 2 inclusive)
- roast_date of type required, non-nullable datetime in
  YYYY-MM-DD hh:mm:ss format
  (between 2020-01-01 00:00:00 and 2026-12-31 23:59:59 inclusive)
- roast_level of type required, non-nullable CoffeeBatchRoastLevel
  enum, default "medium"
- status of type required, non-nullable CoffeeBatchStatus enum,
  default "processing"
- weight_kg of type required, non-nullable float

We should be able to access a specific coffeebatch via
GET /coffeebatch/{id} endpoint

## Subscription
(/subscription endpoint)
Subscription entity has the following fields:
- customer_email of type required, non-nullable string
  (max length 100 inclusive, min length 5 inclusive)
- id of type required, non-nullable string, unique
- monthly_shipment_quantity_kg of type required, non-nullable float
- preferred_roast_level of type required, non-nullable
  SubscriptionRoastPreference enum, default "medium"
- subscription_start_date of type required, non-nullable datetime in
  YYYY-MM-DD hh:mm:ss format
  (between 2020-01-01 00:00:00 and 2026-12-31 23:59:59 inclusive)
- subscription_status of type required, non-nullable
  SubscriptionStatus enum, default "active"
- total_spent of type optional, non-nullable float

We should be able to access a specific subscription via
GET /subscription/{id} endpoint

# Analytics Endpoints

All analytics endpoints must return a JSON object with a single
"result" key, e.g. {"result": 42} or {"result": null} when there
is no data. When there are no matching entities: count returns 0,
sum returns 0, avg/min/max return null.

## coffeebatch cupping_score avg
GET /analytics/coffeebatch/cupping_score/avg endpoint
goes over all possible coffeebatch entities in the data and
calculates their average cupping_score.

## subscription total_spent sum
GET /analytics/subscription/total_spent/sum endpoint
goes over all possible subscription entities in the data and
sums their total_spent.

# Validation Requirements

- Non-nullable fields must reject null values with 400 or 422.
- Required fields must be present in POST requests; omitting them
  should return 400 or 422.
- Reference fields (single or list) must validate that referenced
  entity IDs exist. Return 400 or 422 for invalid references.
- List reference fields default to an empty list [] when not
  provided. They cannot be null.
- Enum fields must only accept the exact values listed
  (case-sensitive). Return 400 or 422 for invalid values.
- Numeric fields (integer/float) must respect minimum/maximum
  constraints. Return 400 or 422 for out-of-range values.
- String fields must respect min_length/max_length constraints.
  Return 400 or 422 for violations.

# Testing Requirements

## Reset Endpoint
The system must implement a POST /reset endpoint for testing
purposes. This endpoint should reset the database to an empty
state, clearing all entities. The automated test suite relies on
this endpoint to ensure test isolation.
Expected response: {"status": "reset"} with status code 200 or 204.
\end{verbatim}
\end{small}

\subsection{Change Description Example}
\label{app:change-example}

After passing the initial turn, the agent receives a change prompt. Here is a real example (continuing the CoffeeBatch/Subscription scenario above):

\begin{small}
\begin{verbatim}
Great work! Now I need you to make the following changes:

Add required non-nullable enum field "grind_preference" of type
SubscriptionGrindPreference (allowed values: whole_bean,
medium_grind, fine_grind, coarse_grind) to entity "Subscription"
with default value "whole_bean"; Rename field "weight_kg" to
"batch_weight_kg" in entity "CoffeeBatch", updating any
dependencies and analytics endpoint paths.; Change constraints on
"CoffeeBatch.origin_country": set max length to 75 inclusive, min
length to 2 inclusive; Add optional nullable field
"last_shipment_date" of type datetime (between 2020-01-01 00:00:00
and 2026-12-31 23:59:59 inclusive) to entity "Subscription"; Add an
analytics endpoint at GET /analytics/coffeebatch/count that
calculates the number of all CoffeeBatch entities (returns int)

Ensure the server still starts with: bash run_server.sh <port>
\end{verbatim}
\end{small}

\subsection{Detailed Feedback Example}
\label{app:feedback-example}

When an agent's implementation fails tests, it receives feedback describing the failures. Here is a real example of detailed feedback (the richest of the three feedback levels):

\begin{small}
\begin{verbatim}
I ran some tests and, according to the tests, there are the
following issues:
Non-nullable test for field CoffeeBatch.roast_level failed (should
reject null) (Non-nullable field roast_level should reject null,
POST /coffeebatch returned 201, expected 400 or 422. Content:
b'{"batch_number":"crvfngvqxaboakj","cupping_score":299.55,
"id":"test_null_reject",
"origin_country":"jalcfalugjcdecpcggfvbbhxmbppbburokz",
"roast_date":"2021-01-03 16:39:47","roast_level":null,"status");
Non-nullable test for field CoffeeBatch.status failed (should
reject null) (Non-nullable field status should reject null, POST
/coffeebatch returned 201, expected 400 or 422. Content:
b'{"batch_number":"ppohps","cupping_score":621.31,
"id":"test_null_reject","origin_country":"agdspwgotyrijkz",
"roast_date":"2025-10-15 13:54:07","roast_level":"dark",
"status":null,"weight_kg":784.8}\n');
Non-nullable test for field Subscription.preferred_roast_level
failed (should reject null) (Non-nullable field
preferred_roast_level should reject null, POST /subscription
returned 201, expected 400 or 422. Content:
b'{"customer_email":"piudmmyzwxlwwbirkyboymbidivkudgbkunhk",
"id":"test_null_reject","monthly_shipment_quantity_kg":422.26,
"preferred_roast_level":null,
"subscription_start_date":"2021-03-07 17:23:31",");
Non-nullable test for field Subscription.subscription_status failed
(should reject null) (Non-nullable field subscription_status should
reject null, POST /subscription returned 201, expected 400 or 422.
Content: b'{"customer_email":"qtjulaorvbugxafofrufyn",
"id":"test_null_reject","monthly_shipment_quantity_kg":514.02,
"preferred_roast_level":"light",
"subscription_start_date":"2025-07-30 13:51:29","subscription)
Note: the issues may be caused by issues that are indirectly
related to the tests, check your code carefully.
Reminder:
- Make sure to validate incoming data.
- Ensure run_server.sh takes care of loading any dependencies if
  there are any.
\end{verbatim}
\end{small}

\noindent Under \textbf{medium} feedback, the agent sees only test names and pass/fail counts, without request/response details:

\begin{small}
\begin{verbatim}
I ran some tests and, according to the tests, there are the
following issues:
2/171 tests failed.
Failed tests:
- test_coffeeproduction_fermentation_type_default_value
- test_coffeeproduction_supplier_deletion_sets_null
Note: the issues may be caused by issues that are indirectly
related to the tests, check your code carefully.
\end{verbatim}
\end{small}

\noindent Under \textbf{minimal} feedback, the agent receives only:

\begin{small}
\begin{verbatim}
I ran some tests and, according to the tests, there are the
following issues:
One or more issues were detected, no further information is
available. Double-check and test carefully.
Note: the issues may be caused by issues that are indirectly
related to the tests, check your code carefully.
\end{verbatim}
\end{small}

\subsection{Programmatic Specification Example}
\label{app:programmatic-spec}

When using the programmatic sampler instead of the LLM-based sampler, entity and field names are generated programmatically rather than following a coherent domain narrative. Here is a real initial spec from a deterministic scenario:

\begin{small}
\begin{verbatim}
# Specifications

The system has 2 entities, 0 connections (relationships) between
entities, and 2 analytics endpoints.

# Enum Types

Enum values are case-sensitive and must match exactly as listed.

- DogPhase: allowed values are in_progress, review, planning
- PageLevel: allowed values are beginner, advanced, intermediate

# Entities

## Dog
(/dog endpoint)
Dog entity has the following fields:
- active_first_secondary of type required, nullable boolean
- deleted_shared_unique of type required, nullable boolean
- enabled_local_raw of type required, nullable boolean
- id of type optional, nullable string, unique
- level_last of type required, nullable integer
- phase of type required, nullable DogPhase enum,
  default "in_progress"
- rank_root_private of type required, nullable integer

We should be able to access a specific dog via
GET /dog/{id} endpoint

## Page
(/page endpoint)
Page entity has the following fields:
- active_unique of type required, nullable boolean
- id of type optional, nullable string, unique
- level of type required, nullable PageLevel enum,
  default "beginner"
- name_external of type required, nullable string
  (max length 100 inclusive)
- priority_raw of type required, nullable integer
  (between 0 and 10000 inclusive)
- public_private of type required, nullable boolean

We should be able to access a specific page via
GET /page/{id} endpoint

# Analytics Endpoints

All analytics endpoints must return a JSON object with a single
"result" key, e.g. {"result": 42} or {"result": null} when there
is no data. When there are no matching entities: count returns 0,
sum returns 0, avg/min/max return null.

## page priority_raw sum
GET /analytics/page/priority_raw/sum endpoint
goes over all possible page entities in the data and sums their
priority_raw.

## page priority_raw max
GET /analytics/page/priority_raw/max endpoint
goes over all possible page entities in the data and finds the
maximum priority_raw.

# Actions

Actions are triggered via POST endpoints and transition an
entity's enum field between values.

## Dog.approve
POST /dog/{id}/approve endpoint
Can be called when the dog's phase is review.
On success, sets phase to in_progress.
Returns 400 if phase is not one of: review.

## Dog.archive
POST /dog/{id}/archive endpoint
Can be called when the dog's phase is in_progress, planning.
On success, sets phase to review.
Returns 400 if phase is not one of: in_progress, planning.

## Page.reopen
POST /page/{id}/reopen endpoint
Can be called when the page's level is beginner.
On success, sets level to advanced.
Returns 400 if level is not one of: beginner.

# Validation Requirements

- Non-nullable fields must reject null values with 400 or 422.
- Required fields must be present in POST requests; omitting them
  should return 400 or 422.
- Reference fields must validate that referenced entity IDs exist.
  Return 400 or 422 for invalid references.
- List reference fields default to an empty list [] when not
  provided. They cannot be null.
- Enum fields must only accept the exact values listed
  (case-sensitive). Return 400 or 422 for invalid values.
- Numeric fields must respect minimum/maximum constraints.
  Return 400 or 422 for out-of-range values.
- String fields must respect min_length/max_length constraints.
  Return 400 or 422 for violations.

# Testing Requirements

## Reset Endpoint
The system must implement a POST /reset endpoint for testing
purposes. This endpoint should reset the database to an empty
state, clearing all entities. The automated test suite relies on
this endpoint to ensure test isolation.
Expected response: {"status": "reset"} with status code 200 or 204.
\end{verbatim}
\end{small}

\noindent A corresponding deterministic change request for this scenario:

\begin{small}
\begin{verbatim}
Add optional non-nullable field "end_external_public" of type
datetime to entity "Dog"; Change constraints on
"Page.name_external": set max length to 90 inclusive; Add new
entity "Comment" with fields: height_legacy (required, nullable
float), credits_previous (required, nullable float),
username_secondary (required, nullable string (max length 255
inclusive)), active_backup_verified (required, nullable bool),
level_backup (required, nullable int (between 0 and 10000
inclusive)); Delete analytics endpoint
"/analytics/page/priority_raw/sum"; Add optional field "comment"
of type reference to Comment, nullable to entity "Page"
\end{verbatim}
\end{small}

\subsection{Failure Category Examples}
\label{app:failure-examples}

This section catalogs concrete examples of each failure category referenced in \cref{sec:analysis} and in the heatmaps of \cref{app:heatmaps}. Each example cites the configuration, the change that triggered the failure, the relevant code excerpt, and the resulting test feedback. Where relevant, we also cite the README instruction (provided to every agent at turn 0) that should have prevented the failure.

\paragraph{Missing feature.} The agent did not implement one or more parts of the requested change. \emph{Kimi CLI + Kimi K2.5, Scenario 2, Turn 2.} The change requested deleting \texttt{Patient.owner\_email}, deleting \texttt{Appointment.priority}, and adding two analytics endpoints. The agent kept validating and storing \texttt{owner\_email}:

\begin{small}
\begin{verbatim}
# Validate owner_email
owner_email, error = validate_string(body.get('owner_email'),
                                     'owner_email', max_length=255)
if error:
    return jsonify({"error": error}), 400

patient = {
    'id': patient_id, 'name': name, 'species': species, 'age_years': age_years,
    'owner_email': owner_email,   # should have been deleted
    'health_status': health_status,
}
\end{verbatim}
\end{small}

The \texttt{priority} field was also still present, and neither analytics endpoint was added. Test feedback: \texttt{Analytics endpoint /analytics/patient/age\_years/avg test failed (returned 404, expected 200); Field deletion test failed for Patient.owner\_email}.

\paragraph{Hallucinated feature.} The agent implements validation or behavior that was never requested---plausible-sounding constraints invented from domain knowledge. \emph{Mini-SWE + Devstral Small 2, Scenario 10, Turn 3.} The spec defines \texttt{ActivityLog.activity\_type} as a plain string and \texttt{duration\_minutes} as a float in $[0.0, 480.0]$. The agent invented an enum restriction and tightened the numeric range:

\begin{small}
\begin{verbatim}
if 'activity_type' not in data:
    errors.append("Missing required field: activity_type")
elif not isinstance(data['activity_type'], str):
    errors.append("Field 'activity_type' must be a string")
elif data['activity_type'] not in ['walk', 'play', 'feed', 'groom']:
    errors.append("Field 'activity_type' must be one of: walk, play, feed, groom")

# ...later in validation:
elif data['duration_minutes'] < 1.0 or data['duration_minutes'] > 120.0:
    errors.append("Field 'duration_minutes' must be between 1.0 and 120.0 inclusive")
\end{verbatim}
\end{small}

Neither constraint appears in the spec. The hallucinated enum persisted across all 11 retry attempts. Test feedback: \texttt{Valid complete data rejected, POST /activity returned 400, expected 200 or 201. Content: \{"errors": ["Field 'activity\_type' must be one of: walk, play, feed, groom", "Field 'duration\_minutes' must be between 1.0 and 120.0 inclusive"]\}}.

\paragraph{Validation too loose.} The agent's validation logic accepts invalid input that should be rejected---most often by conflating ``field omitted'' with ``field explicitly set to null.'' \emph{Kimi CLI + Kimi K2.5, Scenario 1, Turn 0.} The spec says \texttt{FoodItem.category} is required, non-nullable, with default \texttt{"packaged"}. When the field is explicitly sent as \texttt{null}, the agent's validator silently replaces it with the default:

\begin{small}
\begin{verbatim}
def validate_enum(value, field_name, allowed_values, default=None):
    """Validate enum field"""
    if value is None:
        if default is not None:
            return default, None          # BUG: null silently replaced with default
        return None, f"{field_name} is required"
    if value not in allowed_values:
        return None, f"{field_name} must be one of {allowed_values}"
    return value, None

# Called as:
cat_val, err = validate_enum(
    data.get('category'), 'category', FOOD_CATEGORIES,
    default='packaged' if not is_update else None,
)
\end{verbatim}
\end{small}

README instruction ignored: \emph{``Fields can be null if they are nullable, otherwise they cannot be. If a field is non-required and non-nullable, it can be omitted, but when not omitted it cannot be null.''} The README distinguishes ``omitted'' (use default) from ``explicitly null'' (reject if non-nullable); the agent conflated the two. Test feedback: \texttt{Non-nullable field category should reject null, POST /fooditem returned 201, expected 400 or 422}.

\paragraph{Cascade deletion.} When a referenced entity is deleted, the agent fails to update or nullify references pointing to it. \emph{Kimi CLI + Kimi K2.5, Scenario 19, Turn 1.} The agent correctly validates the reference on creation but leaves the delete handler incomplete:

\begin{small}
\begin{verbatim}
@app.route('/glazerecipe/<entity_id>', methods=['DELETE'])
def delete_glazerecipe(entity_id):
    if entity_id not in DB['glazerecipe']:
        return jsonify({"error": "Not found"}), 404
    del DB['glazerecipe'][entity_id]
    # Missing: should set primary_glaze_recipe to null in all
    # CeramicFiringSession entities that reference this GlazeRecipe
    return jsonify({"status": "deleted"}), 200
\end{verbatim}
\end{small}

README instruction ignored: \emph{``Deleting an object from the database should also update any objects which have a reference or reference list referring to the deleted object, with references set to null, and reference lists updated by removing the non-existent reference.''} Test feedback: \texttt{Reference field 'primary\_glaze\_recipe' should be null after deletion, got: ref\_to\_delete}.

\paragraph{Rename failure.} The agent over-applies or misapplies a rename, often renaming fields along with the entity. \emph{Kimi CLI + Kimi K2.5, Scenario 3, Turn 1.} The change requested renaming the \emph{entity} \texttt{Subscription} to \texttt{MealPlan}, with field names preserved. The agent also renamed \texttt{subscription\_tier} to \texttt{mealplan\_tier}:

\begin{small}
\begin{verbatim}
# mealplan_tier - required, non-nullable enum, default "basic"
if 'mealplan_tier' in data:
    if data['mealplan_tier'] is None:
        errors.append("mealplan_tier cannot be null")
    elif data['mealplan_tier'] not in MEALPLAN_TIERS:
        errors.append(f"mealplan_tier must be one of: {', '.join(MEALPLAN_TIERS)}")
    else:
        validated['mealplan_tier'] = data['mealplan_tier']
else:
    validated['mealplan_tier'] = "basic"
\end{verbatim}
\end{small}

README instruction ignored: \emph{``When an entity is renamed, do not rename other entities' fields that refer to the renamed entity. For example, if Dog has a field `cats' referencing Cat, and Cat is renamed to CatSuper, the Dog field stays named `cats' --- only the endpoint changes from /cat to /catsuper.''} Test feedback: \texttt{Default value test for field MealPlan.subscription\_tier failed (expected default: basic)}.

\paragraph{Regression.} The agent breaks previously working functionality while implementing new changes. \emph{Mini-SWE + Devstral 2, Scenario 8, Turn 1.} After renaming \texttt{ClimbingRoute} to \texttt{Route}, a generic catch-all route handler still accepts any entity name:

\begin{small}
\begin{verbatim}
database = {
    'route': {},      # Correctly renamed from 'climbingroute'
    'member': {}
}

@app.route('/<entity>', methods=['GET'])
def list_entities(entity: str):
    """List all entities of a type"""
    entities = database.get(entity, {})
    return jsonify(list(entities.values())), 200   # Returns 200 for ANY path
\end{verbatim}
\end{small}

After the rename, the old \texttt{/climbingroute} endpoint should return 404. But the generic handler returns an empty list with status 200. Test feedback: \texttt{GET /climbingroute returned 200, expected 404 (entity should be deleted)}.

\paragraph{Wrong endpoint.} The agent registers an action or resource at a different URL path than the benchmark expects. \emph{Mini-SWE + Devstral 2, Scenario 0, Turn 8.} The change requested a \texttt{complete} action on \texttt{Order}; the expected endpoint was \texttt{POST /order/\{id\}/complete}. The agent built a dispatcher at a generic path instead:

\begin{small}
\begin{verbatim}
class OrderAction(str, Enum):
    complete = "complete"

class OrderActionRequest(BaseModel):
    action: OrderAction

@app.post("/order/{order_id}/action")
def order_action(order_id: str, action_request: OrderActionRequest):
    if action_request.action == "complete":
        return complete_order(order_id)
    else:
        raise HTTPException(status_code=400, detail=f"Unknown action: ...")
\end{verbatim}
\end{small}

README instruction ignored: \emph{``Actions are POST endpoints at /\{entity\}/\{id\}/\{action\_name\} that transition an enum field between allowed values.''} Test feedback: \texttt{Action complete failed, got 404, content: '\{"detail":"Not Found"\}'}.

\paragraph{Wrong response format.} The agent returns an incorrect HTTP status code or response body. \emph{Mini-SWE + GLM-5, Scenario 1, Turn 8.} For a wrong-state action call, the agent returned 422 instead of the required 400 or 409:

\begin{small}
\begin{verbatim}
@app.route('/fooddonor/<id>/approve_donor', methods=['POST'])
def fooddonor_approve_donor(id):
    if id not in db['fooddonor']:
        return jsonify({"error": "Not found"}), 404
    entity = db['fooddonor'][id]
    if entity.get('verification_status') != 'pending':
        return jsonify({"error": "Can only approve donors with "
                                  "verification_status 'pending'"}), 422
    entity['verification_status'] = 'verified'
    return jsonify(entity), 200
\end{verbatim}
\end{small}

README instruction ignored: \emph{``Status codes must be: 404 --- entity not found; 400 --- the entity's current enum value is not in the allowed `from' values for this action; 409 --- the enum value is valid for the action, but a guard condition is not satisfied.''} Test feedback: \texttt{Expected 400 or 409 for wrong status, got 422}.

\paragraph{Type error.} The agent returns a value with the wrong precision or type. \emph{Mini-SWE + Devstral 2, Scenario 4, Turn 0.} The analytics average endpoint should return the exact float; the agent rounds it:

\begin{small}
\begin{verbatim}
@app.get("/analytics/serviceprovider/average_rating/avg")
def avg_serviceprovider_average_rating():
    providers = database["serviceprovider"].values()
    if not providers:
        return {"result": None}
    ratings = [p["average_rating"] for p in providers
               if p.get("average_rating") is not None]
    if not ratings:
        return {"result": None}
    avg = sum(ratings) / len(ratings)
    return {"result": round(avg, 2)}    # BUG: rounds to 2 decimal places
\end{verbatim}
\end{small}

The spec says ``returns float'' with no mention of rounding. Test feedback: \texttt{Analytics calculation incorrect: got 586.84, expected 586.8433333333334}.

\paragraph{Default value.} The agent treats a field with a default as strictly required, rejecting requests that omit it. \emph{Mini-SWE + Devstral Small 2, Scenario 5, Turn 1.} The change added \texttt{watering\_frequency} as required non-nullable with default \texttt{"daily"}. The agent lists it as strictly required:

\begin{small}
\begin{verbatim}
required_fields = ['id', 'plant_name', 'soil_ph_maximum',
                   'soil_ph_minimum', 'watering_frequency']
for field in required_fields:
    if field not in data:
        errors.append(f"Missing required field: {field}")
\end{verbatim}
\end{small}

README instruction ignored: \emph{``If a required field has a default and is omitted from the POST body, the server should use the default value.''} Test feedback: \texttt{POST /growingguide without watering\_frequency returned 400, expected 200 or 201 (field has default value)}.

\paragraph{Server crash.} An unhandled exception returns 500 instead of a proper validation error---typically a missing type/null check before calling a parser. \emph{Mini-SWE + Devstral 2, Scenario 17, Turn 0.} The agent's datetime validator crashes on \texttt{None} or \texttt{int} input:

\begin{small}
\begin{verbatim}
def validate_datetime(dt_str):
    """Validate datetime format: YYYY-MM-DD hh:mm:ss"""
    if not re.match(r'^\d{4}-\d{2}-\d{2} \d{2}:\d{2}:\d{2}$', dt_str):
        # crashes if dt_str is None or int
        return False
    try:
        datetime.strptime(dt_str, '%
        return True
    except ValueError:
        return False

# Called without null/type check:
if not validate_datetime(data['brew_date']):
    errors.append("brew_date: Must be in YYYY-MM-DD hh:mm:ss format")
\end{verbatim}
\end{small}

README instruction ignored: \emph{``Handle the missing required values, values of incorrect type (e.g.\ an integer cannot be passed where a string is expected), values out of range, references to non-existent objects, etc.\ correctly by returning proper error codes.''} Test feedback: \texttt{POST /batch returned 500, expected 400 or 422. Content: 'TypeError: expected string or bytes-like object, got NoneType'}.

\paragraph{Stuck loop.} The agent enters a repetitive cycle, performing the same or very similar actions without making progress. Common with weaker models on OpenHands, where the stuck detector eventually terminates the session. \emph{OpenHands + Devstral 2, Scenario 18, Turn 1.} While performing field renames, the agent fell into an alternating \texttt{str\_replace} loop on the same region of \texttt{app.py}:

\begin{small}
\begin{verbatim}
# Iter 1: str_replace validate_salestransaction_data -> validate_buyer_data
# Iter 2: str_replace validate_buyer_data -> validate_salestransaction_data
# Iter 3: str_replace validate_salestransaction_data -> validate_buyer_data
# ... repeats indefinitely
\end{verbatim}
\end{small}

OpenHands' stuck detector eventually caught the pattern: \texttt{Alternating Action, Observation loop detected}. All 11 retries re-entered the same loop; the scenario completed only 1/101 turns.

\paragraph{Self-kill \texttt{pkill}.} The agent runs a broad \texttt{pkill} or \texttt{killall} pattern intending to restart its server, and the pattern also kills its own harness process (whose command line contains \texttt{server.py}). \emph{OpenHands + Devstral Small 2, Scenario 19.} The agent issued:

\begin{small}
\begin{verbatim}
{"command": "pkill -f \"python server.py\""}
\end{verbatim}
\end{small}

The OpenHands agent process was launched with a task prompt quoting \texttt{server.py} (e.g., ``Implement a REST API server\ldots''), so \texttt{pkill -f} matched the agent's own command line and killed it (exit code 143 = SIGTERM). README instruction ignored: \emph{``When killing processes with pkill, use exact PID instead of regexes whenever possible to avoid killing your own coding agent process.''} OpenHands' \texttt{NeverConfirm} policy re-executed the pending action on resume, so all retries immediately died.

\paragraph{Invalid tool call.} The model emits tool calls with missing required parameters or malformed arguments, and may loop on failed calls. \emph{QwenCode + Qwen3-Coder-Next, Scenario 3, Turn 4.} After deleting \texttt{server.py} intending to rewrite it, the agent repeatedly issued malformed tool calls:

\begin{small}
\begin{verbatim}
// Attempt 1: write_file without 'content' parameter
{"type":"tool_use", "name":"write_file",
 "input":{"file_path":"/workspace/server.py"}}
{"type":"tool_result", "is_error":true,
 "content":"params must have required property 'content'"}

// Attempts 2--N: run_shell_command with empty input
{"type":"tool_use", "name":"run_shell_command", "input":{}}
{"type":"tool_result", "is_error":true,
 "content":"params must have required property 'command'"}
// ... repeats until session timeout
\end{verbatim}
\end{small}

This pattern repeated across all 11 retry attempts of turn 4. Session stats show 79 total tool calls, 16 failures (7 \texttt{write\_file}, 9 \texttt{run\_shell\_command}). Since \texttt{server.py} was never recreated, the server could not start and the scenario terminated.

\section{Agent Harness Details}
\label{app:agents}

All harnesses run inside isolated Docker containers with full shell access. Unless noted, harnesses were used without modification. Separately from the test-feedback retry budget $R$, we retry each agent call up to \Nharnessretries{} times if the harness process itself fails (e.g.\ due to connection issues, a malformed tool call, or the agent accidentally \texttt{pkill}-ing its own process). These infrastructure-level retries do not count against $R$.

We use \textbf{OpenCode}~\citep{opencode} and \textbf{OpenHands}~\citep{openhands} as-is. We modify \textbf{Mini-SWE}~\citep{miniswe}---a lightweight single-tool agent that issues shell commands via one tool call---to add context compression (compressing conversation history when token count exceeds a configurable threshold) so that long scenarios do not overflow the context window. Where available, we additionally test \textbf{model-provider agents}, used as-is: QwenCode~\citep{qwencode} (Alibaba) for Qwen models, Kimi CLI~\citep{kimicli} (Moonshot) for Kimi K2.5, and Mistral Vibe~\citep{mistralvibe,mistral2025devstral2} for Devstral models.

\cref{tab:harness-versions} lists the exact harness versions used in our experiments.

\begin{table}[h]
\centering
\small
\caption{Harness versions used in all experiments.}
\label{tab:harness-versions}
\begin{tabular}{ll}
\toprule
Harness & Version \\
\midrule
OpenCode    & 1.4.3 \\
Mini-SWE    & 2.0.0a1 \\
OpenHands   & 1.12.1 \\
Mistral Vibe & 2.7.6 \\
Kimi CLI    & 1.34.0 \\
QwenCode    & 0.10.5 \\
\bottomrule
\end{tabular}
\end{table}

\section{Additional Results}
\label{app:additional-results}

\begin{table}[h]
\centering
\small
\caption{Avg turns passed $\pm$ SE (pass rate \%) per model--harness combination, @$R=10$ retries. --- = not run.}
\label{tab:oss-coverage-budget11}
\begin{tabular}{lcccc}
\toprule
Model & OpenCode & Mini-SWE & OpenHands & \shortstack{Model Provider\\Agent} \\
\midrule
Devstral 2 & \phantom{0}9.1\,${\scriptstyle\pm\,1.2}$ (\phantom{0}0\%) & \phantom{0}8.9\,${\scriptstyle\pm\,1.2}$ (\phantom{0}0\%) & 3.5\,${\scriptstyle\pm\,0.6}$ (0\%) & 13.7\,${\scriptstyle\pm\,3.3}$ (\phantom{0}0\%) \\
Devstral Small 2 & 14.8\,${\scriptstyle\pm\,3.6}$ (\phantom{0}0\%) & \phantom{0}4.8\,${\scriptstyle\pm\,0.9}$ (\phantom{0}0\%) & 3.5\,${\scriptstyle\pm\,0.7}$ (0\%) & 17.1\,${\scriptstyle\pm\,4.7}$ (\phantom{0}0\%) \\
GLM-5 & 82.5\,${\scriptstyle\pm\,7.3}$ (65\%) & 17.2\,${\scriptstyle\pm\,2.0}$ (\phantom{0}0\%) & 9.2\,${\scriptstyle\pm\,1.3}$ (0\%) & --- \\
Kimi K2.5 & 49.5\,${\scriptstyle\pm\,6.9}$ (10\%) & 19.1\,${\scriptstyle\pm\,3.7}$ (\phantom{0}0\%) & 7.5\,${\scriptstyle\pm\,1.2}$ (0\%) & 57.1\,${\scriptstyle\pm\,8.5}$ (30\%) \\
Nemotron Super & \phantom{0}5.0\,${\scriptstyle\pm\,1.2}$ (\phantom{0}0\%) & \phantom{0}2.8\,${\scriptstyle\pm\,0.7}$ (\phantom{0}0\%) & 1.8\,${\scriptstyle\pm\,0.2}$ (0\%) & --- \\
Qwen3-Coder-Next & 14.3\,${\scriptstyle\pm\,3.1}$ (\phantom{0}0\%) & \phantom{0}9.2\,${\scriptstyle\pm\,2.3}$ (\phantom{0}0\%) & 4.0\,${\scriptstyle\pm\,0.6}$ (0\%) & \phantom{0}5.8\,${\scriptstyle\pm\,0.9}$ (\phantom{0}0\%) \\
Qwen3.5-122B & 61.0\,${\scriptstyle\pm\,7.9}$ (35\%) & 52.6\,${\scriptstyle\pm\,9.4}$ (25\%) & 9.6\,${\scriptstyle\pm\,1.1}$ (0\%) & 49.0\,${\scriptstyle\pm\,7.5}$ (10\%)
 \\
\bottomrule
\end{tabular}
\end{table}

\begin{table}[h]
\centering
\small
\caption{Avg turns passed $\pm$ SE (pass rate \%) per model under three feedback levels (OpenCode, @$R=2$ retries).}
\label{tab:feedback-ablation}
\begin{tabular}{lccc}
\toprule
Model & Detailed & Medium & Minimal \\
\midrule
Devstral 2 & \phantom{0}4.8\,${\scriptstyle\pm\,0.8}$ (\phantom{0}0\%) & \phantom{0}2.0\,${\scriptstyle\pm\,0.6}$ (\phantom{0}0\%) & \phantom{0}0.3\,${\scriptstyle\pm\,0.1}$ (0\%) \\
Devstral Small 2 & \phantom{0}5.5\,${\scriptstyle\pm\,2.0}$ (\phantom{0}0\%) & \phantom{0}2.5\,${\scriptstyle\pm\,0.9}$ (\phantom{0}0\%) & \phantom{0}1.4\,${\scriptstyle\pm\,0.6}$ (0\%) \\
GLM-5 & 57.0\,${\scriptstyle\pm\,8.6}$ (25\%) & 32.4\,${\scriptstyle\pm\,6.7}$ (10\%) & 10.7\,${\scriptstyle\pm\,3.2}$ (0\%) \\
Kimi K2.5 & 26.8\,${\scriptstyle\pm\,5.5}$ (\phantom{0}0\%) & 18.1\,${\scriptstyle\pm\,3.6}$ (\phantom{0}0\%) & \phantom{0}4.6\,${\scriptstyle\pm\,1.0}$ (0\%) \\
Nemotron Super & \phantom{0}2.9\,${\scriptstyle\pm\,1.0}$ (\phantom{0}0\%) & \phantom{0}1.6\,${\scriptstyle\pm\,0.6}$ (\phantom{0}0\%) & \phantom{0}0.3\,${\scriptstyle\pm\,0.2}$ (0\%) \\
Qwen3-Coder-Next & \phantom{0}7.6\,${\scriptstyle\pm\,1.8}$ (\phantom{0}0\%) & \phantom{0}3.9\,${\scriptstyle\pm\,0.8}$ (\phantom{0}0\%) & \phantom{0}1.8\,${\scriptstyle\pm\,0.4}$ (0\%) \\
Qwen3.5-122B & 39.4\,${\scriptstyle\pm\,7.1}$ (10\%) & 12.1\,${\scriptstyle\pm\,3.8}$ (\phantom{0}0\%) & \phantom{0}2.8\,${\scriptstyle\pm\,0.9}$ (0\%)
 \\
\bottomrule
\end{tabular}
\end{table}

\subsection{Results Variance Analysis}
\label{app:variance}

Running several repetitions of every configuration would be prohibitively costly. Instead, we re-run OpenCode + Qwen3.5-122B five times on the same 20 scenarios (only the model's sampling randomness differs between runs) and treat the resulting spread as an estimate of measurement noise for the rest of the paper. \cref{tab:variance} reports the results.

\begin{table}[h]
\centering
\small
\caption{Variance across five runs of OpenCode + Qwen3.5-122B with detailed feedback on the same 20 scenarios. We report the average number of turns passed under two attempt budgets: \textbf{@3} (the default setting, $R=\Nretries$ retries per turn) and \textbf{@11} (the maximum budget we evaluated, 11 attempts per turn). Per-run values show $\pm$1 SE across scenarios; the Mean row shows $\pm$1 SD across runs.}
\label{tab:variance}
\begin{tabular}{lcccc}
\toprule
Run & Avg Turns (@3) & Avg Turns (@11) & Avg First Failure & Pass Rate \\
\midrule
Run 1 & 29.9\,${\scriptstyle\pm\,5.3}$ & 48.4\,${\scriptstyle\pm\,7.5}$ & 0.8\,${\scriptstyle\pm\,0.3}$ & 15\%\,${\scriptstyle\pm\,8\%}$ \\
Run 2 & 39.4\,${\scriptstyle\pm\,7.1}$ & 61.0\,${\scriptstyle\pm\,7.9}$ & 1.1\,${\scriptstyle\pm\,0.4}$ & 35\%\,${\scriptstyle\pm\,11\%}$ \\
Run 3 & 32.5\,${\scriptstyle\pm\,6.6}$ & 53.9\,${\scriptstyle\pm\,8.0}$ & 0.9\,${\scriptstyle\pm\,0.2}$ & 25\%\,${\scriptstyle\pm\,10\%}$ \\
Run 4 & 37.0\,${\scriptstyle\pm\,7.6}$ & 64.0\,${\scriptstyle\pm\,7.9}$ & 0.8\,${\scriptstyle\pm\,0.4}$ & 35\%\,${\scriptstyle\pm\,11\%}$ \\
Run 5 & 23.3\,${\scriptstyle\pm\,3.8}$ & 58.3\,${\scriptstyle\pm\,9.2}$ & 1.1\,${\scriptstyle\pm\,0.4}$ & 30\%\,${\scriptstyle\pm\,11\%}$ \\
\midrule
Mean & 32.4\,${\scriptstyle\pm\,6.3}$ & 57.1\,${\scriptstyle\pm\,6.2}$ & 1.0\,${\scriptstyle\pm\,0.1}$ & 28\%\,${\scriptstyle\pm\,8\%}$
 \\
\bottomrule
\end{tabular}
\end{table}

\subsection{Failure-Type Heatmaps}
\label{app:heatmaps}

\paragraph{Failure categories.} The failure axis in each heatmap uses the following categories (concrete code-level examples for each are in \cref{app:failure-examples}):
\begin{itemize}
    \item \textbf{Missing feature.} The agent did not implement one or more parts of the requested change.
    \item \textbf{Hallucinated feature.} The agent implements behavior that was never requested---plausible-sounding features invented from domain knowledge.
    \item \textbf{Data Validation error.} The agent's validation accepts invalid input that should be rejected (often by conflating ``omitted'' with ``explicitly null''), or rejects input that should be accepted (e.g. having an overly strict regex for emails).
    \item \textbf{Cascade deletion.} When a referenced entity is deleted, references pointing to it are not updated or nullified.
    \item \textbf{Rename failure.} A rename is over- or mis-applied (e.g.\ entity-level rename leaks into field names).
    \item \textbf{Regression.} New changes break previously working functionality.
    \item \textbf{Wrong endpoint.} An action or resource is registered at a different URL path than the spec requires.
    \item \textbf{Type error.} A returned value has the wrong precision or type (e.g.\ rounding a float that must be exact).
    \item \textbf{Default value.} An error involving field with a declared default, e.g. value is treated as strictly required, rejecting requests that omit it.
    \item \textbf{Enum handling.} The agent mishandles enum-typed fields---e.g.\ accepts values outside the declared set, fails to update accepted values after a rename, or rejects valid values due to case mismatch.
    \item \textbf{Server crash.} Server crashed due to an internal error.
    \item \textbf{Stuck loop.} The agent enters a repetitive cycle of near-identical actions without making progress.
    \item \textbf{Self-kill \texttt{pkill}.} A broad \texttt{pkill}/\texttt{killall} pattern matches the agent's own harness process and kills it.
    \item \textbf{Invalid tool call.} The model emits tool calls with missing required parameters or malformed arguments.
\end{itemize}

\begin{figure}[p]
\centering
\includegraphics[width=\linewidth]{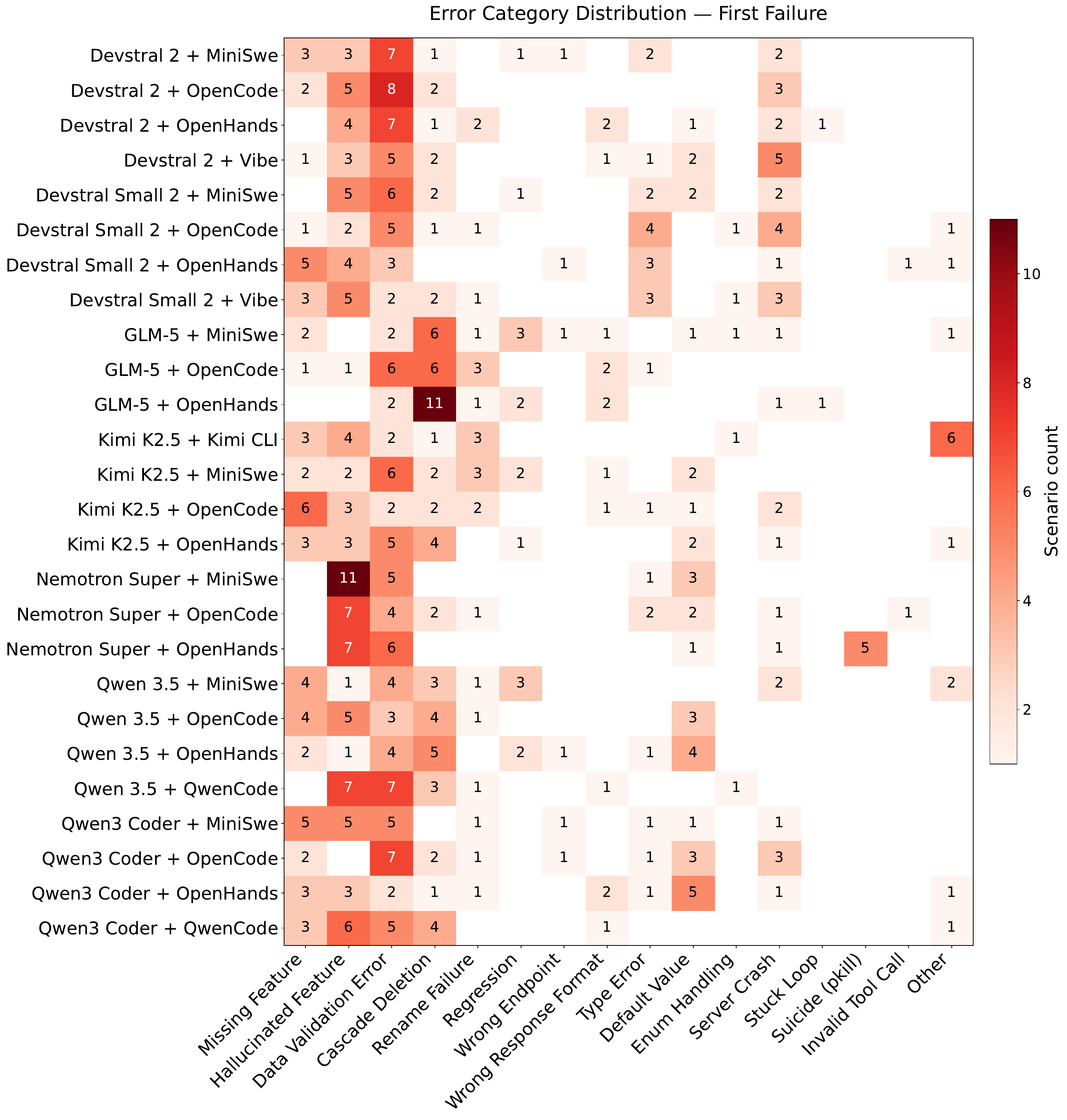}
\caption{Distribution of failure types at the turn each scenario \emph{first} fails. Rows are model--harness configurations, columns are failure categories, and each cell reports the number of scenarios (out of 20) in which that category caused the first failure.}
\label{fig:heatmap-first-failure}
\end{figure}

\begin{figure}[p]
\centering
\includegraphics[width=\linewidth]{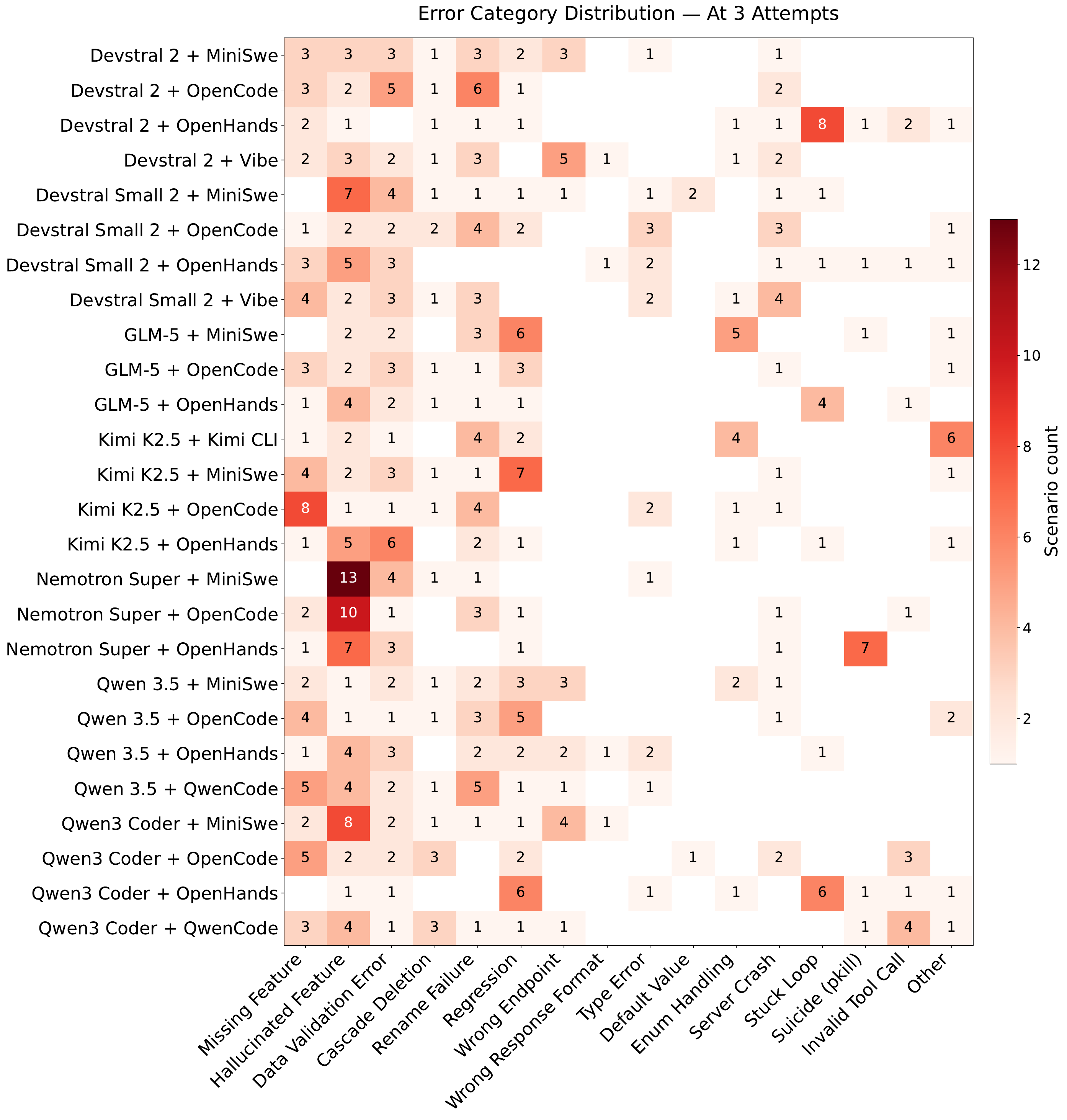}
\caption{Distribution of failure types across all failed turns under the default retry budget ($R=\Nretries$). Rows are model--harness configurations, columns are failure categories, and each cell reports the number of scenarios (out of 20) in which that category appeared at least once.}
\label{fig:heatmap-3-attempts}
\end{figure}

\begin{figure}[p]
\centering
\includegraphics[width=\linewidth]{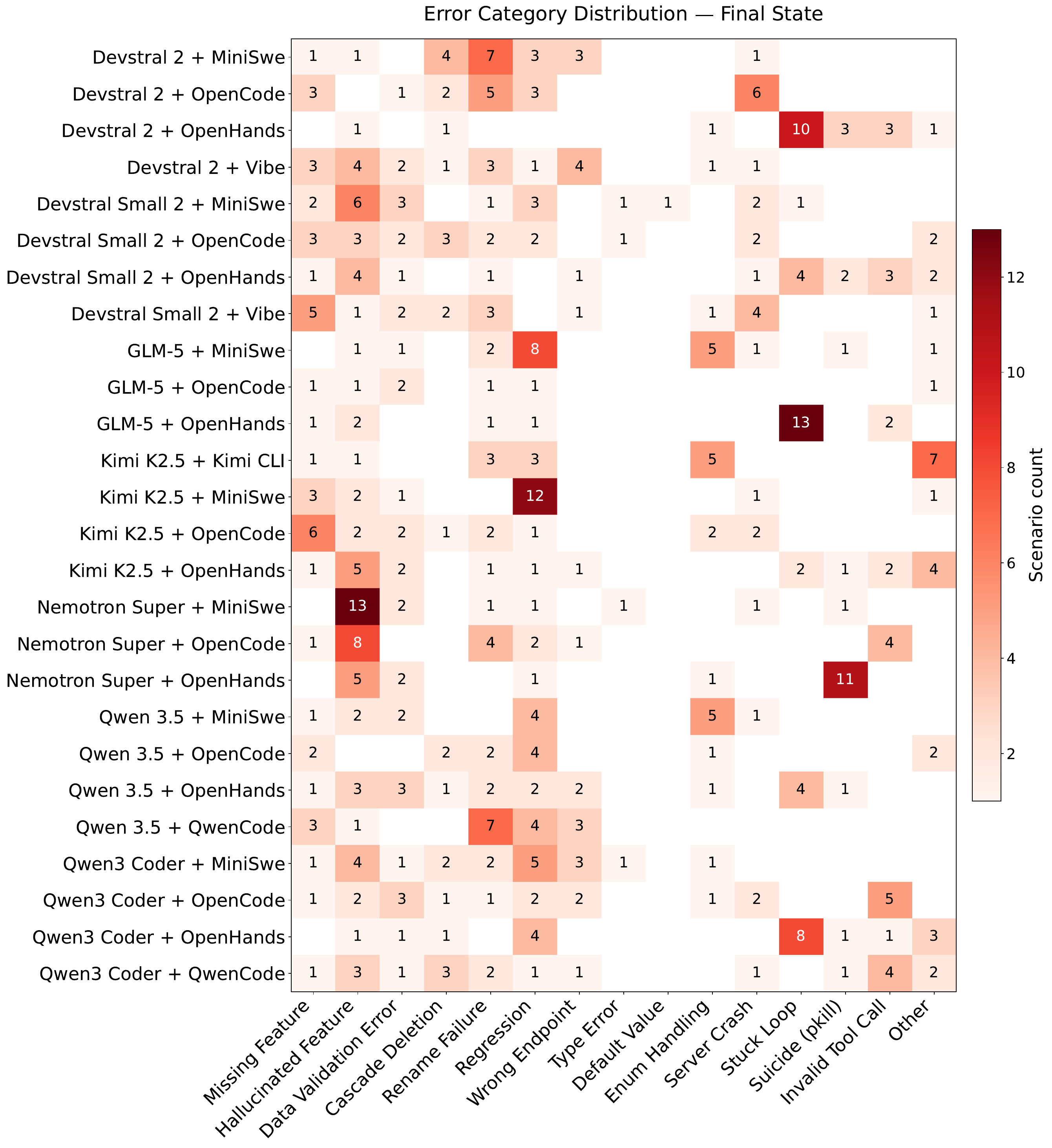}
\caption{Distribution of failure types across all failed turns under an extended retry budget of 11 attempts per turn. Comparing with \cref{fig:heatmap-3-attempts} shows which failure categories a larger retry budget eliminates (cells that drop) versus which ones persist.}
\label{fig:heatmap-final}
\end{figure}

\crefrange{fig:heatmap-first-failure}{fig:heatmap-final} decompose failures by type across every model--harness combination. In each heatmap, rows are model--harness configurations and columns are failure categories; each cell reports the number of scenarios (out of 20) in which that configuration hit that failure category. The three views differ only in the set of runs counted: (i)~failures observed at the turn each scenario \emph{first} fails, (ii)~all failures encountered under the default $R=\Nretries$ retry budget, and (iii)~all failures under an extended budget of 11 attempts per turn. Together they make it clear which failure modes are universal. Notably, at first failure and with the default 3-attempt budget, the dominant failures are \emph{implementation} issues---incomplete renames, cascade-deletion bugs, and missing analytics endpoints---whereas at 11 attempts, \emph{infrastructure} failures---agents killing their own harness process, tool-loop detection triggering, and malformed tool calls---become comparatively more visible, reflecting the longer runs that reach points where these modes can occur.

\makeatletter
  \ifpreprint
  \else
    \clearpage  %
    \input{checklist}
  \fi
\makeatother

\end{document}